\documentclass[11pt]{article}

\usepackage[matrix,arrow]{xy}

\usepackage{epsfig}

\usepackage[mathscr]{euscript}

\usepackage{amssymb,amsthm}
\usepackage{amsmath}
\usepackage{amscd}
\usepackage{latexsym}
\usepackage{graphics}
\usepackage{color}

\usepackage{graphicx}

\input{xy}
\xyoption{all}

\def\beq{\begin{equation}}
\def\eeq{\end{equation}}
\def\bea{\begin{eqnarray}}
\def\eea{\end{eqnarray}}

\catcode`@=11
\renewcommand{\section}{\@startsection{section}{1}{0pt}{\medskipamount}
{\medskipamount}{\large\bf}}
\numberwithin{equation}{section}
\catcode`@=12
\def\a{\alpha}
\def\b{\beta}
\def\g{\gamma}

\def\de{\delta}
\def\ve{\varepsilon}

\def\t{\theta}
\def\Th{\Theta}

\def\m{\mu}
\def\n{\nu}
\def\r{\rho}
\def\s{\varsigma}
\def\p{\phi}

\def\vk{\varkappa}

\def\ome{\omega}
\def\Om{\Omega}
\def\La{\Lambda}

\def\Ow{\widehat\Omega}
\def\ot{\widehat\omega}
\def\1{{\bar 1}}
\def\2{{\bar 2}}
\def\3{{\bar 3}}
\def\4{{\bar 4}}

\def\lrc{\,\lrcorner\,}
\def\hra{\,\hookrightarrow\,}

\newcommand{\yb}{\bar{y}}
\newcommand{\ab}{{\bar{\alpha}}}
\newcommand{\bb}{{\bar{\beta}}}

\newcommand{\zb}{\bar{z}}
\newcommand{\zeb}{\bar{\zeta}}

\newcommand{\HQ}{{\rm H}}

\newcommand{\su}{{{\rm SU}(2)}}
\newcommand{\suL}{{{\mathfrak{su}}(2)}}
\newcommand{\sutL}{{{\mathfrak{su}}(3)}}
\newcommand{\sut}{{{\rm SU}(3)}}
\newcommand{\uo}{{{\rm U}(1)}}
\newcommand{\uoL}{{{\mathfrak u}(1)}}

\newcommand{\suk}{{{\rm SU}(N)}}
\newcommand{\sukL}{{{\mathfrak{su}}(N)}}
\newcommand{\glrm}{{{\rm GL}}}

\newcommand{\urm}{{{\rm U}}}
\newcommand{\urmL}{{{\mathfrak u}}}
\newcommand{\surm}{{{\rm SU}}}
\newcommand{\surmL}{{{\mathfrak{su}}}}
\newcommand{\sorm}{{{\rm SO}}}

\newcommand{\slc}{{{\rm SL}(2,\C)}}

\newcommand{\sltc}{{{\rm SL}(3,\C)}}
\newcommand{\sltcL}{{{\mathfrak{sl}}(3,\C)}}

\newcommand{\mfrak}{{\mathfrak{m}}}
\newcommand{\gfrak}{{\mathfrak{g}}}
\newcommand{\hfrak}{{\mathfrak{h}}}

\newcommand{\tfrak}{{\mathfrak{t}}}

\newcommand{\Hom}{{\rm Hom}}
\newcommand{\End}{{\rm End}}

\newcommand{\one}{{\bf 1}}

\newcommand{\C}{\mathbb C}
\newcommand{\R}{\mathbb R}
\newcommand{\F}{\mathbb F}
\newcommand{\X}{\mathbb X}
\newcommand{\Z}{\mathbb Z}

\newcommand{\Acal}{{\cal A}}
\newcommand{\Ncal}{{\cal N}}
\newcommand{\Hcal}{{\cal H}}
\newcommand{\Fcal}{{\cal F}}
\newcommand{\Ecal}{{\cal E}}
\newcommand{\J}{{\cal J}}

\newcommand{\T}{{\cal T}}
\newcommand{\Pcal}{{\cal P}}
\newcommand{\Vcal}{{\cal V}}

\newcommand{\Scal}{{\cal S}}

\newcommand{\Rbar}{{\overline{R}}}
\newcommand{\Qbar}{{\overline{Q}}}
\newcommand{\Psfbar}{{\overline{\sf P}}}
\newcommand{\phibar}{{\overline{\phi}}}
\newcommand{\psibar}{{\overline{\psi}}}

\newcommand{\Csf}{{\sf C}}
\newcommand{\Psf}{{\sf P}}
\newcommand{\Rep}{{\mathscr{R}}}
\newcommand{\Gscr}{{\mathscr{G}}}
\newcommand{\Uscr}{{\mathscr{U}}}
\newcommand{\qvec}{{\vec q}}

\newcommand{\with}{{\qquad{\rm with}\qquad}}

\def\im{{\rm i}}
\def\N2{$N{=}2$}
\def\pa{\mbox{$\partial$}}
\def\diff{\mbox{d}}
\def\tr{{\rm tr}}
\def\sfrac#1#2{{\textstyle\frac{#1}{#2}}}
\def\>{\rangle}
\def\<{\langle}
\def\+{\dagger}
\def\={\ =\ }
\def\und{\qquad\textrm{and}\qquad}
\def\and{\quad\textrm{and}\quad}
\def\for{\quad\textrm{for}\quad}
\def\with{\quad\textrm{with}\quad}

\newcommand{\mbf}[1]{{\boldsymbol {#1} }}

\begin{document}
\begin{titlepage}
\setcounter{page}{0}
\begin{flushright}
HWM--10--31\\
EMPG--10--15\\
\end{flushright}

\vskip 2.5cm

\begin{center}

{\Large\bf Double quiver gauge
  theory \\[10pt] and nearly K\"ahler flux compactifications
}

\vspace{15mm}

{\large Alexander D. Popov${}^{1}$}
\ \ and \ \ {\large Richard J. Szabo${}^2$}
\\[5mm]
\noindent ${}^1${\em Bogoliubov Laboratory of Theoretical Physics, JINR\\
141980 Dubna, Moscow Region, Russia}
\\[5mm]
\noindent ${}^2${\em Department of Mathematics, Heriot-Watt University\\
Colin Maclaurin Building, Riccarton, Edinburgh EH14 4AS, U.K.}\\
and \\
{\em Maxwell Institute
  for Mathematical Sciences, Edinburgh, U.K.}\\[5mm]
{Email: {\tt popov@theor.jinr.ru, R.J.Szabo@ma.hw.ac.uk}}

\vspace{15mm}

\begin{abstract}
\noindent
We consider $G$-equivariant dimensional reduction of Yang-Mills theory
with torsion
on manifolds of the form $M\times G/H$ where $M$ is a smooth
manifold, and $G/H$
is a compact six-dimensional homogeneous space provided with a never integrable almost complex
structure and a family of SU(3)-structures which includes a nearly
K\"ahler structure. We establish an equivalence between
$G$-equivariant pseudo-holomorphic vector bundles on $M\times G/H$ and
new quiver bundles on $M$ associated to the double of a quiver $Q$,
determined by the $\sut$-structure, with relations ensuring the
absence of oriented cycles in $Q$. When $M=\R^2$, we describe an equivalence between
$G$-invariant solutions of Spin(7)-instanton equations on $M\times G/H$
and solutions of new quiver vortex equations on $M$. It is shown that generic invariant
Spin(7)-instanton configurations correspond to quivers $Q$ that
contain non-trivial oriented cycles.
\end{abstract}

\end{center}

\end{titlepage}

\section{Introduction and summary}

Realistic scenarios in string theory require compactification from
ten-dimensional spacetime to four dimensions along a compact
six-dimensional internal space $\X^6$. Initially, the cases where
$\X^6$ is a coset space $G/H$ were considered, usually with K\"ahler
structure. Subsequently, in the phenomenologically more interesting
theories of heterotic strings, Calabi-Yau spaces $\X^6$ were utilized.

However, it has been realized in recent years that Calabi-Yau
compactifications suffer from the presence of many massless moduli
fields in the resulting four-dimensional field theories; K\"ahler
cosets also lead to unrealistic effective field theories. This problem
can be cured at least partially by allowing for $p$-form fluxes on
$\X^6$. String vacua with $p$-form fields along the extra dimensions are
called ``flux compactifications'' and have been intensively studied in
recent years~\cite{10}. In particular, fluxes in heterotic string
theory were analysed in~\cite{9,11,12,13}. The allowed internal
manifolds $\X^6$ in this case are not K\"ahler, but include
quasi-K\"ahler and nearly K\"ahler manifolds~\cite{13}, and more
general almost hermitian manifolds with $\sut$-structure.

The analysis of Yang-Mills theory on product manifolds of the form
$M\times\X^6$, where $\X^6$ is a quasi-K\"ahler six-manifold, is of
great interest in this context. After constructuing a vacuum solution
in heterotic string theory, one should consider Kaluza-Klein
dimensional reduction of heterotic supergravity over $\X^6$ by expansion
around this background, and in particular describe the massless sector
of induced Yang-Mills-Higgs fields on $M$. The $G$-equivariant
dimensional reduction of Yang-Mills theory on K\"ahler homogeneous
spaces $G/H$, for a compact Lie group $G$ with a closed subgroup $H$,
was considered in~\cite{GP1,A-CG-P1,A-CG-P2}; the induced
Yang-Mills-Higgs theory on $M$ in this case is a quiver gauge
theory. This formalism was further developed and applied in a variety
of contexts in~\cite{PS1,LPS2,LPS3,Popov1,DS}.\footnote{Another way in
which quiver gauge theories arise as low-energy effective field
theories in string theory is through considering orbifolds $\X^6$ and
placing D-branes at the orbifold singularities~\cite{quiverG}.}

In this paper we will extend the formalism of $G$-equivariant
dimensional reduction from the K\"ahler case to the quasi-K\"ahler
case, with the coset space $G/H$ endowed with a never integrable
almost complex structure. We shall describe in detail the new quivers
with relations associated to $G$-equivariant pseudo-holomorphic vector
bundles on $G/H$, and establish a categorical equivalence between the
corresponding quiver bundles on $M$ and invariant pseudo-holomorphic
bundles on $M\times G/H$; this generalizes results of~\cite{A-CG-P1}
from the K\"ahler case and the example of nearly K\"ahler dimensional
reduction considered in~\cite{21}. We will spell out explicitly the
properties of these new types of quivers, quiver gauge theories, and
quiver vortex equations. These quiver gauge theories can have
applications in flux compactifications of heterotic string theory,
though these are left for future work.

For many explicit calculations we work on the six-dimensional complete
flag
manifold $G/H=\sut/\uo\times
\uo=:\F_3$ with its associated $\sut$-equivariant pseudo-holomorphic
homogeneous vector bundles. The requisite geometry and representation
theory are best understood for this case, and detailed results can be
obtained by following the formalism developed in~\cite{LPS3}. Moreover, in~\cite{LNP} it is shown that, out of the four
known compact nearly K\"ahler spaces in six dimensions, only $\F_3$
produces non-trivial heterotic string vacua. We shall demonstrate how
the $\sut$-equivariant gauge theory derived in~\cite{LPS3} changes
when one considers dimensional reduction over $\F_3$ with a
quasi-K\"ahler structure rather than a K\"ahler structure. For
instance, quiver vortex equations associated to the nearly K\"ahler
flag manifold $\F_3$ exhibit many qualitative differences compared to
those associated with the K\"ahler geometry of $\F_3$~\cite{LPS3}, as
first pointed out in~\cite{21}. 

To the nearly K\"ahler manifold $\F_3$ (as well as to the other three
compact nearly K\"ahler manifolds in six dimensions) one can associate
quivers and quiver gauge theories which differ from those in the
K\"ahler case. In particular, one can introduce two distinct almost
complex structures with corresponding fundamental two-forms
$\omega$. The Yang-Mills lagrangian after equivariant dimensional
reduction does not depend on the choice of (integrable or
non-integrable) almost complex structure; in a real frame, the
lagrangian and gauge field solutions of the equivariance conditions
retain the same form. We obtain in this way two distinct quiver gauge
theories by introducing an almost complex structure (and pulling it
back to the gauge bundles), and writing all fields in a complex
frame. Then the lagrangian can be rewritten as a ``Bogomol'ny square''
in two distinct ways using either a K\"ahler or a nearly K\"ahler
two-form $\omega$, leading to two different sets of first-order
hermitian Yang-Mills equations for the vacuum states of the quiver
gauge theory. Moreover, one can consider equivariant dimensional
reduction of Yang-Mills theory on homogeneous eight-manifolds
$\X^8=\R^2\times G/H$, with and without SU(4)-structure, to a quiver
gauge theory on $\R^2$; the corresponding reductions of the BPS
Yang-Mills equations have sharply different forms in the two
cases. Solutions of these equations in the former case can be used to
construct explicit vacua for heterotic supergravity and Yang-Mills
theory in flux compactifications on nearly K\"ahler (and other almost
hermitian) manifolds.

The organisation of the remainder of this paper is as follows. In
section~\ref{Homgen} we collect various geometrical facts concerning
coset spaces and $\sut$-structure manifolds in six dimensions. In
section~\ref{F3Kahler} we work out these geometrical structures
explicitly in the case of the six-manifold $\F_3$. In
section~\ref{Pseudo} we recall the construction of quivers with
relations and quiver gauge theories associated with holomorphic
homogeneous vector bundles over complex flag varieties $G/H$, and
discuss how they are modified in the pseudo-holomorphic case
corresponding to a never integrable almost complex structure on
$G/H$. In section~\ref{Doublereps} we describe in detail the new
quivers and relations from a purely algebraic perspective. In
section~\ref{EGTQB} we describe the corresponding quiver bundles and
connections. In section~\ref{QGT} we construct natural corresponding
quiver gauge theories and analyse their vacuum field equations. In
section~\ref{vortex} we consider BPS-type gauge field equations on
eight-manifolds $\R^2\times G/H$ with SU(4)-structure, and their
equivariant dimensional reduction to quiver vortex equations in two
dimensions, comparing with the analogous reductions of the usual
K\"ahler cases.

\section{Homogeneous spaces and $\mbf{{\rm SU}(n)}$-structure manifolds\label{Homgen}}

\noindent
{\bf Coset space geometry. \ } In this paper we will study dimensional
reduction of gauge theories over certain coset spaces which can be
used in flux compactifications of string theory. Let $G/H$ be a reductive homogeneous
manifold of dimension $h=\dim G-\dim H$, where $H$ is a closed subgroup of a compact semisimple Lie
group $G$ which contains a maximal
abelian subgroup of $G$. Then the Lie algebra $\mathfrak{g}$ of $G$ admits a decomposition
\begin{equation}\label{2.45}
{\mathfrak{g}}={\mathfrak{h}}\oplus{\mathfrak{m}}\ ,
\end{equation}
where $\mathfrak{h}$ is the Lie algebra of $H$, 
$\mathfrak{m}$ is an $H$-invariant subspace of $\gfrak$, i.e. 
$[{\mathfrak{h}},{\mathfrak{m}}]\subset{\mathfrak{m}}$, and we choose
$\mathfrak{m}$ so that it is
orthogonal to $\mathfrak{h}$ with respect to the Cartan-Killing
form. With a slight abuse of notation, we identify $\mfrak$ throughout
with its dual $\mfrak^*$, i.e. the cotangent space $T_0^*(G/H)$ at the identity.

We choose a basis set $\{I_A\}$ for $\mathfrak{g}$ with
$A=1,\dots,\dim G$ in such a way that $I_a$ for $a=1,\dots,h$
form a basis for the subspace
$\mathfrak{m}\subset\mathfrak{g}$ and $I_i$ for $i=h+1,\dots,\dim G$ yield a basis for
the holonomy subalgebra
$\mathfrak{h}$. The structure constants $f^C_{AB}$ are defined by the
Lie brackets
\begin{equation}\label{2.47}
[I_A, I_B]=f^C_{AB}\, I_C \ \with g_{AB}:= f^D_{AC}\, f^C_{DB}=\de_{AB}\ ,
\end{equation}
where we have further chosen the basis so that it is orthonormal with respect to the
Cartan-Killing form on $\gfrak$. Then $f_{ABC}:=f^D_{AB}\, \de_{DC}$
is totally antisymmetric in $A,B,C$.
For a reductive homogeneous space, the relations (\ref{2.47}) can be rewritten as
\begin{equation}\label{2.48}
[I_i, I_j]=f^k_{ij}\, I_k\ ,\qquad [I_i, I_a]=f^b_{ia}\, I_b\und [I_a, I_b]=f^c_{ab}\, I_c+f^i_{ab}\, I_i \ .
\end{equation}
The invariant Cartan-Killing metric $g_{AB}$ is given by
\beq
g_{ab}=2f_{ad}^i\,
f_{ib}^d+f_{ad}^c\, f_{cb}^d= \de_{ab} \ ,
\label{gabCartan}\eeq
\beq
g_{ij}=f^a_{ib}\, f^b_{aj}+f^k_{il}\, f^l_{kj}=\de_{ij} \und g_{ia}=0
\ .
\label{gijgia}\eeq

The basis elements $I_a,I_i$ induce invariant one-forms $e^a,e^i$
on $G/H$, where $\{e^a\}$ form an invariant local orthonormal frame
for the cotangent bundle $T^*(G/H)\cong G\times_H\mathfrak{m}$, while
$\{e^i\}$ define the canonical connection $a^0= e^i\, I_i$ which is
the unique $G$-invariant connection
on the principal $H$-bundle $G\to G/H$. They obey the Maurer-Cartan equations
\begin{equation}\label{2.55}
\diff e^a =-f^a_{ib}\, e^i\wedge e^b-\mbox{$\frac{1}{2}$} \,
f^a_{bc}\, e^b\wedge e^c\und \diff e^i = -\mbox{$\frac{1}{2}$} \, f^i_{bc}\,
e^b\wedge e^c-\mbox{$\frac{1}{2}$} \,
f^i_{jk}\, e^j\wedge e^k \ ,
\end{equation}
and the $G$-invariant metric on $G/H$, 
lifted from the metric (\ref{gabCartan}) on $\mfrak$, is given by
\beq
g= \delta_{ab}\, e^a\otimes e^b \ .
\label{metricGH}\eeq
The first equation in (\ref{2.55}) can be interpreted as a Cartan
structure equation
\beq
\diff e^a +\Gamma^a_b\wedge e^b= T^a = \mbox{$\frac12$}\,T^a_{bc}\,
e^b\wedge e^c \ ,
\label{Cartaneq}\eeq
where the metric connection $\Gamma^a_b=e^i\, f_{ib}^a$ is the canonical connection $a^0$ acting on one-forms via the adjoint action of $\mathfrak{h}$ on $\mathfrak{m}$, and $T^a_{bc}=-f^a_{bc}$ is its torsion. This connection has structure group $H$. In the sequel we will also consider more general metric connections with torsion tensor components $\vk\, f^a_{bc}$ for $\vk\in\R$.

\smallskip

\noindent
{\bf Quasi-K\"ahler manifolds. \ }
An $H$-structure on a smooth orientable manifold $\X^m$ of dimension
$m$ is a reduction of the structure group $\glrm(m,\R)$ of the tangent bundle
$T\X^m$ to a closed subgroup $H$. Choosing an
orientation and riemannian metric $g$ defines an $\sorm(m)$-structure on $\X^m$. We assume that $m=2n$ is even and
that $(\X^{2n},g)$ is an almost hermitian manifold. Then there exists
an almost complex structure $\J\in\End(T\X^{2n})$, with
$\J^2=-{\bf 1}_{T\X^{2n}}$, which is compatible with the metric
$g$, i.e. $g(\J W,\J Z)=g(W,Z)$ for all $W,Z\in T\X^{2n}$. This
defines a $\urm(n)$-structure on $\X^{2n}$, and one introduces the
fundamental two-form $\omega$ with
\beq
\omega(W,Z):= g(\J W,Z)
\label{oemgaWZdef}\eeq
for $W,Z\in T\X^{2n}$, i.e. $\omega$ is an almost K\"ahler form of
type $(1,1)$ with respect to $\J$.

The three-form $\diff\omega$ generally has
$(3,0)$+$(0,3)$ and $(2,1)$+$(1,2)$ components with respect to
$\J$. The almost hermitian manifold $\X^{2n}$ is called
\emph{quasi-K\"ahler} or \emph{$(1,2)$-symplectic} if only the $(3,0)$+$(0,3)$ components of
$\diff\omega$ are non-vanishing~\cite{3,3a}. It is called \emph{nearly
  K\"ahler} if in addition the Nijenhuis tensor $\mathfrak N$ of the canonical
hermitian connection on $T\X^{2n}$ is totally antisymmetric; in this
case $\mathfrak N$ is the real part of a $(3,0)$-form proportional to
$\diff\omega$. Thus in the quasi-K\"ahler case there exists a global
complex three-form $\Omega$ on $\X^{2n}$. In particular, for $n=3$
the $(3,0)$-form $\Omega$ trivializes the canonical bundle of $\X^6$
and reduces the $\urm(3)$ holonomy group to $\surm(3)$.

\smallskip

\noindent
{\bf $\mbf{{\rm SU}(n)}$-structures. \ }
An almost hermitian manifold $(\X^{2n},g,\J)$ with topologically trivial canonical bundle allows an $\surm(n)$-structure. 
An $\surm(n)$-structure
on an oriented riemannian manifold $(\X^{2n}, g)$ is determined by a pair $(\ome , \Om )$, where $\ome$ is a non-degenerate real
two-form (an almost symplectic structure) and $\Om$ is a decomposable complex
$n$-form such that
\begin{equation}\label{2.28}
\ome\wedge\Om
=0\und\Om\wedge\overline\Om=\mbox{$\frac{(2\,\im)^n}{n!}$}\
\ome^{\wedge n}\ .
\end{equation}
The complex $n$-form
\begin{equation}\label{2.29}
\Om=\Th^1\wedge\cdots \wedge \Th^n
\end{equation}
determines an almost complex structure $\J$ on $\X^{2n}$ such that
\begin{equation}\label{2.30}
\J\,\Th^\a=\im\,\Th^\a \ \for \a=1,\dots,n\ ,
\end{equation}
i.e. the forms $\Th^\a$ span the space of forms of type $(1,0)$ with respect to $\J$. The $(n,0)$-form
$\Om$ is a global section of the (trivial) canonical bundle of $\X^{2n}$,
so that $c_1(\X^{2n})=0$. 

The fundamental two-form $\ome$ is of type $(1,1)$ with respect
to $\J$ by virtue of (\ref{2.28}), and $g=\ome\circ \J$
is an almost hermitian metric. We choose
\begin{equation}\label{2.31}
g=\sum_{\a=1}^n\, \Th^\a\otimes \Th^{\bar\a}\und
\ome=\frac{\im}{2}\,\sum_{\a=1}^n\, \Th^\a\wedge\Th^{\bar\a} \ ,
\end{equation}
where $\Theta^{\bar\a}:=\overline{\Theta^\a}$. Their non-vanishing components are therefore given by
\begin{equation}\label{2.32}
g_{\a\bb}=\mbox{$\frac{1}{2}$}\, \de_{\a\bb}\ ,\quad
g^{\a\bb}=2\de^{\a\bb}\und
\ome_{\a\bb}=\mbox{$\frac{\im}{2}$}\, \de_{\a\bb}\ ,\quad
\ome^{\a\bb}=-2\,\im\, \de^{\a\bb}
\end{equation}
with respect to this basis. For $n>3$, these manifolds include
Calabi-Yau torsion spaces for which the Nijenhuis tensor is the real
part of a $(2,1)$-form.

\smallskip

\noindent
{\bf Nearly K\"ahler structures on six-manifolds. \ } A six-manifold
$(\X^6,g)$ with $\sut$-structure $(\omega,\Omega)$ is nearly K\"ahler if the two-form $\ome$
and three-form $\Om$ satisfy
\begin{equation}\label{2.40}
\diff\ome = \mbox{$\frac{3}{2}$}\,W_1\,{\rm Im}\,\Om\und
\diff\Om = W_1\,\ome\wedge\ome \ ,
\end{equation}
with a constant $W_1\in\R$~\cite{2,3,4}. These backgrounds solve the Einstein equations
with positive cosmological constant, and they yield a
metric-compatible connection with totally
  antisymmetric (intrinsic) torsion and $\sut$-holonomy which admits covariantly
  constant spinor fields without coupling to other gauge fields.
There are only four known examples of compact
nearly K\"ahler six-manifolds, and they are all coset spaces
\begin{eqnarray}
{\rm SU(3)/U(1)}\times{\rm U(1)}=\F_3 &,& 
{\rm Sp(2)/Sp(1)}\times{\rm U(1)}= \C P^3 \ , \nonumber \\[4pt]
G_2{\rm /SU(3)}=S^6 &\and&
{\rm SU(2)}^3/{\rm SU(2)}=S^3\times S^3\ .
\label{2.44}\end{eqnarray}
Here ${\rm Sp}(1)\times\uo$ is chosen to be a non-maximal subgroup of
Sp(2)~\cite{3}, while in the final quotient $\su$ is the diagonal
subgroup of $\su^3$.
The cosets $G/H$ in (\ref{2.44}) are all reductive, and in each case
the subgroup $H$ of $G$ can be embedded in $\sut$.

The $\sut$-structure on a coset space in (\ref{2.44}) is determined
entirely from the Lie algebra $\mathfrak{g}$. The subgroup $H$ is the
fixed point set of an automorphism of $G$ of order three, which
induces a $3$-symmetry $s_*:\gfrak\to \gfrak$ such that
$s_*^3=\one_\mfrak$ on the tangent space
$\mathfrak{m}$ and $s_*=\one_\hfrak$ on the Lie subalgebra $\hfrak$~\cite{3} (see also~\cite{HILP}). The eigenspace decomposition of $s_*$ gives a 
complexification of the splitting~(\ref{2.45}),
\beq
\mathfrak{g}^\C=\mathfrak{h}^\C\oplus\big(\mathfrak{m}^+\oplus
\mathfrak{m}^- \big) \ ,
\label{SU3complexdecomp}\eeq
which satisfies
\beq
\big[\mathfrak{m}^\pm\,,\, \mathfrak{m}^\pm\big]\subset \mathfrak{m}^\mp \ , \qquad
\big[\mathfrak{h}\,,\,\mathfrak{m}^\pm\big] \subset \mathfrak{m}^\pm \und \big[\mathfrak{m}^+\,,\,
\mathfrak{m}^-\big]\subset \mathfrak{h}^\C \ .
\label{gChCmpm}\eeq
We define an $H$-invariant almost complex structure $\J$ on $G/H$ by identifying $\mathfrak{m}^+=\bigwedge^{1,0}T_0^*(G/H)$ and $\mathfrak{m}^-= \bigwedge^{0,1}T_0^*(G/H)$. Note
that $\J$ is a never integrable almost complex structure due to the
first property of (\ref{gChCmpm}).

We choose a basis of the complexified Lie algebra $\mathfrak{g}^\C$, given by
\beq
-\im\, I_i \ , \qquad I_\alpha^-=\mbox{$\frac12$}\,(I_{2\a-1}-\im\,I_{2\a}) \und
I_{\bar\a}^+=\mbox{$\frac12$}\,(I_{2\a-1}+\im\, I_{2\a})
\label{Icomplexbasis}\eeq
with $i=7,\dots,\dim G$ and $\a=1,2,3$, such that $\mathfrak{m}^\pm={\rm
  span}_\C\big\{I_{\alpha,\bar\alpha}^\pm\big\}$. Choosing an
invariant local orthonormal basis $\{e^a\}$ of the cotangent bundle
$T^\ast(G/H)$ with $a=1,\dots,6$ as above, the forms
\begin{equation}\label{2.34}
\Theta^\a= e^{2\a-1}+\im\, e^{2\a}
\end{equation}
are of type $(1,0)$ with respect to $\J$. The metric $g$ in
(\ref{2.31}) coincides in this case with the natural $G$-invariant metric
(\ref{metricGH}) on the coset, while the fundamental two-form $\ome$ in (\ref{2.31}) 
reads
\begin{equation}\label{2.35}
\ome = e^1\wedge e^2 + e^3\wedge e^4 + e^5\wedge e^6\ .
\end{equation}
Furthermore, the three-form $\Omega=\Theta^1\wedge\Theta^2\wedge\Theta^3$ in (\ref{2.29}) reads
\begin{equation}\label{2.36}
 \Om ={\rm Re}\,\Om +
\im\, {\rm Im}\,\Om = \big( e^{135}+e^{425}+e^{416}+e^{326} \big) +
\im \,\big(e^{136}+e^{426}+e^{145}+e^{235}\big)\ ,
\end{equation}
where $e^{a_1\dots a_k}:=e^{a_1}\wedge\cdots\wedge e^{a_k}.$ Then the
pair $(\omega,\Omega)$ defines an invariant $\sut$-structure on the
coset spaces $G/H$ given in (\ref{2.44}).

In what follows we will lower all algebra indices $a,b,\dots$ and
$i,j,\dots$ with the help of the Cartan-Killing metric from
(\ref{gabCartan})--(\ref{gijgia}). For all four nearly K\"ahler coset spaces in (\ref{2.44}) one can
choose the non-vanishing structure constants $\{f_{abc}\}$ such that
\begin{equation}\label{fabcnot0}
f_{135}=f_{425}=f_{416}=f_{326}=-\mbox{$\frac{1}{2\sqrt{3}}$} \ .
\end{equation}
This choice for the generators $\{I_a\}$ is correlated with
(\ref{2.36}) and the Maurer-Cartan equations. In particular,
\begin{equation}\label{2.51}
{\rm Re}\,\Om = \mbox{$\frac{1}{3!}$}\,({\rm Re}\,\Om )_{abc}\,
e^{abc} \ \with ({\rm Re}\,\Om )_{abc}=-2\sqrt{3}\, f_{abc} \ ,
\end{equation}
which coincides with the statement that
\begin{equation}\label{2.54}
T_{abc}= - f_{abc}
\end{equation}
is a totally antisymmetric torsion and
$W_1=\frac{1}{\sqrt3}$~\cite{HILP}. The remaining structure constants are
constrained by the identities~\cite{HILP}
\begin{equation}\label{2.50}
f_{aci}\, f_{bci}=f_{acd}\, f_{bcd}=\mbox{$\frac{1}{3}$}\, \de_{ab}\ ,
\end{equation}
\beq
\omega_{cd}\,f_{adi}=\omega_{ad}\, f_{cdi} \und \omega_{ab}\, f_{abi}=0 \ ,
\label{Jfids}\eeq
where the first equality in (\ref{Jfids}) expresses invariance of the
Lie bracket of $\gfrak$ under the $3$-symmetry $s_*$ while the last identity expresses the embedding $H\subseteq \sut$.

Using the $(1,0)$-forms $\Th^\a$, we rewrite the Maurer-Cartan
equations (\ref{2.55}) as
\bea
\diff\Th^\a &=& -\im\,C^\a_{i\b}\, e^i\wedge\Th^\b-\im\,C^\a_{i\bb} \,
e^i\wedge\Th^{\bb}-\mbox{$\frac{1}{2}$}\,C^\a_{\b\g}\, \Th^\b\wedge\Th^\g -
C^\a_{\b\bar\g}\, \Th^\b\wedge\Th^{\bar\g}
-\mbox{$\frac{1}{2}$} \,C^\a_{\bb\bar\g}\,
\Th^{\bb}\wedge\Th^{\bar\g}\ ,\nonumber \\[4pt]
\im\,\diff e^i &=& -\mbox{$\frac{1}{2}$}\,C^i_{\b\g}\, \Th^\b\wedge\Th^\g -
C^i_{\b\bar\g}\,\Th^\b\wedge\Th^{\bar\g}
-\mbox{$\frac{1}{2}$}\,C^i_{\bb\bar\g}\, \Th^{\bb}\wedge\Th^{\bar\g}-\mbox{$\frac{1}{2}$} \,
C^i_{jk}\, e^j\wedge e^k \ ,
\label{2.60}\eea
where we denote by $C^\a_{i\b}$, $C^i_{\b\g}$, etc. the structure
constants of $\gfrak^\C$
in the basis (\ref{Icomplexbasis}). From (\ref{fabcnot0}) 
we obtain
\begin{equation}\label{Cnot0}
C^{\1}_{23}=C^{\2}_{31}=C^{\3}_{12}=C^{1}_{\2\3}=C^{2}_{\3\1}
=C^{3}_{\1\2}=-\mbox{$\frac{1}{2\sqrt{3}}$} \ .
\end{equation}
Recalling that $g_{\a\bb}=\sfrac{1}{2}\, \de_{\a\bb}$, from (\ref{Cnot0})
it follows that
\begin{equation}\label{2.62}
C_{\a\b\g}=g_{\a\ab}\, C^{\ab}_{\b\g}=-\mbox{$\frac{1}{4\sqrt{3}}$}
\,\ve_{\a\b\g} \und C_{\ab\bb\bar\g}=g_{\a\ab}\, C^{\a}_{\bb\bar\g}
=-\mbox{$\frac{1}{4\sqrt{3}}$} \,\ve_{\ab\bb\bar\g} \ .
\end{equation}
The first equation of (\ref{2.60}) can again be interpreted as
structure equations
\begin{equation}\label{2.38}
\diff\Th^\a+\Gamma^\a_\b\wedge\Th^\b=T^\a\und
\diff\Th^{\ab}+\Gamma^{\ab}_{\bb}\wedge\Th^{\bb}=T^{\ab}\ ,
\end{equation}
where $\Gamma =(\Gamma^\a_\b)$ is the (torsionful) connection with
holonomy $\hfrak$ and the last term defines the Nijenhuis tensor (torsion) with components
$T^\a_{\bb\bar\g}$ and their complex conjugates,
\begin{equation}\label{2.39}
T^\a =\mbox{$\frac{1}{2}$}\, T^\a_{\bb\bar\g}\, \Th^{\bb}\wedge\Th^{\bar\g}
\und
T^{\ab} =\mbox{$\frac{1}{2}$}\, T^{\ab}_{\b\g}\, \Th^{\b}\wedge\Th^{\g}\ .
\end{equation}
Using (\ref{2.62}) one has
\begin{equation}\label{2.42}
T_{\a\b\g}=\mbox{$\frac{1}{4\sqrt3}$}\, \ve_{\a\b\g}\und
T_{\ab\bb\bar\g}= \mbox{$\frac{1}{4\sqrt3}$}\, \ve_{\ab\bb\bar\g}\ .
\end{equation}

\smallskip

\noindent
{\bf A three-parameter family of SU(3)-structures. \ } Nearly K\"ahler six-manifolds
form a special subclass of six-manifolds with SU(3)-structure. The generic case
is classified by intrinsic torsion~\cite{2} which can be characterized by
five irreducible SU(3)-modules with torsion classes
$W_1,\dots,W_5$.\footnote{Here $W_1$ is complex, but one can always choose it
to be real for nearly K\"ahler six-manifolds. Calabi-Yau manifolds correspond to the vanishing of all
five intrinsic
torsion classes.} Here $W_1$ is a complex function, $W_2$ is a
$(1,1)$-form with $\omega\lrc W_2=0$, $W_3$ is
the real part of a $(2,1)$-form with
$\omega\lrc W_3=0$, and $W_4$
and $W_5$ are real one-forms. The symbol $\lrc$ denotes contraction
which is defined in terms of the Hodge duality operator $\star$ by $u\lrc
v=\star(u\wedge \star v)$. The five torsion classes are determined
by the 
decompositions
\bea
\diff\ome&=& \mbox{$\frac{3\, \im}{4}$}\, \big(W_1\, \overline\Om - \overline W_1\,
\Om \big) + W_3
+\ome\wedge W_4\ , \nonumber \\[4pt]
\diff\Om&=& W_1\,\ome\wedge\ome + \ome\wedge W_2 + \Om\wedge W_5\ .
\label{2.63}\eea
{}From (\ref{2.40}) one sees that for nearly K\"ahler manifolds only $W_1\ne 0$.
We obtain from them more general almost hermitian SU(3)-manifolds with also $W_2\ne 0$ by
rescaling the one-forms $\Th^\a$ by constants $\s_\alpha\in\R$ as
\begin{equation}\label{2.64}
\Th^\a\ \longmapsto\ \widetilde\Th^\a=\mbox{$\frac1{2\sqrt3}$}\, \s_\a^{-1}\, \Th^\a
\end{equation}
for $\a=1,2,3$. Such manifolds are quasi-K\"ahler~\cite{3,3a}.

After the rescaling (\ref{2.64}), the structure constants (\ref{Cnot0}) are
rescaled as
\begin{equation}\label{Ctildenot0}
\widetilde C^{\ab}_{\b\g}=\mbox{$2\sqrt3~ \frac{\s_\b\,\s_\g}{\s_\a}$}
\,C^{\ab}_{\b\g}=
-\mbox{$\frac{\s_\b\,\s_\g}{
  \s_\a}$}\,\ve^{\ab}_{\b\g}\ , \qquad \widetilde C^i_{jk}=C^i_{jk} \ ,
\nonumber\eeq
\beq
\widetilde C^\a_{i\b}=\mbox{$\frac{\s_\b}{\s_\a}$}\, C_{i\b}^\a \ , \qquad
\widetilde C^\a_{i\bar\b}=\mbox{$\frac{\s_\b}{\s_\a}$}\, C_{i\bb}^\a\ ,\qquad
\widetilde C^i_{\b\g}=12{\s_\b}\, \s_\g \,C^i_{\b\g} \ ,
\qquad
\widetilde C^i_{\b\bar\g}=12{\s_\b}\, \s_\g \,C^i_{\b\bar\g}\ ,
\end{equation}
plus their complex conjugates. The metric and the fundamental two-form become
\begin{equation}\label{2.66}
\widetilde g=\widetilde\Th^1\otimes \widetilde\Th^{\1}+\widetilde\Th^2\otimes \widetilde\Th^{\2}+
\widetilde\Th^3\otimes \widetilde\Th^{\3}
\und
\widetilde\ome=\sfrac{\im}{2}\, \big(\, \widetilde\Th^1\wedge\widetilde\Th^{\1}+
\widetilde\Th^2\wedge\widetilde\Th^{\2}+\widetilde\Th^3\wedge\widetilde\Th^{\3}\,
\big)\ .
\end{equation}
The Maurer-Cartan equations for the $(1,0)$-forms $\widetilde\Th^\a$
are given by (\ref{2.60}) but with
the rescaled structure constants (\ref{Ctildenot0}).

\section{SU(3)-structures on $\F_3$\label{F3Kahler}}

\noindent
{\bf Homogeneous spaces of SU(3). \ } In this section we study in detail the
first coset space in the list (\ref{2.44}), whereby the geometry described in
section~\ref{Homgen} can be made very explicit. The projective plane
$\C P^2$ and the complete flag manifold
$\F_3$ on $\C^3$ are related through the fibrations
\beq
\xymatrix{
 & \sut \ar[dl]_{\pi_3} \ar[dr]^{\pi_1} & \\
\F_3 \ \ar[rr]_{\pi_2}  & & \ \C P^2
}
\label{2.1}\eeq
with fibres $\uo\times\su$,
$\su/\uo$ and $\uo\times\uo$ for the bundle projections
$\pi_1, \pi_2$ and $\pi_3$, respectively. A representative element of
the coset space $\C P^2=\sut/\uo\times\su$ is a local section of the
principal fibre bundle
$\pi_1$ given
by the $3\times 3$ matrix
\begin{equation}\label{2.2}
V=\g^{-1}\, \begin{pmatrix}1&- T^{\+}\, \\T&W\end{pmatrix} \ \in \ {\rm SU}(3)\ ,
\end{equation}
where
\begin{equation}\label{2.3}
T:=\begin{pmatrix}\yb^{\2}\, \\y^1\end{pmatrix} \ , \qquad
W:=\g\, {\bf 1}_2 - \frac{1}{\g +1}\, T\, T^{\+}\und
\g = \sqrt{1+ T^{\+}\, T}=\sqrt{1+y^\a\, \yb^{\ab}}
\end{equation}
obey
\begin{equation}\label{2.4}
 W^{\+}=W\ ,\quad W\, T=T\und W^2=\g^2\, {\bf 1}_2- T\, T^{\+} \ ,
\end{equation}
and therefore
$V^{\+}\, V=V\,V^{\+}={\bf 1}_3$. Here $y^1, y^2$ are local complex
coordinates on $\C P^2$.

A representative element of the coset space $\C P^1\cong\su/\uo\cong S^2$ is a local section of the Hopf fibration $S^3\to S^2$
given by the matrix
\begin{equation}\label{2.12}
h=\frac{1}{\sqrt{1+\zeta\, \overline\zeta}}\,
\begin{pmatrix}1&-\overline\zeta \ \\ \zeta & 1\end{pmatrix}
\ \in \ {\rm SU}(2)\cong S^3\ ,
\end{equation}
where $\zeta$ is a local complex coordinate on $\C P^1$. Then a representative element for the
coset space $\F_3=\sut/\uo\times \uo$ is a local section
of the principal torus bundle $\pi_3$ given by the $3\times 3$ matrix~\cite{1}
\begin{equation}\label{2.13}
\widehat V=V\, \widehat h=\g^{-1}\, \begin{pmatrix}1&-
  T^{\+}\, \\T&W\end{pmatrix} \,
\begin{pmatrix}1&0\\0&h\end{pmatrix} \ \in \ 
{\rm SU}(3)\ ,
\end{equation}
with $V$ given in (\ref{2.2}). The sphere bundle $\pi_2$
describes $\F_3$ as the twistor space of $\C P^2$. 

\smallskip

\noindent
{\bf Monopole and instanton fields on $\C P^2$. \ } Let us
introduce a flat connection on the trivial vector bundle
$\C P^2\times\C^3$ given by the $\urmL(3)$-valued one-form
\begin{equation}\label{2.5}
\Acal = V^{-1}\, \diff V =: \begin{pmatrix}2b&-\sfrac{1}{2\La}\,
  \t^{\+}\, \\
\, \sfrac{1}{2\La}\, \t&B\end{pmatrix}\ ,
\end{equation}
where the real parameter $\La$ characterizes the ``size'' of the coset
$\C P^2$ and from (\ref{2.2}) one obtains
\begin{equation}\label{2.6}
 b=\frac{1}{4\g^2}\,\big(T^{\+}\, \diff T - \diff T^{\+}\, T\big)\und
B=\frac{1}{\g^2}\, \big(W\, \diff W + T\, \diff T^{\+}-
\mbox{$\frac{1}{2}$}\, \diff T^{\+}\, T -
\mbox{$\frac{1}{2}$}\,T^{\+}\, \diff T \big)\ ,
\end{equation}
and
\begin{equation}\label{2.7}
\t=\frac{2\La}{\g^2}\, W\,\diff T= \begin{pmatrix}\t^{\2}\, \\
\t^1\end{pmatrix}=\frac{2\La}{\g}\,
\begin{pmatrix}\diff\yb^{\2}\, \\ \diff y^1\end{pmatrix}-
\frac{2\La}{\g^2\, (\g +1)}\, \begin{pmatrix}\yb^{\2}\, \\
y^1\end{pmatrix}\,\big(\yb^{\1}\, \diff y^{1} + y^2\, \diff\yb^{\2}\, \big)\ .
\end{equation}
Here $\t^1$ and $\t^2$ form a local $\sut$-equivariant orthonormal basis of $(1,0)$-forms on
$\C P^2$.
The flatness condition, $\diff\Acal + \Acal\wedge\Acal =0$, leads to
the component equations
\begin{equation}\label{2.8}
 f^-:=\diff b=\frac{1}{8\La^2}\,\t^{\+}\wedge\t = -\frac{1}{8\La^2}\,
\big (\t^{1}\wedge\t^{\1}-\t^{2}\wedge\t^{\2}\, \big)
\end{equation}
and
\begin{equation}\label{2.9}
 F^{+}=\diff B^{+} + B^{+}\wedge B^{+}=-\frac{1}{8\La^2}\,
\begin{pmatrix}\t^{1}\wedge\t^{\1}+\t^2\wedge\t^{\2}&2\t^{\1}
\wedge\t^{\2}\\-2\t^1\wedge\t^{2}&-\t^{1}\wedge\t^{\1}
-\t^2\wedge\t^{\2}\end{pmatrix}\ ,
\end{equation}
where
\begin{equation}\label{2.10}
 B^{+}= \begin{pmatrix}a_+&-\overline b_+\\
b_+&-a_+\end{pmatrix}=B  + b\, {\bf 1}_2
\end{equation}
and
\begin{equation}\label{2.11}
 F=\diff B + B\wedge B=\frac{1}{4\La^2}\,\t\wedge\t^{\+}=
-\frac{1}{4\La^2}\,
\begin{pmatrix}\t^2\wedge\t^{\2}&\t^{\1}\wedge\t^{\2}\\
-\t^1\wedge\t^{2}&-\t^{1}\wedge\t^{\1}\end{pmatrix}=:
F^{+}-f^-\, {\bf 1}_2\ .
\end{equation}
{}From (\ref{2.8}) and (\ref{2.9}) it follows that $\star f^-=-f^-$ and
$\star F^+=F^+$, where $\star$ is the Hodge duality
operator on $\C P^2$ with respect to the Fubini-Study metric
$g=\theta^1\otimes\t^{\1}+\t^2\otimes \t^{\2}$ for $\t^\a$ given in (\ref{2.7}), i.e. $b$ is an
anti-self-dual $\uoL$-connection (monopole potential) on a complex line bundle over $\C P^2$
and $B^+$ is a self-dual $\suL$-connection (instanton potential) on a complex vector
bundle of rank two over $\C P^2$. The gauge potential $B$ is the (canonical)
$\urmL(2)$-valued Levi-Civita connection on the tangent bundle of the
coset space~$\C P^2$.

\smallskip

\noindent
{\bf Generalized monopole and instanton fields on $\F_3$. \ } Consider the trivial complex
vector bundle $\F_3\times\C^3$ endowed with a flat connection
\begin{equation}\label{2.14}
\widehat\Acal = \widehat V^{-1}\, \diff\widehat V =\widehat h^\+\, \Acal\, \widehat h +
\widehat h^\+\, \diff\widehat h=:
 \begin{pmatrix}2b&-\sfrac{1}{2\La}\,\widehat\t\,^{\+}\, \\
\, \sfrac{1}{2\La}\, \widehat\t&\widehat B\end{pmatrix} \ ,
\end{equation}
where
\begin{equation}\label{2.15}
 \widehat\t = h^{\+}\, \t =\frac{1}{\sqrt{1+\zeta\, \zeb}}\, 
\begin{pmatrix}
\t^{\2}+\overline\zeta\, \t^1\\ \t^1-\zeta\, \t^{\2}
\end{pmatrix}
=:
\begin{pmatrix}
 \widehat\t\,^{\2}\, \\ \widehat\t\,^1
\end{pmatrix}\ ,\qquad
\widehat\t\,^{\+}=\t^{\+}\, h=\big(\, \widehat\t\,^2\ \ \widehat\t\,^{\1}\, \big)
\end{equation}
and
\begin{equation}\label{2.16}
 \widehat B = h^{\+}\, B\,h +h^{\+}\,\diff h=\widehat B^+-b\, {\bf 1}_2=:
\begin{pmatrix}
 \widehat a_+& -\sfrac{1}{2R}\, \widehat\t\,^{\3}\, \\ \, \sfrac{1}{2R}\,\widehat\t\,^{3}&-\widehat a_+
\end{pmatrix} -b\, {\bf 1}_2 \ ,
\end{equation}
with
\begin{equation}\label{2.17}
\widehat a_+ =\frac{1}{1+\zeta\, \zeb}\,\Big( \big(1-\zeta\, \zeb\,\big)\,
a_+ + \zeb\, b_+ -
\zeta\, \overline b_+ + \mbox{$\frac{1}{2}$}\, \big(\,\zeb\, \diff\zeta
-\zeta\, \diff\zeb\, \big)\Big) 
\end{equation}
and
\begin{equation}\label{2.18}
\widehat\t\,^{3} =\frac{2R}{1+\zeta\, \zeb}\, \left(\diff\zeta + b_+ - 2\zeta\, a_+ +
\zeta^2\, \overline b_+\right)\ .
\end{equation}
Here $b$, $a_+$ and $b_+$ are given in (\ref{2.10}), while $R$ is the
radius of the fibre two-sphere $S^2\cong\C P^1$.
The curvature of $\widehat\Acal$ is given by
\begin{equation}\label{2.19}
\widehat\Fcal = \diff\widehat \Acal +\widehat\Acal\wedge\widehat\Acal  =
 \begin{pmatrix}
2\, \diff b -\frac{1}{4\La^2}\,\widehat\t\,^{\+}\wedge\widehat\t&
-\frac{1}{2\La}\,\big(\diff\widehat\t\,^{\+}+\widehat\t\,^{\+}\wedge
\widehat B{-}2\widehat\t\,^{\+}\wedge b \big)
\\[2mm]
\, \frac{1}{2\La}\,\big(\diff\widehat\t+\widehat B\wedge\widehat
\t{-}2b\wedge\widehat\t\, \big)&
\widehat F^+ -\diff b\, {\bf 1}_2
-\frac{1}{4\La^2}\,\widehat\t\wedge\widehat\t\,^{\+}
\end{pmatrix} \ ,
\end{equation}
and therefore
from the flatness condition $\widehat\Fcal=0$ we obtain
\begin{equation}\label{2.20}
 \widehat f^-=f^-=\diff b=-\frac{1}{8\La^2}\,\big(\widehat\t\,^1\wedge\widehat\t\,^{\1}-
 \widehat\t\,^2\wedge\widehat\t\,^{\2}\, \big)=
-\frac{1}{8\La^2}\,\big(\t^1\wedge\t^{\1}-\t^2\wedge\t^{\2}\, \big)
\end{equation}
and
\begin{equation}\label{2.21}
\widehat F^+ =\diff\widehat B^++\widehat B^+\wedge \widehat B^+ = -\frac{1}{8\La^2}\,
 \begin{pmatrix}
\widehat\t\,^1\wedge\widehat\t\,^{\1}+\widehat\t\,^2\wedge\widehat\t\,^{\2}&
2\widehat\t\,^{\1}\wedge\widehat\t\,^{\2}\\
-2\widehat\t\,^{1}\wedge\widehat\t\,^{2}&-\widehat\t\,^1\wedge\widehat\t\,^{\1}-
\widehat\t\,^2\wedge\widehat\t\,^{\2}
 \end{pmatrix} \ ,
\end{equation}
along with
\begin{equation}\label{2.22}
\diff\widehat\t + \big(\widehat B - 2b\, {\bf 1}_2 \big)\wedge\widehat\t =0\ .
\end{equation}
The gauge fields $\widehat f^-:=\pi^\ast_2f^-$ and $\widehat F^+:=\pi^\ast_2F^+$
are pull-backs of the monopole and instanton gauge fields $f^-$ and $F^+$ on $\C P^2$ to the
flag manifold $\F_3$ by the
twistor fibration $\pi_2$ from~(\ref{2.1}).

In particular, the abelian gauge field $\widehat f^-$ satisfies the hermitian Yang-Mills equations on $\F_3$
for both the K\"ahler and nearly K\"ahler geometries described below~\cite{1}. In the nearly K\"ahler case the abelian gauge field $\diff\widehat a_+$ also satisfies the hermitian Yang-Mills equations. These monopole-type fields will be used later on in our constructions of quiver gauge theories.

\smallskip

\noindent
{\bf K\"ahler geometry of $\F_3$. \ } The metric and an almost K\"ahler
structure on $\F_3$ read
\begin{equation}\label{2.23}
\widehat g = \widehat\t\,^1\otimes\widehat\t\,^{\1} + \widehat\t\,^2\otimes\widehat\t\,^{\2} + \widehat\t\,^3\otimes\widehat\t\,^{\3}\und
\widehat\ome =\mbox{$\frac{\im}{2}$}\, \big(\, \widehat\t\,^1\wedge\widehat\t\,^{\1}+\widehat\t\,^2\wedge\widehat\t\,^{\2}
+ \widehat\t\,^3\wedge\widehat\t\,^{\3}\, \big)\ ,
\end{equation}
where $\widehat\t\,^\a$ with $\a =1,2,3$ are given in (\ref{2.15}) and (\ref{2.18}).
The $\sut$-invariant one-forms $\widehat\t\,^\a$ define an
integrable almost complex structure $\J_+$ on $\F_3$ such that
\begin{equation}\label{2.24}
\J_+\widehat\t\,^\a =\im\, \widehat\t\,^\a\ ,
\end{equation}
i.e. $\widehat\t\,^\a$ are $(1,0)$-forms with respect to $\J_+$.
{}From (\ref{2.19})--(\ref{2.22}) we obtain the structure equations
\begin{equation}\label{2.25}
\diff\begin{pmatrix}\widehat\t\,^1\\ \widehat\t\,^2\\ \widehat\t\,^3\end{pmatrix}+
\begin{pmatrix}-\widehat a_+-3b&0&-\sfrac{1}{2R}\, \widehat\t\,^{\2}\, \\[1mm]
0&-\widehat a_++3b&\sfrac{1}{2R}\, \widehat\t\,^{\1}\\[1mm]
\sfrac{R}{4\La^2}\,\widehat\t\,^2& -\sfrac{R}{4\La^2}\,\widehat\t\,^1& -2\widehat a_+
\end{pmatrix}\wedge\begin{pmatrix}\widehat\t\,^1\\ \widehat\t\,^2\\
  \widehat\t\,^3\end{pmatrix}=0 \ ,
\end{equation}
which define the Levi-Civita connection $\widehat\Gamma =(\widehat\Gamma^\a_\b)$ on
the tangent bundle of $\F_3$
by the formula
\begin{equation}\label{2.26}
\diff\widehat\t\,^\a + \widehat\Gamma^\a_\b\wedge\widehat\t\,^\b =0\ .
\end{equation}

{}From (\ref{2.25}) it follows that $\widehat \ome$ is K\"ahler, i.e. $\diff\widehat \ome =0$,
if and only if
\begin{equation}\label{2.27}
R^2=2\La^2\ .
\end{equation}
Then the connection matrix $\widehat\Gamma$ in (\ref{2.25}) takes values in
the Lie algebra $\mathfrak{u}(3)$, i.e. the holonomy group is
$\urm(3)$.
The non-vanishing structure constants $\widehat C_{AB}^C$ of the Lie algebra $\sutL$
for the complex basis of one-forms $\widehat\t\,^{\a}$ adapted to the K\"ahler structure on $\F_3$ and
the structure equations (\ref{2.25}) are given by
\begin{equation}\label{2.61a}
\begin{matrix}
\widehat C^{\1}_{2\3}=\widehat C^{\2}_{1\3}=-\frac{1}{2\sqrt{6}}\ ,\qquad
\widehat C^{3}_{12}=-\frac{1}{\sqrt{6}}\ ,\\[4mm]
\widehat C^{7}_{1\1}=\widehat C^{7}_{2\2}=\widehat C^{7}_{3\3}=-\frac{1}{4\sqrt{3}}\ ,\qquad
\widehat C^{8}_{1\1}=\frac{1}{4}\und \widehat C^{8}_{2\2}=-\frac{1}{4} \ ,
\end{matrix}
\end{equation}
and their complex conjugates, plus
\beq
\begin{matrix}
\widehat C^{1}_{71}=\widehat C^{2}_{72}=\frac{1}{2\sqrt{3}}\ ,
\qquad \widehat C^{3}_{73}=-\frac{1}{\sqrt{3}}\ ,\qquad
\widehat C^{1}_{81}=-\frac{1}{2}\ ,
\qquad \widehat C^{2}_{82}=\frac{1}{2}\ ,\\[4mm]
\widehat C^\ab_{7\ab}=-\widehat C^\a_{7\a} \und \widehat C^\ab_{8\ab}=-\widehat C^\a_{8\a}
\end{matrix}
\label{2.61b}\eeq
for $\a=1,2,3$. Here we have chosen $R^2=2\La^2=6$.

\smallskip

\noindent
{\bf Nearly K\"ahler geometry of $\F_3$. \ } We have introduced above an integrable
almost complex structure $\J_+$ and a K\"ahler structure on the flag
manifold $\F_3$, defined via
the $(1,0)$-forms $\widehat\t\,^\a$. Let us now introduce the forms
\begin{equation}\label{2.33}
 \Theta^1:=\widehat\t\,^1\ ,\qquad \Theta^2:=\widehat\t\,^2\und\Theta^3:=\widehat\t\,^{\3}\ ,
\end{equation}
which are of type $(1,0)$ with respect to an almost complex structure
$\J = \J_-$~\cite{3a}, $\J_-\,\Theta^\a = \im\,\Theta^\a$, defined in
(\ref{2.30}). 
The almost complex structure $\J_-$ is obtained
from $\J_+$ by changing its sign along the $\C P^1$-fibres of the twistor bundle $\pi_2$, i.e.
$\J_{\pm}\Th^{1,2}=\im\, \Th^{1,2}, \
\J_{\pm}\Th^{3}=\mp\,\im\,\Th^{3}$. It is never integrable.

Using the redefinition (\ref{2.33}), we obtain from (\ref{2.25}) the structure
equations
\begin{equation}\label{2.37}
\diff\begin{pmatrix}\Th^1\\ \Th^2\\ \Th^3\end{pmatrix}+
\begin{pmatrix}-\widehat a_+-3b&0&0\\
0&-\widehat a_++3b&0\\
0& 0& 2\widehat a_+
\end{pmatrix}
\wedge
\begin{pmatrix}\Theta^1\\ \Theta^2\\ \Theta^3\end{pmatrix}
=\frac{1}{2R}\, 
\begin{pmatrix}\Theta^{\2}\wedge\Theta^{\3}\\ \Theta^{\3}\wedge\Theta^{\1}
\\ \frac{R^2}{\La^2}\, \Theta^{\1}\wedge\Theta^{\2}\end{pmatrix}\ .
\end{equation}
The first term here defines the (torsionful) connection
$\Gamma =(\Gamma^\a_\b)$, with $\urmL(1)\oplus \urmL(1)$ holonomy, whose components are obtained by comparing (\ref{2.37})
with (\ref{2.38}), and the last term defines the Nijenhuis tensor (torsion) with components
$T^\a_{\bb\bar\g}$ and their complex conjugates. We also have
\begin{equation}\label{2.39a}
\diff{b}=-\frac{1}{8\La^2}\, \big(\Th^1\wedge\Th^{\1} -\Th^2\wedge\Th^{\2}\, \big)
\end{equation}
and
\begin{equation}\label{2.39b}
\diff \widehat a_+=-\frac{1}{8\La^2}\, \big(\Th^1\wedge\Th^{\1} +\Th^2\wedge\Th^{\2}\, \big)
+\frac{1}{4R^2}\,\Th^3\wedge\Th^{\3}
\end{equation}
for the abelian gauge fields on $\F_3$.

The pair of forms $(\omega,\Omega)$ defined by (\ref{2.29}) and
(\ref{2.31}) for $n=3$ defines a one-parameter family of invariant
$\sut$-structures on $\F_3$, parametrized by the ratio
$\frac{R^2}{\Lambda^2}$. From (\ref{2.37}) it follows that
the conditions (\ref{2.40}) for the coset space $\F_3$ to be nearly K\"ahler yield
\begin{equation}\label{2.41}
R^2=\La^2 \ .
\end{equation}
We fix the scales of $\C P^1$ and $\C P^2$ in $\F_3$ so that
\begin{equation}\label{2.53}
R=\La =\sqrt{3} \ .
\end{equation}
Then the connection $\Gamma$ coincides with the canonical connection
which was used
throughout section~\ref{Homgen}.
Notice that the K\"ahler
and nearly K\"ahler structures correspond not only to different choices of
almost complex structures $\J_+$ and $\J_-$ on $\F_3$,
but also to metrics
\begin{equation}\label{2.43}
\widehat g=\Th^1\otimes\Th^{\1}+\Th^2\otimes\Th^{\2}+2\Th^3\otimes \Th^{\3}\und
g=\Th^1\otimes \Th^{\1}+\Th^2\otimes \Th^{\2}+\Th^3\otimes\Th^{\3}
\end{equation}
which differ by a factor of 2 along the fibre direction $\C P^1\hra\F_3$. Both $\widehat g$ and $g$
are Einstein metrics. 

\smallskip

\noindent
{\bf Basis for SU(3)-generators. \ } The non-vanishing structure constants of
$\mathfrak{su}(3)$ which conform with the nearly K\"ahler structure
(\ref{fabcnot0})--(\ref{Jfids}) are given by
\beq
f_{135}=f_{425}=f_{416}=f_{326}=-\mbox{$\frac{1}{2\sqrt{3}}$} \ ,
\nonumber\eeq
\begin{equation}\label{2.49}
f_{127}=f_{347}=\mbox{$\frac{1}{2\sqrt{3}}$} \ ,\qquad
f_{128}=-f_{348}=-\mbox{$\frac{1}{2}$} \und f_{567}=-\mbox{$\frac{1}{\sqrt{3}}$} \ .
\end{equation}
Correspondingly, we choose the basis for
$3\times3$ matrices of the antifundamental representation of $\mathfrak{su}(3)$ given by
\begin{equation}\nonumber
I_1=\frac{1}{2\sqrt{3}}\,\begin{pmatrix}0&0&-1\\0&0&0\\1&0&0\end{pmatrix}
\ ,\qquad
I_2=\frac{1}{2\sqrt{3}}\,\begin{pmatrix}0&0&\im\\0&0&0\\\im&0&0\end{pmatrix}
\ ,\qquad
I_3=\frac{1}{2\sqrt{3}}\,\begin{pmatrix}0&1&0\\-1&0&0\\0&0&0\end{pmatrix}\ ,
\end{equation}
\begin{equation}\nonumber
I_4=\frac{1}{2\sqrt{3}}\,\begin{pmatrix}0&\im&0\\\im&0&0\\0&0&0\end{pmatrix}
\ ,\qquad
I_5=\frac{1}{2\sqrt{3}}\,\begin{pmatrix}0&0&0\\0&0&1\\0&-1&0\end{pmatrix}
\ ,\qquad
I_6=\frac{1}{2\sqrt{3}}\,\begin{pmatrix}0&0&0\\0&0&\im\\0&\im&0\end{pmatrix}\ ,
\end{equation}
\begin{equation}\label{2.46}
I_7=\frac{\im}{2\sqrt{3}}\,\begin{pmatrix}0&0&0\\0&-1&0\\0&0&1\end{pmatrix}
\und
I_8=\frac{\im}{6}\,\begin{pmatrix}2&0&0\\0&-1&0\\0&0&-1\end{pmatrix}\ .
\end{equation}
Then the flat connection (\ref{2.14}) can be written as
\beq \label{hatAcaleI}
\widehat\Acal= e^i\, I_i +e^a\, I_a \ ,
\eeq
where $e^a$ from (\ref{2.34})--(\ref{2.36}) form an orthonormal frame for
the cotangent bundle $T^\ast\F_3$, and $e^i=\{e^7, e^8\}$ are two $\urmL(1)$-valued gauge potentials
on two line bundles of degree one over $\F_3$ whose curvatures
generate the cohomology group $\HQ^2(\F_3;\Z)\cong\Z\oplus
\Z$. Written in the form (\ref{hatAcaleI}), the flatness condition
$\widehat\Fcal= \diff\widehat \Acal +\widehat\Acal\wedge\widehat\Acal  =0$ is equivalent to the
Maurer-Cartan equations (\ref{2.55}). In the case
at hand, the group
$H=\uo\times \uo$ is abelian and therefore $f^i_{jk}=0$.

Equivalently, we can write (\ref{hatAcaleI}) as
\bea \label{2.56}
\widehat\Acal = \im\, e^i\, (- \im\,I_i)+\Th^\a\, I_\a^- +\Th^{\ab}\, I_{\ab}^+ = 
\frac{1}{2\sqrt{3}}\, \begin{pmatrix}
\frac{2}{\sqrt{3}}\,\im\,e^8&\Th^2& -\Th^{\1}\\
-\Th^{\2}&-\im\,e^7-\frac{1}{\sqrt{3}}\,\im\,e^8&\Th^3\\
\Th^1&-\Th^{\3}&\im\,e^7-\frac{1}{\sqrt{3}}\,\im\,e^8
\end{pmatrix}\ ,
\eea
where the matrices
\bea\nonumber
I_1^-:=\mbox{$\frac{1}{2}$}\,(I_1-\im\, I_2)=\frac{1}{2\sqrt{3}}\,
\begin{pmatrix}0&0&0\\0&0&0\\1&0&0\end{pmatrix} &,& 
I_{\1}^+:=\mbox{$\frac{1}{2}$}\,(I_1+\im\, I_2)=\frac{1}{2\sqrt{3}}\,
\begin{pmatrix}0&0&-1\\0&0&0\\0&0&0\end{pmatrix}\ , \\[4pt]
\nonumber
I_2^-:=\mbox{$\frac{1}{2}$}\,(I_3-\im\, I_4)=\frac{1}{2\sqrt{3}}\,
\begin{pmatrix}0&1&0\\0&0&0\\0&0&0\end{pmatrix}
&,& 
I_{\2}^+:=\mbox{$\frac{1}{2}$}\,(I_3+\im\, I_4)=\frac{1}{2\sqrt{3}}\,
\begin{pmatrix}0&0&0\\-1&0&0\\0&0&0\end{pmatrix}\ , \\[4pt]
\nonumber
I_3^-:=\mbox{$\frac{1}{2}$}\,(I_5-\im \, I_6)=\frac{1}{2\sqrt{3}}
\begin{pmatrix}0&0&0\\0&0&1\\0&0&0\end{pmatrix}  &,&
I_{\3}^+:=\mbox{$\frac{1}{2}$} \,(I_5+\im\, I_6)=\frac{1}{2\sqrt{3}}\,
\begin{pmatrix}0&0&0\\0&0&0\\0&-1&0\end{pmatrix}\ , \\[4pt]
-\im\,I_{7}=\frac{1}{2\sqrt{3}}\,
\begin{pmatrix}0&0&0\\0&-1&0\\0&0&1\end{pmatrix} &\and& 
-\im\,I_{8}=\frac{1}{6}\,
\begin{pmatrix}2&0&0\\0&-1&0\\0&0&-1\end{pmatrix}
\label{2.58}\eea
form a basis for the complexified Lie algebra $\sltcL$ in the
antifundamental representation. Here complex conjugation acts by
interchanging barred and unbarred indices. From (\ref{2.14})
and (\ref{2.56}) it follows that
\begin{equation}\label{2.59}
e^7=2\sqrt{3}\,\im\,\widehat a_+\und e^8=-6\,\im\, b\ .
\end{equation}
Explicit expressions for $e^a$ in terms of $\sut$-invariant gauge potentials and for $\Th^\a$ in terms of local coordinates on
$\F_3$ can also be easily extracted from~(\ref{2.13})--(\ref{2.18}).

Comparing (\ref{2.60}) with (\ref{2.37})--(\ref{2.39b}), the
non-vanishing structure
constants of $\sltcL$
in the basis (\ref{2.58}) are given by
\begin{equation}\label{2.61}
\begin{matrix}
C^{\1}_{23}=C^{\2}_{31}=C^{\3}_{12}=C^{1}_{\2\3}=C^{2}_{\3\1}
=C^{3}_{\1\2}=-\frac{1}{2\sqrt{3}}\ ,\\[4mm]
C^{1}_{71}=C^{2}_{72}=-C^{\1}_{7\1}=-C^{\2}_{7\2}=\frac{1}{2\sqrt{3}}\ ,
\qquad C^{3}_{73}=-C^{\3}_{7\3}=-\frac{1}{\sqrt{3}}\ ,\\[4mm]
C^{1}_{81}=-C^{\1}_{8\1}=-\frac{1}{2}\ ,
\qquad C^{2}_{82}=-C^{\2}_{8\2}=\frac{1}{2}\ ,\\[4mm]
C^{7}_{1\1}=C^{7}_{2\2}=-\frac{1}{4\sqrt{3}}\ ,\qquad
C^{7}_{3\3}=\frac{1}{2\sqrt{3}}\ ,
\qquad C^{8}_{1\1}=\frac{1}{4} \und C^{8}_{2\2}=-\frac{1}{4}\ .
\end{matrix}
\end{equation}
After the rescaling (\ref{2.64}), the structure constants (\ref{2.61}) are
rescaled as
\begin{equation*}
\widetilde C^{\ab}_{\b\g}=2\sqrt3\ \mbox{$\frac{\s_\b\,\s_\g}{\s_\a}$}\,C^{\ab}_{\b\g}= \mbox{$
-\frac{\s_\b\,\s_\g}{\s_\a}$} \,\ve^{\ab}_{\b\g}\ ,
\end{equation*}
\begin{equation*}
\widetilde C^1_{71}=C^1_{71}=\mbox{$\frac{1}{2\sqrt{3}}$}\ ,\qquad
\widetilde C^2_{72}=C^2_{72}=\mbox{$\frac{1}{2\sqrt{3}}$}\ ,\qquad
\widetilde C^3_{73}=C^3_{73}=-\mbox{$\frac{1}{\sqrt{3}}$} \ ,
\end{equation*}
\begin{equation}
\widetilde C^1_{81}=C^1_{81}=-\mbox{$\frac{1}{2}$} \ ,\qquad
\widetilde C^2_{82}=C^2_{82}=\mbox{$\frac{1}{2}$} \ ,
\label{2.65}\end{equation}
\begin{equation*}
\widetilde C^7_{1\1}={12\s_1^2}\,C^7_{1\1}=-\sqrt3\,\s_1^2 \ ,
\qquad
\widetilde C^7_{2\2}={12\s_2^2}\,C^7_{2\2}=-\sqrt3\, \s_2^2 \ ,
\qquad
\widetilde C^7_{3\3}={12\s_3^2}\,C^7_{3\3}=2\sqrt3\,\s_3^2 \ ,
\end{equation*}
\begin{equation*}
\widetilde C^8_{1\1}={12\s_1^2}\,C^8_{1\1}=3\s_1^2 \ ,\qquad
\widetilde C^8_{2\2}={12\s_2^2}\,C^8_{2\2}=-3\s_2^2 \ ,
\end{equation*}
plus their complex conjugates. 
In particular, by setting $\s_1=\s_2=\s_3=2\sqrt3\, R^{-1}$ we can restore the
$\C P^1$ radius $R$, and for $\s_1=\s_2=2\sqrt3\, \La^{-1}$, $\s_3=2\sqrt3\, R^{-1}$ we
can restore both of our original size
parameters $\La$ and~$R$.

\section{Pseudo-holomorphic equivariant vector bundles\label{Pseudo}}

\noindent
{\bf Equivariant bundles. \ } In this paper we are interested
in Yang-Mills theory with torsion and $G$-equivariant
gauge fields on manifolds of the form
\begin{equation}\label{3.1}
\X^{d+h} =M\times G/H\ ,
\end{equation}
where $M$ is a smooth manifold of real dimension $d$ and
$G/H$ is a reductive homogeneous manifold of real dimension $h=\dim
G-\dim H$. The group $G$ acts
trivially on $M$ and in the standard way by isometries of the coset
space $G/H$. By standard induction and reduction, there is an equivalence between smooth
$G$-equivariant vector bundles $\Ecal$ over $\X^{d+h}$ and smooth
$H$-equivariant vector bundles $E$ over $M$, where $H$ acts trivially on~$M$; a smooth $H$-equivariant bundle $E\to M$ induces a smooth
$G$-equivariant bundle $\Ecal\to\X^{d+h}$ by the fibred product
\beq
\Ecal=G\times_H E \ .
\label{indbundle}\eeq

For each $p\in M$, the restriction $\Ecal_p$ of $\Ecal$ to the coset
$G/H \cong\{p\}\times G/H\hookrightarrow \X^{d+h}$ is a homogeneous
vector bundle on $G/H$ which is in correspondence, via (\ref{indbundle}), with the
representation $E_p$ of $H$ on the fibres of the complex vector bundle
$E\to M$. Let $T$ be a maximal torus of $G$ such that $T\subseteq H$,
and let $\tfrak$ be its Lie algebra. If $V_\lambda$ is the
irreducible representation of $H$ with weight vector $\lambda\in\tfrak^*$,
then the fibred product
\beq
\Vcal_\lambda:=G\times_H V_\lambda
\label{Vcallambda}\eeq
is the induced irreducible smooth homogeneous vector bundle on
$G/H$ whose associated principal bundle has structure group $H$. Then every smooth $G$-equivariant complex vector bundle $\Ecal\to\X^{d+h}$
can be equivariantly decomposed as~\cite{A-CG-P1}
\beq
\Ecal= \bigoplus_{\lambda\in W}\,E_\lambda\boxtimes\Vcal_\lambda \
,
\label{Ecaldecomp}\eeq
where $E_\lambda\to M$ are smooth complex vector bundles with trivial
$H$-action, and $W\subset\tfrak^*$ is the finite set of eigenvalues for the action of $H$
on $E$.

\smallskip

\noindent
{\bf Holomorphic bundles. \ } On any coset space $G/H$ which is a flag
manifold, one can introduce an integrable almost complex structure and
(almost) K\"ahler structure. Then
the action of $G$ on $G/H$ is symplectic and $G/H$ is diffeomorphic to the projective variety $G^\C/P$ with the
canonical complex structure, where $P$ is a parabolic subgroup of the
complexification $G^\C$ of $G$. We assume
henceforth that $M$ is a complex manifold (in particular, its
dimension $d$ is even), and denote by $H^\C$ the
universal complexification of $H$. Then there is a one-to-one
correspondence between $G^\C$-invariant holomorphic structures on the
$G$-equivariant vector bundle (\ref{indbundle}), and
$H^\C$-equivariant holomorphic structures on the
$H$-equivariant vector bundle $E$ together with extensions of the
$H^\C$-action to a holomorphic $P$-action on $E$. At the level of
(complex) Lie algebras, the
extension of the reductive Levi subgroup $H^\C$ to $P$ is described by a decomposition
$\mathfrak{p}=\mathfrak{h}^\C\oplus\mathfrak{u}$,
where the nilpotent radical $\mathfrak{u}=\bigwedge^{0,1}T_0^*(G/H)$ for the canonical complex
structure is an $H$-invariant \emph{subalgebra} of $\mathfrak{g}^\C$,
i.e.~$[\mathfrak{u},\mathfrak{u}]\subset \mathfrak{u}$ and
$[\hfrak,\mathfrak{u}]\subset \mathfrak{u}$.

By restriction and Corollary~1.13
of~\cite{A-CG-P1}, in this case there is a functorial equivalence
between the category of holomorphic homogeneous vector bundles on
$G/H$ and the category of finite-dimensional representations of a
certain bounded quiver $Q$
satisfying a set of relations $R$. The set of vertices $Q_0$ of $Q$ is
the set of weights $W$ for the corresponding $H^\C$-action and the set
of arrows $Q_1$ is given by the actions of the generators of the nilpotent radical
$\mathfrak{u}$ on the weight vectors, while the set of relations $R$
express commutativity of the quiver diagram through the Lie algebra
relations of $\mathfrak{u}$ represented on weight vectors. The relations
$R$ in this geometric framework appear as the condition for
holomorphicity of the homogeneous bundles $\Vcal$ over $G/H$, i.e. as
the condition $\Fcal^{0,2}=\overline{\partial}_\Acal\wedge \overline{\partial}_\Acal =0$
for integrability of the Dolbeault operator $\overline{\partial}_\Acal$
on~$\Vcal$~\cite{A-CG-P1} with respect to a $G$-invariant connection $\Acal$ on
$\Vcal$ and the canonical complex structure. In this case, the
relations $R$ are in one-to-one correspondence with integrability of
the canonical $(0,1)$-distribution on $G/H$ and with a proper subalgebra
$\mathfrak{u}$ of $\gfrak^\C$.

In the example $G/H=\sut/\uo\times\uo=\F_3$, an integrable almost complex structure and
(almost) K\"ahler structure is described by
(\ref{2.23})--(\ref{2.27}). In this case $P$ is a Borel subgroup of
$\sltc$ and the Levi decomposition is the usual root space
decomposition. In the basis (\ref{2.58}), the
generators of $\mathfrak{u}$ are $I_{\bar 1}^+$, $I_2^-$ and $I_3^-$ which (after
rescaling) close to the three-dimensional Heisenberg algebra with
central element $I_{\bar 1}^+$. Thus the arrows $Q_1$ translate weight
vectors by the set of positive roots of $\sltcL$, while the relations $R$ are
induced by the Heisenberg commutation relations which express
commutativity of the corresponding quiver diagrams~\cite{A-CG-P1,LPS3}. 

\smallskip

\noindent
{\bf Pseudo-holomorphic bundles. \ } In the following we will describe how these quivers and their
representations are modified when on $G/H$ we consider instead a
family of SU(3)-structures, as described in section~\ref{Homgen}. In
this case the canonical complex structure on $G/H$ is replaced with a never
integrable almost complex structure, and the K\"ahler structure is replaced
by a quasi-K\"ahler structure. Instead of homogeneous
\emph{holomorphic} bundles $\Vcal$ over $G/H$, we now consider
\emph{pseudo-holomorphic} bundles. Pseudo-holomorphicity of the
homogeneous bundle is again
defined by a connection $\Acal$ on $\Vcal$ whose curvature
$\Fcal=\diff\Acal+\Acal\wedge \Acal$ is of type $(1,1)$ with respect
to the chosen almost complex structure, i.e.
$\Fcal^{0,2}=0=\Fcal^{2,0}$~\cite{4}. In sections~\ref{QGT} and~\ref{vortex}
we will demonstrate that these equations lead to the
appropriate set of relations on the pertinent quiver. We will establish a functorial equivalence between the
category of pseudo-holomorphic homogeneous vector bundles on $G/H$ and a category
of certain finite-dimensional ``unitary'' representations of a new
quiver $\Qbar$ with new relations $\Rbar$. Using (\ref{Ecaldecomp}),
the extension from homogeneous pseudo-holomorphic bundles $\Vcal\to
G/H$ to $G$-equivariant pseudo-holomorphic bundles $\Ecal\to\X^{d+h}$ is a straightforward
technical task. In that case we obtain a functorial equivalence
between the category of $G$-equivariant pseudo-holomorphic bundles
over $\X^{d+h}=M\times G/H$ and a category of ``hermitian''
$\big(\,\Qbar\,,\, \Rbar\, \big)$-bundles on $M$. We shall show in
later sections that
this correspondence follows from using equivariant dimensional reduction of Yang-Mills
theory on $\X^{d+h}$ to derive a quiver gauge theory on $M$ with relations arising from
hermitian Yang-Mills equations on $\X^{d+h}$.

In section~\ref{Doublereps} we describe in detail the
basic properties of the new quivers $\Qbar$ and relations $\Rbar$ from
a purely algebraic perspective. Since the $G$-equivariance conditions
do not depend on an (either integrable or non-integrable) almost
complex structure on $G/H$, the fundamental isotopical decomposition
(\ref{Ecaldecomp}) always holds. Thus the set of vertices of the quiver $\Qbar$
is the same as that of the holomorphic setting and coincides with the
set of weights $\lambda\in\tfrak^*$ occuring in the decomposition of the given $H$-module
into irreducible representations $V_\lambda$. Now, however, the
vertices are connected together via the generators $I_\alpha^-$ of
(\ref{Icomplexbasis}) which do not close a subalgebra of $\gfrak^\C$
due to (\ref{gChCmpm}). With the
given choice of almost complex structure, this just reflects the
non-integrability of the $(0,1)$-distribution on $G/H$ in this case. This means that in the
pseudo-holomorphic case one should consider, in a certain sense,
representations of the full path algebra generated by the unoriented
weight diagram $W$. The purpose of section~\ref{Doublereps} is to make this statement precise.

\section{Double quiver representations\label{Doublereps}}

\noindent
{\bf Reduction of $\mbf G$-modules. \ } We are interested in
$G$-equivariant complex vector bundles $\widehat\Vcal$ over a reductive homogeneous
manifold $G/H$ induced by $H$-modules which
descend from some finite-dimensional
irreducible representation of $G$ on $\widehat V\cong \C^q$. After restriction to
$H\subset G$
this representation decomposes into irreducible representations of $H$ such that
\begin{equation}\label{3.2}
\widehat V=\bigoplus^n_{r=1}\, V_{q_r} \ \with \sum\limits^n_{r=1}\, q_r=q\und\widehat I_i=
\begin{pmatrix}
I^{q_1}_i&0&\dots&0\\
0&\ddots&&\vdots\\
\vdots&&\ddots&0\\
0&\dots&0&I^{q_n}_i
\end{pmatrix}\ ,
\end{equation}
where $I^{q_r}_i$ with $i=h+ 1,\dots,\dim G$ are the generators of $q_r\times q_r$ irreducible
representations $V_{q_r}$ of~$H$. Correspondingly, since we assume that $H$ contains a maximal abelian subgroup of $G$, the remaining generators $\widehat I_a$ of
$G$ in this representation have the off-diagonal form
\begin{equation}\label{3.3}
\widehat I_a=
\begin{pmatrix}0&I^{q_{12}}_a&\dots&I^{q_{1n}}_a \\
I^{q_{21}}_a&0&\ddots&\vdots\\
\vdots&\ddots&\ddots&I^{q_{n-1\,n}}_a \\
I^{q_{n1}}_a&\dots&I^{q_{n\,n-1}}_a&0
\end{pmatrix}\ ,
\end{equation}
where $I_a^{q_{rs}}$ with $a=1,\dots,h$ are $q_r\times q_s$ matrices. Depending
on the representation of $G$, some $I_a^{q_{rs}}$ can be zero matrices.

From the commutation relations (\ref{2.48}) one finds
\begin{eqnarray}\label{3.4}
I_i^{q_{r}}\, I_a^{q_{rs}}-I_a^{q_{rs}}\, I_i^{q_{s}} &=& f_{ia}^b\, I_b^{q_{rs}}\ ,
\\[4pt] \label{3.5}
\sum_{r\ne s}\, \big(I_a^{q_{sr}}\, I_b^{q_{rs}}-I_b^{q_{sr}}\, I_a^{q_{rs}}\big) &=&
f_{ab}^i\, I_i^{q_{s}}\ ,
\\[4pt] \label{3.6}
\sum_{r\ne s,l}\, \big(I_a^{q_{sr}}\, I_b^{q_{rl}}-I_b^{q_{sr}}\, I_a^{q_{rl}}\big) &=&
f_{ab}^c\, I_c^{q_{sl}}\ ,
\end{eqnarray}
for $r,s,l=1,\dots,n$. (There are
no sums over $r$ and $s$ in (\ref{3.4}).)
Additional constraints follow from the definition of the Lie algebra of
$G$. For example, in the case of unitary groups $G$ one has
\begin{equation}\label{3.7}
\big(I_a^{q_{rs}}\big)^\dagger =- I_a^{q_{sr}}\ .
\end{equation}

For the coset space $\F_3=\sut/\uo\times\uo$ with
its never integrable almost complex structure, we can make this
decomposition very explicit. For each fixed pair of
non-negative integers $(k,l)$ there is an irreducible representation
$\widehat V^{k,l}$ of $\sut$ of dimension 
\beq
q^{k,l}=\mbox{$\frac12$}\,(k+1)\,(l+1)\,(k+l+2) \ .
\label{dimkl}\eeq
The integer $k$ is the
number of fundamental representations $\widehat V^{1,0}$ and $l$
the number of conjugate representations $\widehat V^{0,1}$
appearing in the usual tensor product construction of
$\widehat V^{k,l}$, i.e. the Young diagram of
$\widehat V^{k,l}$ contains $k+l$ boxes in its first row and $l$
boxes in its second row.

In this case $n^{k,l}=q^{k,l}$ since all irreducible $H$-modules are
one-dimensional, and the collection of weight vectors of $H=\uo\times\uo$ in $\sut$ label
points in the weight diagram $W^{k,l}$ for $\widehat V^{k,l}$. We denote
them by $(q,m)_n$, where $q$ and $m$ are respectively isospin and
hypercharge eigenvalues, and the label by the total isospin integer
$n$ is used to keep track of multiplicities of states in the weight
diagram. They may be conveniently
parameterized by a pair of independent $\su$ spins $j_\pm$,
with $2j_+=0,1,\dots,k$ and $2j_-=0,1,\dots,l$, and the corresponding
component spins $m_\pm\in\{-j_\pm,-j_\pm+1,\dots,j_\pm-1,j_\pm\}$ as
\beq
q=2(m_++m_-) \ , \qquad
m=6(j_+-j_-)-2(k-l) \qquad \mbox{and} \qquad n=2(j_++j_-) \ .
\label{mnqjpm}\eeq
The $\su$ spin $j_+$ (resp.~$j_-$) is the value of the isospin contributed by the upper (resp.~lower) indices of the $\sut$
tensor corresponding to the irreducible module $\widehat V^{k,l}$. The
integers $(q,m)_n$ all have the same even/odd parity. 

To explicitly represent the coset generators in (\ref{3.2})--(\ref{3.3}) in
this case, we
will use the Biedenharn basis for the irreducible representation $\widehat
V^{k,l}$ of $\sut$~\cite{LPS3}. The generators of $H^\C=(\C^\times)^2$ for the irreducible
module corresponding to the weight vector $(q,m)_n$ in this basis are given by
\beq
-\im\,I_{7}^{(q,m)_n}=\mbox{$\frac{1}{4\,\sqrt3}$}\, (q-m) \und
-\im\,I_{8}^{(q,m)_n} =
\mbox{$\frac1{12}$}\, (q+3m) \ ,
\label{I78Bied}\eeq
while the non-vanishing off-diagonal matrix elements of the remaining
generators of $\sltc$ are
\bea
I_1^{- \ (q-1,m-3)_{n\pm1}\,(q,m)_n}&=& \sqrt{\mbox{$\frac{n\pm
      q+1\pm1}{24(n+1\pm1)}$}} ~ \lambda^\mp_{k,l}(n\pm1,m-3) \ ,
\nonumber \\[4pt] I_2^{- \ (q+2,m)_n\, (q,m)_n} &=& \sqrt{\mbox{$\frac{(n-q)\,
        (n+q+2)}{48}$}} \ , \nonumber\\[4pt]
I_3^{- \ (q-1,m+3)_{n\pm1}\, (q,m)_n}&=& \sqrt{\mbox{$\frac{n\mp
      q+1\pm1}{24(n+1)}$}} ~ \lambda^\pm_{k,l}(n,m) \ ,
\label{offdiagBied}\eea
where
\bea
\lambda_{k,l}^+(n,m)&=&\mbox{$\frac1{\sqrt{n+2}}~
\sqrt{\big(\frac{k+2l}3+\frac n2+\frac
    m6+2\big)\,\big(\frac{k-l}3+\frac n2+\frac m6+1\big)\,\big(
\frac{2k+l}3-\frac n2-\frac m6\big)}$} \ , \nonumber\\[4pt]
\lambda_{k,l}^-(n,m)&=&\mbox{$\frac1{\sqrt n}~
\sqrt{\big(\frac{k+2l}3-\frac n2+\frac
    m6+1\big)\,\big(\frac{l-k}3+\frac n2-\frac m6\big)\,\big(
\frac{2k+l}3+\frac n2-\frac m6+1\big)}$} \ .
\label{lambdaklnm}\eea
The latter constants are defined for $n>0$ and we set
$\lambda_{k,l}^-(0,m):=0$. The analogous relations for $I_{\bar\alpha}^+$ can be derived by hermitian conjugation
of (\ref{offdiagBied}) using the property (\ref{3.7}).

\smallskip

\noindent
{\bf Quivers and path algebras. \ } The unoriented graph $W$ associated to an irreducible
$G$-module $\widehat V$ is composed of $n$ vertices, one
associated to each of the $H$-modules $V_{q_r}$ appearing in the
decomposition (\ref{3.2}). After switching to a
suitable basis (\ref{Icomplexbasis}) for the Lie algebra of the complexified group $G^\C$,
an edge between two vertices $v_r$ and $v_s$ of the
graph is associated to each pair of non-zero matrices $I_\alpha^{-\,
  q_{rs}},I_{\bar\alpha}^{+\, q_{rs}}$. By giving the graph $W$ an
orientation, we turn it into a quiver $Q=(Q_0,Q_1)$ with $Q_0$ the set
of vertices $v_r$ of $W$, and $Q_1$ the set of arrows $a_{rs}:v_r\to
v_s$ associated with the non-zero generators $I_\alpha^{-\,
  q_{rs}}$. The quiver comes equiped with head and tail maps
$h,t:Q_1\rightrightarrows Q_0$, which for an arrow $a:v\to v'$ are defined by
$h(a)=v'$ and $t(a)=v$. A path in $Q$ of length $\ell$ is a sequence of
$\ell$ arrows in $Q_1$ which compose. If $h(a)=t(a'\,)$ for $a,a'\in Q_1$, then we may
produce a path $a'\,a$ defined by $\bullet \xrightarrow{a\ } \bullet
\xrightarrow{a'} \bullet$, and so on. Each arrow $a\in Q_1$ itself is a path of
length one. To each vertex $v\in Q_0$ we associate the trivial path
$1_v$ of length zero with $h(1_v)=t(1_v)=v$. More generally, an
oriented $\ell$-cycle in $Q$ is a path $p$ of length $\ell$ with
$h(p)=t(p)$. Throughout this section we use various facts from the representation theory of quivers; see e.g.~\cite{Quiver} for details.

The combinatorial data encoded by the quiver $Q$ can be studied
algebraically by introducing
the path algebra $\C Q$, the vector space over $\C$ spanned by
all paths together with multiplication given by concatenation of open
oriented paths. If two paths do not compose then their product is
defined to be $0$. The trivial paths $1_v$ for $v\in Q_0$ are
idempotents in this algebra and thereby define a system of mutually
orthogonal projectors on the associative
$\C$-algebra $\C Q$, i.e. $1_v^2=1_v$ and $1_v\, 1_{v'}=0$ for $v\neq
v'$. For any arrow $a:v\to v'$ in $Q_1$, one has $a\, 1_v=a$ and
$1_{v'}\, a=a$. Since $Q$ is a finite quiver, and every path starts and ends at
some vertex in $Q_0$, it follows that the algebra $\C Q$ is unital with identity
element
\beq
1_{\C Q}=\sum_{v\in Q_0}\, 1_v \ .
\label{CQunit}\eeq
The quiver $Q$ and its path algebra $\C Q$ are the basic building blocks of our ensuing
constructions. However, for the reasons explained in section~\ref{Pseudo}, we must
instead work with a certain ``completion'' of this quiver in a sense
that we explain precisely below. 

\smallskip

\noindent
{\bf The quiver $\mbf{Q^{k,l}}$. \ } Let us look at the quiver
$Q^{k,l}=\big(Q_0^{k,l}\,,\,Q_1^{k,l} \big)$ associated to an irreducible
$\sut$-representation $\widehat V^{k,l}$.
In this case the graph $W^{k,l}$
is simply the weight diagram of $\widehat V^{k,l}$. The set of vertices
$Q_0^{k,l}$ consists of weights $(q,m)_n$ constructed according to the
formula (\ref{mnqjpm}). The arrows $Q_1^{k,l}$ are built according to
the non-vanishing matrix elements (\ref{offdiagBied}), which give the
action of the off-diagonal coset generators on the weight vectors as
\bea
I_1^-\,:\,(q,m)_n&\longmapsto&(q-1,m-3)_{n\pm1} \ , \nonumber\\[4pt]
I_2^-\,:\,(q,m)_n&\longmapsto&(q+2,m)_n \ ,
\nonumber\\[4pt] I_3^-\,:\,(q,m)_n&
\longmapsto&(q-1,m+3)_{n\pm1} \ .
\label{Q3arrows}\eea
Compared to the quivers which arise in the holomorphic
case~\cite{LPS3}, the directions of arrows associated to $I_1^-$ are
reversed, corresponding to the change in sign of the almost complex
structure along the $\C P^1$-fibre direction of $\F_3$. Note that there can be multiple arrows emanating between two
vertices due to degenerate weight vectors $(q,m)_n$ and $(q,m)_{n'}$
with $n\neq n'$.

Let us consider some explicit constructions. For the three-dimensional fundamental representation $\widehat V^{1,0}$, this prescription
gives the quiver
\beq
Q^{1,0} \ : \ \qquad 
\xymatrix{
(-1,1)_1 \ \ar[rr]|{ \ a_2 \ } & & \ (1,1)_1 \ar[dl]|{ \ a_1 \ } \\
 & (0,-2)_0 \ar[ul]|{ \ a_3 \ } &
}
\label{Q10quiver}\eeq
while the conjugate representation $\widehat V^{0,1}$ yields
\beq
Q^{0,1} \ : \ \qquad 
\xymatrix{
& (0,2)_0 \ar[dl]|{ \ a_1 \ } & \\
(-1,-1)_1 \ \ar[rr]|{ \ a_2 \ } & & \ (1,-1)_1 \ar[ul]|{ \ a_3 \ }
}
\label{Q01quiver}\eeq
From the tensor product construction of $\widehat V^{k,l}$, it follows
that arbitrary quivers $Q^{k,l}$ can be built by gluing the fundamental triangles of
the quivers (\ref{Q10quiver})--(\ref{Q01quiver}) together in appropriate ways
such that coinciding edges have the same orientation. For instance,
for the six-dimensional representation $\widehat V^{2,0}$ this
prescription gives
\beq
Q^{2,0} \ : \ \qquad
\xymatrix{
(-2,2)_2 \ \ar[rr]|{ \ c_2 \ } & & \ (0,2)_2 \ \ar[rr]|{ \ b_2 \ }
\ar[dl]|{ \ c_1 \ }
& & \ (2,2)_2
\ar[dl]|{ \ b_1 \ } \\
 & (-1,-1)_1 \ \ar[ul]|{ \ c_3 \ } \ar[rr]|{ \ a_2 \ } & & \ (1,-1)_1
 \ar[ul]|{ \ b_3 \ }
 \ar[dl]|{ \ a_1 \ } & \\ 
 & & (0,-4)_0 \ar[ul]|{ \ a_3 \ } & & 
}
\label{Q20quiver}\eeq
while for the eight-dimensional adjoint representation $\widehat V^{1,1}$
we obtain
\beq
Q^{1,1} \ : \ \qquad
\xymatrix{
 & (-1,3)_1 \ \ar[rr]|{ \ c_2 \ } \ar[dl]|{ \ d_1 \ } & & \ (1,3)_1
 \ar@/_/[dl]|{ \ c_1 \ }
 \ar@/^/[dl]|{ \ b_1 \ } &  \\
(-2,0)_2 \ \ar[rr]|{ \ d_2 \ } & & \ (0,0)_{0,2} \ \ar[rr]|{ \ a_2 \ }
\ar@/^/[ul]|{ \ d_3 \ } \ar@/_/[ul]|{ \ c_3 \ } \ar@/^/[dl]|{ \ f_1 \ }
\ar@/_/[dl]|{ \ e_1 \ } & & \ (2,0)_2 \ar[dl]|{ \ a_1 \ } \ar[ul]|{ \ b_3
  \ } \\
 & (-1,-3)_1 \ \ar[rr]|{ \ f_2 \ } \ar[ul]|{ \ e_3 \ } & & \ (1,-3)_1
 \ar@/^/[ul]|{ \ f_3 \ } \ar@/_/[ul]|{ \ a_3 \ }
}
\label{Q11quiver}\eeq

\smallskip

\noindent
{\bf The double quiver $\mbf\Qbar$. \ } Double quivers often arise as
a means of translating combinatorial problems in graph theory into the
algebraic framework of quivers; once an orientation is chosen on a
graph, the ``doubling'' is inevitably necessary to
retain the generic data of the original unoriented graph. A classic example is the McKay quiver which is a double
quiver of an affine Dynkin diagram of type ADE describing the
relationship between finite subgroups of $\su$ and kleinian
singularities. In our case, the graph of interest is the weight
diagram $W$ associated to the decomposition of an irreducible $G$-module
$\widehat V$ into $H$-modules. The double $\Qbar$ of the quiver $Q$ is the
quiver obtained from $Q$ with the same vertex set $\Qbar_0=Q_0$ by
adjoining to each arrow $a\in Q_1$ an arrow $a^*$ with the opposite
orientation, i.e. $h(a^*)=t(a)$ and $t(a^*)=h(a)$. For example, the
double of the antifundamental quiver (\ref{Q01quiver}) is
\beq
\Qbar\,^{0,1} \ : \ \qquad 
\xymatrix{
 & \ (0,2)_0 \ \ar@/^/[dl]|{ \ a_1 \
} \ar@/^/[dr]|{ \ a^*_3 \ } & \\
(-1,-1)_1 \ \ar@/^/[rr]|{ \ a_2 \ } \ar@/^/[ur]|{ \ a^*_1 \
} & & \ (1,-1)_1 \ar@/^/[ll]|{ \ a_2^* \ }
  \ar@/^/[ul]|{ \ a_3 \ }
}
\label{Qbar01quiver}\eeq
while the double of the quiver (\ref{Q20quiver}) is
\beq
\Qbar\,^{2,0} \ : \ \qquad
\xymatrix{
(-2,2)_2  \ \ar@/^/[rr]|{ \ c_2 \ } \ar@/^/[dr]|{ \ c^*_3 \ } & & \  (0,2)_2 \ 
\ \ar@/^/[dr]|{ \ b_3^* \ } \ar@/^/[rr]|{ \ b_2 \ } \ar@/^/[ll]|{ \ c_2^* \ }  \ar@/^/[dl]|{ \ c_1 \ }
& &  \ (2,2)_2
\ar@/^/[dl]|{ \ b_1 \ } \ar@/^/[ll]|{ \ b^*_2 \ } \\
 &  \ (-1,-1)_1 \ \ar@/^/[ur]|{ \ c^*_1 \ } \ar@/^/[dr]|{ \ a^*_3 \ } \ar@/^/[ul]|{ \ c_3 \ }
 \ar@/^/[rr]|{ \ a_2 \ } & &  \ (1,-1)_1 \ar@/^/[ul]|{ \ b_3 \ } \ar@/^/[ll]|{ \ a^*_2 \ } \ar@/^/[ur]|{ \ b^*_1 \ } 
 \ar@/^/[dl]|{ \ a_1 \ } & \\
 & &  \ (0,-4)_0 \ \ar@/^/[ul]|{ \ a_3 \ } \ar@/^/[ur]|{ \ a^*_1 \ } & & 
}
\label{Qbar20quiver}\eeq

Since the graph $W$ does not contain any edge-loops (edges joining a
vertex to itself), the path algebra $\C\, \Qbar$ of the double quiver is canonically
isomorphic to the path algebra $\C W$, which is precisely the property
we were after. The path algebra $\C\, \Qbar$ is unital with the same
identity element~(\ref{CQunit}). It has the natural structure
of an involutive algebra with conjugate-linear anti-algebra involution $\iota:\C\,\Qbar \to\C\,\Qbar$
defined by $\iota(a)=a^*$ and $\iota(a^*)=a$ for $a\in Q_1$. The inclusion
of quivers $Q\subset \Qbar$ makes $\C Q$ a subalgebra of $\C\,\Qbar$,
whereas mapping each arrow $a^*\in\Qbar_1\setminus Q_1$ to $0$ gives
a surjective algebra homomorphism $\C\,\Qbar \to \C Q$.

\smallskip

\noindent
{\bf Representations of $\mbf \Qbar$. \ } A representation of the
quiver $Q$ in a category $\Csf$ is given by a collection
$V=(V_v)_{v\in Q_0}$ of objects of $\Csf$ associated to each vertex
together with a collection of morphisms $\phi=(\phi_a:V_{t(a)}\to
V_{h(a)})_{a\in Q_1}$ for each arrow. A morphism
$f:(V,\phi)\to(V',\phi'\,)$ between two representations is a family of
morphisms $f=(f_v:V_v\to V'_v\,)_{v\in Q_0}$ such that $\phi_a'\,
f_{t(a)}= f_{h(a)}\, \phi_a$ for all $a\in Q_1$. In this way, the representations
of $Q$ form a category. Any path $p=a_1\cdots a_\ell$ induces a
morphism $\phi(p)=\phi_{a_1}\cdots\phi_{a_\ell}:V_{t(p)}\to
V_{h(p)}$. The morphism induced by the trivial path $1_v$ at $v\in
Q_0$ is $\phi(1_v)={\bf 1}_{V_v}:V_v\to V_v$. Similarly, one defines
representations of the double quiver $\Qbar$ in~$\Csf$.

The fundamental case is when $\Csf$ is the abelian category of complex vector
spaces. In this case a representation of $Q$ is called a linear
representation or a $Q$-module. The category of linear representations
of $Q$ is equivalent to the category of left modules over its path
algebra $\C Q$~\cite{Quiver}. Hence a $Q$-module $(V,\phi)$ can be simply described
as a $\C Q$-module $V$, with the canonical identifications $V_v=1_v\, V$
for $v\in Q_0$. Under this equivalence, the representations of the
trivial paths $\phi(1_v)$ are orthogonal projections $\Pi_v$
from
the $\C Q$-module
\beq
V=\bigoplus_{v\in Q_0}\, V_v
\label{VoplusVv}\eeq
onto the subspace $V_v$ for $v\in Q_0$.

Any linear representation of the double quiver $\Qbar$
is also a linear representation of $Q$ by restricting the
corresponding module over $\C\, \Qbar$ to the subalgebra $\C
Q\subset\C\, \Qbar$. Conversely, given a $Q$-module $(V,\phi)$, one
naturally induces a $\Qbar$-module $(V,\phibar\,)$ by choosing
hermitian inner products on each
of the complex vector spaces $V_v$ for $v\in Q_0$, and setting
$\phibar_a=\phi_a:V_{t(a)}\to
V_{h(a)}$ and $\phibar_{a^*}=\phi_a^\dag:V_{h(a)}\to
V_{t(a)}$ the hermitian conjugate of $\phi_a$ for $a\in
Q_1$. Such a module will be refered to as a \emph{unitary
  representation} of the double quiver $\Qbar$; it defines an
involutive representation of the path algebra $\C\, \Qbar$ with the
involution $\iota$.

The $\C$-vector space of linear representations of the quiver $Q$ with fixed
$V=(V_v)_{v\in Q_0}$ is
\beq
\Rep(Q,V)= \bigoplus_{a\in Q_1}\, \Hom(V_{t(a)},V_{h(a)}) \ .
\label{RepQV}\eeq
Upon choosing bases for $V_v\cong\C^{q_v}$ for each $v\in Q_0$, this
space may be identified with the affine variety $\prod_{a\in Q_1}\,
\Hom(\C^{q_{t(a)}}, \C^{q_{h(a)}})\cong\C^r$ where $r=\sum_{a\in
  Q_1}\, q_{t(a)}\, q_{h(a)}$. The complex gauge group
\beq
\Gscr(V)=\Big(\, \prod_{v\in Q_0}\, {\rm GL}(V_v)\, \Big) \ \Big/ \ \C^\times
\label{GVgroup}\eeq
acts naturally by conjugating elements of (\ref{RepQV}) as bifundamental fields, i.e. $\phi_a\mapsto g_{h(a)}\,
\phi_a\, g_{t(a)}^{-1}$ for each $a\in Q_1$ and $g=(g_v)_{v\in
  Q_0}\in\Gscr(V)$. The corresponding gauge orbits are
precisely the isomorphism classes of $Q$-modules with dimension vector
$\qvec=(q_v)_{v\in Q_0}$. 

The vector space $\Rep(\,\Qbar,V)$ of double quiver representations in $V$
may be naturally identified with the \emph{cotangent bundle} on
$\Rep(Q,V)$ through the trace pairing
\bea
\Rep(\, \Qbar,V)&=&\bigoplus_{a\in
  Q_1}\,\Hom(V_{t(a)},V_{h(a)}) \ \oplus \ \bigoplus_{a\in Q_1}\, \Hom(V_{t(a^*)},V_{h(a^*)}) \nonumber\\[4pt] &=& \Rep(Q,V)\oplus \Rep(Q,V)^* \ = \ T^*\Rep(Q,V) \ .
\label{RepQbarV}\eea
In particular, it has a canonical $\Gscr(V)$-invariant symplectic structure defined by the
two-form
\beq
\omega_{\Qbar}(\,\phibar\,,\,\psibar\,)=\sum_{a\in
  Q_1}\,\tr\big(\phi_a\,\psi_{a^*}-\phi_{a^*}\,\psi_a \big)
\label{omegaQbar}\eeq
for $\phibar=(\phi_a,\phi_{a^*})_{a\in Q_1},\psibar =(\psi_a,\psi_{a^*})_{a\in Q_1}\in\Rep(\, \Qbar,V)$.\footnote{Here we identify $\Rep(\,
\Qbar,V)$ with its tangent space at any point.} Let
$\mathfrak{g}(V)=\big( \bigoplus_{v\in Q_0}\,\mathfrak{gl}(V_v)\big) \ominus\C$. Then the
linear $\Gscr(V)$-action on $\Rep(\, \Qbar,V)$ is hamiltonian and the
corresponding $\Gscr(V)$-equivariant moment map
$\mu_V=(\mu_{V,v})_{v\in Q_0}:\Rep(\,
\Qbar,V) \to \mathfrak{g}(V)^*$ is given
by
\beq
\mu_{V,v}(\,\phibar\,) = \sum_{a\in h^{-1}(v)}\, \phi_a\, \phi_{a^*}-
\sum_{a\in t^{-1}(v)}\, \phi_{a^*}\, \phi_a
\label{muVv}\eeq
for $\phibar\in\Rep(\, \Qbar,V)$.\footnote{Here we use the trace pairing to identify the dual space of
the Lie algebra $\mathfrak{g}(V)$ with the traceless endomorphisms in $\mathfrak{g}(V)$.} A choice of hermitian metric on $V$
naturally makes $\Rep(\, \Qbar,V)$ into a flat hyper-K\"ahler space. For further details, see e.g.~\cite{CB1}.

For any collection of complex numbers $\lambda=(\lambda_v)_{v\in
  Q_0}$, the points of $\mu_V^{-1}\big((\lambda_v\, {\bf 1}_{V_v})_{v\in
  Q_0}\big)$ in $\Rep(\,\Qbar,V)$
can be identified in a gauge invariant way with modules of
dimension vector $\qvec$ over the deformed preprojective algebra~\cite{CBH}
\beq
{\cal P}_\lambda=\C\, \Qbar \ \Big/ \ \Big\langle \,\sum\limits_{a\in Q_1}\,
  [a,a^*] - \sum\limits_{v\in Q_0}\, \lambda_v\, 1_v\, \Big\rangle \ .
\label{calPlambda}\eeq
The corresponding Marsden-Weinstein symplectic
quotient $\mu_V^{-1}\big((\lambda_v\, {\bf 1}_{V_v})_{v\in Q_0}\big)/\!/
\Gscr(V)$ is called an affine quiver variety. These gauge orbits classify the isomorphism classes of semi-simple
representations of ${\cal P}_\lambda$ of dimension $\qvec$. The
preprojective algebras ${\cal P}_0$ are of fundamental importance in
diverse areas of representation theory, geometry, and quantum groups
(see e.g.~\cite{Ringel} for an overview). Their deformations commonly
occur in problems of noncommutative geometry; for the example of the
McKay quivers, they are related to algebras of functions on
noncommutative deformations of kleinian
singularities~\cite{CBH}. However, since our quivers $Q$ satisfy the
hypotheses of Proposition~8.2.2 of~\cite{CBEG}, the center of the
preprojective algebra ${\cal P}_0$ is trivial in this case, $Z({\cal
  P}_0)=\C\, 1_{\C Q}$.

Let us look at some explicit examples of this hamiltonian reduction associated to
unitary $\Qbar\,^{k,l}$-modules induced by representations
$(V^{k,l},\phi)$ of the quiver $Q^{k,l}$ associated to an irreducible
$\sut$-representation as above. For the antifundamental quiver
(\ref{Q01quiver}), the moments maps are
\bea
\mu_{V^{0,1},(-1,-1)_1}&=& \phi_1\, \phi_1^\dag -\phi_2^\dag\,\phi_2 \
, \label{muV011} \\[4pt] \mu_{V^{0,1},(1,-1)_1}&=& \phi_2\, \phi_2^\dag -\phi_3^\dag\,\phi_3 \
, \label{muV012} \\[4pt] \mu_{V^{0,1},(0,2)_0}&=& \phi_3\, \phi_3^\dag -\phi_1^\dag\,\phi_1 \
, 
\label{muV013}\eea
while for the quiver (\ref{Q20quiver}) one finds
\bea
\mu_{V^{2,0},(-2,2)_2}&=& \xi_3\, \xi_3^\dag- \xi_2^\dag\, \xi_2 \ ,
\label{muV201}\\[4pt]
\mu_{V^{2,0},(0,2)_2}&=& \xi_2\, \xi_2^\dag- \xi_1^\dag\, \xi_1
+\psi_3\, \psi_3^\dag- \psi_2^\dag\, \psi_2 \ , 
\label{muV202}\\[4pt]
\mu_{V^{2,0},(2,2)_2}&=& \psi_2\, \psi_2^\dag- \psi_1^\dag\, \psi_1 \ , 
\label{muV203}\\[4pt]
\mu_{V^{2,0},(-1,-1)_1}&=& \xi_1\, \xi_1^\dag- \xi_3^\dag\, \xi_3
+\phi_3\, \phi_3^\dag- \phi_2^\dag\, \phi_2 \ , 
\label{muV204}\\[4pt]
\mu_{V^{2,0},(1,-1)_1}&=& \psi_1\, \psi_1^\dag- \psi_3^\dag\, \psi_3
+\phi_2\, \phi_2^\dag- \phi_1^\dag\, \phi_1 \ , 
\label{muV205}\\[4pt]
\mu_{V^{2,0},(0,-4)_0}&=& \phi_1\, \phi_1^\dag- \phi_3^\dag\, \phi_3 \ .
\label{muV206}\eea
Here we have abbreviated $\phi_\alpha:=\phi_{a_\alpha}$,
$\psi_\alpha:=\phi_{b_\alpha}$, and $\xi_\alpha:= \phi_{c_\alpha}$ for
$\alpha=1,2,3$. 

For a generic quiver $Q^{k,l}$, this construction can be
straightforwardly generalized by using the structure of the
corresponding weight diagram. Denoting the quiver module morphisms
representing the adjoints of the arrows (\ref{Q3arrows}) respectively by
$\phi^{1,3 \ (\pm)}_{(q,m)_n}$, $\phi^2_{(q,m)_n}$, the associated
moment maps are given by
\bea
\mu_{V^{k,l},(q,m)_n}&=&
\sum_\pm\,\Big(\phi^{1 \ (\pm)}_{(q+1,m+3)_{n\mp 1}}\, {}^\dag \,
\phi^{1 \ (\pm)}_{(q+1,m+3)_{n\mp 1}}-
\phi^{1 \ (\pm)}_{(q,m)_{n}}\, \phi^{1 \ (\pm)}_{(q,m)_{n}}{}^\dag \Big)
\nonumber\\
&& +\, \Big(\phi^2_{(q-2,m)_n}{}^\dag \, \phi^2_{(q-2,m)_n} -
\phi^2_{(q,m)_n}\, \phi^2_{(q,m)_n}{}^\dag\Big) \nonumber \\
&&+\, \sum_\pm\,\Big(\phi^{3 \ (\pm)}_{(q+1,m-3)_{n\mp 1}}\, {}^\dag \,
\phi^{3 \ (\pm)}_{(q+1,m-3)_{n\mp 1}}-
\phi^{3 \ (\pm)}_{(q,m)_{n}}\, \phi^{3 \ (\pm)}_{(q,m)_{n}}{}^\dag \Big)
\ ,
\label{muVkl}\eea
with the morphisms understood to be zero whenever the corresponding
vertex labels map outside the range dictated by (\ref{mnqjpm}). The
associated modules over the preprojective algebra are then
described by the moment map equations
$\mu_{V^{k,l},(q,m)_n}=\lambda_{(q,m)_n}\, {\bf 1}_{V_{(q,m)_n}}$.

In section~\ref{EGTQB} we will construct, via equivariant
dimensional reduction, natural representations of the quiver $Q$ in
the case when $\Csf$ is the
category of complex vector bundles over a smooth manifold~$M$. Such a representation is called a quiver bundle or
$Q$-bundle. A quiver bundle on $M$ can be regarded as a family of
quiver modules (the fibres of the quiver bundle), parametrized by
points of $M$; hence many of the above constructions and results
concerning quiver modules extend to this more general setting. In
particular, a quiver bundle can be regarded from a more algebraic
perspective as a locally free sheaf of modules over the path algebra
bundle of the quiver~\cite{A-CG-P2,GK1}. 

By equiping a $Q$-bundle with a hermitian structure, one
naturally induces a unitary representation of the double quiver
$\Qbar$ in the category $\Csf$. We call such a quiver bundle a
hermitian $\Qbar$-bundle. In this case, the gauge group
(\ref{GVgroup}) reduces to its unitary subgroup $\Uscr(V)$. In
section~\ref{vortex} we will see that the natural gauge theory
equations associated with a hermitian $\Qbar$-bundle through dimensional reduction have an analogous
moment map interpretation by combining the moment map (\ref{muVv})
(understood now in the category of complex vector bundles) with the
moment map associated to the canonical $\Uscr(V)$-invariant
symplectic form on the space of unitary connections on a hermitian
vector bundle over a K\"ahler manifold $M$~\cite{A-CG-P2}. In particular, the
equations naturally distinguish between the quasi-K\"ahler, nearly
K\"ahler and K\"ahler reductions over the coset space $G/H$,
corresponding to unitary representations of different preprojective algebras
(\ref{calPlambda}).

\smallskip

\noindent
{\bf The relations $\mbf \Rbar$. \ } The quivers associated to
holomorphic homogeneous vector bundles over $G/H$ have the further crucial
property that they contain no oriented $\ell$-cycles of length
$\ell>0$; equivalently, the corresponding path algebra is finite-dimensional. The absence of such cycles follows from parabolicity of the
subgroup $P\subset G^\C$ and it is of fundamental importance for
obtaining nicely behaved quiver gauge theory moduli spaces. The crux
of this condition is that it enables one to introduce total orderings on the vertex
sets $Q_0$. This allows one to formulate and study the appropriate
notion of stable quiver bundles and to relate them to solutions of
quiver vortex equations, as the ordering can be exploited to construct natural $G$-equivariant filtrations of holomorphic
homogeneous vector bundles by holomorphic sub-bundles over
$G/H$~\cite{A-CG-P1}. It also implies that
all quiver representations with the same dimension vector are gauge
equivalent~\cite{LPS2}.

In contrast, the quivers $Q$ associated to
pseudo-holomorphic homogeneous vector bundles over $G/H$ contain
non-trivial oriented cycles; for instance, the quivers $Q^{k,l}$ all
contain $\ell$-cycles with $\ell\geq3$. This will be the generic
situation we encounter later on for the quiver bundles with
connections that arise through equivariant dimensional reduction of
Spin(7)-instanton equations; the Spin(7)-instanton moduli space thus has a
very complicated stack structure that must be dealt with using
appropriate stability conditions, as in~\cite{A-CG-P2}. We will now define a generating set of relations on the corresponding double quiver
$\Qbar$ which eliminates all oriented cycles in the quiver
diagram of $\Qbar$, and hence of $Q$. The corresponding quiver bundles come from hermitian Yang-Mills equations and have better behaved moduli schemes, as in the holomorphic case.

A finite set of relations $R$ of the quiver $Q$ corresponds to a two-sided ideal of the path
algebra $\C Q$, denoted $\langle R\rangle$, i.e. a collection of
$\C$-linear combinations $r=\sum_i\, c_i\, p_i$ of paths $p_i$. In this paper we deal only
with \emph{admissible relations}, i.e. the paths $p_i$ all
have the same head and tail vertices. The bounded quiver with relations $(Q,R)$ is
described algebraically by the factor algebra $\C Q/\langle
R\rangle$. For the quivers associated to the $G$-module decompositions
(\ref{3.2})--(\ref{3.3}), the collection of relations
comes from the commutation relations (\ref{3.4})--(\ref{3.6}) among
the block generators, written in an appropriate complex basis (\ref{Icomplexbasis}). In the
pseudo-holomorphic case, these relations involve both $I_\alpha^{-\,
  q_{rs}}$ and $I_{\bar\alpha}^{+\, q_{rs}}= \big( I_\alpha^{-\,
  q_{sr}} \big)^\dag$, and therefore must be understood as relations
$\Rbar$ on the double quiver $\Qbar$. Let us demonstrate this
procedure explicitly on the quivers $Q^{k,l}$ above.

We start with the antifundamental double quiver
(\ref{Qbar01quiver}), and consider the set of relations
$\Rbar\,^{0,1}=\{r_1,r_2,r_3\}$ given by
\beq
r_1=a_1-a^*_2\, a^*_3 \ , \qquad r_2=a_2-a^*_3\, a_1^* \und
r_3=a_3-a_1^*\, a_2^* \ .
\label{R01}\eeq
In the corresponding factor algebra
$\C\,\Qbar\,^{0,1}/\langle\, \Rbar\,^{0,1} \rangle$, we have $r_\a=0$ for
$\a=1,2,3$. By demanding that the involutive algebra structure of
$\C\,\Qbar\,^{0,1}$ descend to the factor algebra under the canonical
projection $\C\,\Qbar\,^{0,1}\to \C\,\Qbar\,^{0,1}/\langle\, \Rbar\,^{0,1} \rangle$,
we also have the conjugate relations $r_\a^*:=\iota(r_\a)$ given by
\beq
r_1^*=a^*_1-a_3\, a_2 \ , \qquad r^*_2=a^*_2-a_1\, a_3 \und
r^*_3=a^*_3-a_2\, a_1 \ .
\label{R01conj}\eeq
The relations (\ref{R01})--(\ref{R01conj}) express commutativity of
the quiver diagram (\ref{Qbar01quiver}) along the primitive paths of $\Qbar\,^{0,1}$.

Let us look at the implications of the relations $r_\a=0=r_\a^*$ in
the factor algebra. From (\ref{R01})--(\ref{R01conj}) we obtain the
relations
\beq
\big(1_{(-1,-1)_1}-p_1\big)\, a_1=0 \ , \qquad \big(1_{(1,-1)_1}-p_2\big)\, a_2=0 \und
\big(1_{(0,-2)_0}-p_3\big)\, a_3=0 \ ,
\label{eparels}\eeq
with
\beq
p_1:=a_1\, a_3\, a_2 \ , \qquad p_2:=a_2\, a_1\, a_3 \und p_3:= a_3\,
a_2 \, a_1 \ ,
\label{padefs}\eeq
as well as relations
\beq
a_1^*\, a_1=p_3= a_3\, a_3^* \ , \qquad a_2^*\, a_2=p_1=a_1\, a_1^*
\und a_3^*\, a_3=p_2=a_2\, a_2^* \ ,
\label{astarastar}\eeq
and
\beq
a_1\, \big(1_{(0,-2)_0}-p_3\big)=0 \ ,\qquad a_2\,
\big(1_{(-1,-1)_1}-p_1\big)=0 \und a_3\, \big(1_{(1,-1)_1}-p_2\big)=0
\ .
\label{aeps}\eeq
From (\ref{eparels}) and (\ref{aeps}) it follows that the oriented
three-cycles (\ref{padefs}) of the original quiver (\ref{Q01quiver})
are equivalent to trivial paths in the factor algebra,
\beq
p_1=1_{(-1,-1)_1} \ , \qquad p_2=1_{(1,-1)_1} \und p_3=1_{(0,-2)_0} \
.
\label{perels}\eeq
Furthermore, from (\ref{astarastar}) it follows that the oriented
two-cycles $a_\a^*\, a_\a$ and $a_\a\, a_\a^*$ for $\a=1,2,3$ in the
double quiver (\ref{Qbar01quiver}) are also all equivalent to trivial
paths in $\C\,\Qbar\,^{0,1}/\langle\, \Rbar\,^{0,1} \rangle$.

It is readily checked that the basic relations (\ref{R01}) (and all
those implied by them above) are satisfied by the fundamental
representation basis (\ref{2.58}) for $\sltcL$. Similarly, one writes
down relations analogous to (\ref{R01}) for the fundamental double
quiver of (\ref{Q10quiver}), but with the signs in (\ref{R01})
reversed due to the opposite orientations between the quiver diagrams (\ref{Q10quiver})--(\ref{Q01quiver}). Arbitrary quivers $\Qbar\,^{k,l}$ are constructed by gluing
fundamental triangles together. The corresponding relations
$\Rbar\,^{0,1}$ and $\Rbar\,^{1,0}$ combine to give primitive
commutativity conditions on the double quiver diagram. From an
algebraic perspective, the relations
around elementary parallelograms in the quiver diagram mimick the
$\sltcL$ Lie algebra relations in higher irreducible representations $\widehat V^{k,l}$ for the basis given by (\ref{I78Bied})--(\ref{offdiagBied}).

Let us consider a particular subdiagram of an arbitrary double quiver
$\Qbar\,^{k,l}$ with the generic structure
\beq
\xymatrix{
(q-1,m+3)_{n\pm 1}  \ \ar@/^/[rr]|{ \ c_2 \ } \ar@/^/[dr]|{ \ c^*_3 \ } & & \  (q+1,m+3)_{n\pm1} \ 
\ \ar@/^/[dr]|{ \ b_3^* \ } \ar@/^/[ll]|{ \ c_2^* \ }  \ar@/^/[dl]|{ \ c_1 \ }
& \\
 &  \ (q,m)_n \ \ar@/^/[ur]|{ \ c^*_1 \ } \ar@/^/[dr]|{ \ a^*_3 \ } \ar@/^/[ul]|{ \ c_3 \ }
 \ar@/^/[rr]|{ \ a_2 \ } & &  \ (q+2,m)_n \ar@/^/[ul]|{ \ b_3 \ } \ar@/^/[ll]|{ \ a^*_2 \ } 
 \ar@/^/[dl]|{ \ a_1 \ } \\
 & &  \ (q+1,m-3)_{n\pm1} \ \ar@/^/[ul]|{ \ a_3 \ } \ar@/^/[ur]|{ \ a^*_1 \ } &
}
\label{Qklbarquiver}\eeq
For clarity we do not indicate multiple arrows from the original
quiver $Q^{k,l}$ involving degenerate weight vectors
$(q',m'\,)_{n\pm1}$; they contribute identical relations to those given
below. The primitive relations of $\Rbar\,^{k,l}$ expressing
equivalences of paths between opposite vertices connected by diagonal
arrows in the two
parallelograms of (\ref{Qklbarquiver}) are given by
\beq
s_1=c_1+c_3^*\, c_2^*-a_2^*\, b_3^* \und s_2=a_2+a_1^*\, a_3^*
-b_3^*\, c_1^* \ ,
\label{srels}\eeq
plus their conjugates $s_1^*:=\iota(s_1)$ and $s_2^*:=\iota(s_2)$. In general, the subset of relations $\Rbar\,^{k,l}$ in any elementary plaquette of
$\Qbar\,^{k,l}$ are always of the generic form (\ref{srels}), with
arrows set to $0$ whenever the corresponding head or tail vertices lie
outside the range (\ref{mnqjpm}). In particular, arrows on the
boundary of the original quiver $Q^{k,l}$ are always subject to
triangular relations such as those of
(\ref{R01})--(\ref{R01conj}). Note that commutativity is expressed
\emph{only} for paths between vertices that are joined by diagonal
arrows of $Q_1^{k,l}$; in fact, one easily checks that other
commutativity relations are inconsistent with the relations of
(\ref{R01})--(\ref{R01conj}) and (\ref{srels}). Thus the double quiver
$\Qbar$ is not described by a commutative quiver diagram after
imposing the relations $\Rbar$. For example, the complete set of
relations $\Rbar\,^{2,0}$ for the double quiver (\ref{Qbar20quiver})
comprise the three-term relations (\ref{srels}) plus an additional one
involving the $b_3$ arrow, and
six triangular relations of the form (\ref{R01}) involving the arrows
$a_1$, $a_3$, $b_1$, $b_2$, $c_2$ and $c_3$, together with their
conjugates obtained as images under the involution $\iota$ of~$\C\,\Qbar\,^{2,0}$.

Generally, one again establishes exactly as
before that all oriented $\ell$-cycles with $\ell\geq2$ in
$\Qbar\,^{k,l}$ (and hence with $\ell\geq3$ in $Q^{k,l}$) are
equivalent to trivial cycles in the corresponding factor algebra $\C\,
\Qbar\,^{k,l}/\langle\, \Rbar\,^{k,l} \rangle$ with the induced $*$-involution
$\iota$. Below we show that the relations $\Rbar$ for the
double quiver $\Qbar$ constructed in this way are in fact equivalent
to the minimal requirement for triviality of all oriented cycles on the \emph{original}
quiver $Q$. This means that the generating set of relations $\Rbar$ is
complete.

A representation $(V,\phi)$ of $Q$ is a representation of the quiver
with relations $(Q,R)$ if the collection of morphisms $\phi$ satisfy all relations
$r=\sum_i\, c_i\, p_i$ in the ideal $\langle R\rangle\subset\C Q$ of
the path algebra,
i.e. $\phi(r)= \sum_i\, c_i\, \phi(p_i)=0$. The category of linear
representations of $(Q,R)$ is equivalent to the category of left
modules over the factor algebra $\C Q/\langle R\rangle$. As before, we do not
distinguish between a $(Q,R)$-module and the corresponding $\C
Q/\langle R\rangle$-module. Since $R$ is a set of admissible relations for the
quiver $Q$, it defines an affine subvariety $\Rep(Q,R,V)\subset \Rep(Q,V)$ of the
representation variety (\ref{RepQV}) cut out by the polynomial
equations $\phi(r)=0$ coming from $R$. In subsequent sections, we will consider
unitary representations of the double quiver with relations $\big(\,
\Qbar\,,\, \Rbar\,\big)$, which thus have to satisfy all relations
spelled out above.

\smallskip

\noindent
{\bf The path groupoid $\mbf{\Psfbar\,Q}$. \ } We complete our description of the relations $\Rbar$
by demonstrating that the minimal requirement for absence of non-trivial cycles in the
original quiver $Q$ implies the generating set of relations $\Rbar$
constructed above for the double quiver $\Qbar$. For this, we need an
alternative way to study equivalence relations on the paths of our
quivers. We will use a categorical approach.

Let $\Psf Q$ be the additive path
category of the quiver $Q$; its objects are the vertices $Q_0$ and its
morphisms are paths in $Q$. Let $R$ be a generating set of admissible relations on
$Q$ which renders all oriented cycles equivalent to trivial paths in the
factor algebra $\C Q/\langle R\rangle$. Let $(R)$ be the minimal
equivalence relation on $\Psf Q$ containing $R$ such that if $r\in
(R)$, then $p\,r\, p'\in(R)$ for any pair of morphisms $p,p'$ of
$\Psf Q$ with $t(p)=h(r)$ and $h(p'\,)=t(r)$. Let $\Psf Q/(R)$ be the
corresponding factor category.

On the path category $\Psf\, \Qbar$ of the
double quiver of $Q$, we introduce an
equivalence relation by identifying $a\,a^*$ and $a^*\, a$ with the trivial
paths $1_{t(a)}$ and $1_{h(a)}$, respectively, for any arrow $a\in Q_1$. The
corresponding factor category is a groupoid $\Psfbar\, Q$ over the
base $Q_0$, called the
path groupoid of the original quiver $Q$, with source and target
maps $t,h:\Psfbar\,Q\rightrightarrows Q_0$; the equivalence classes
$[1_v]$ for $v\in Q_0$ are partial identities for the concatenation
product, and the inverse of a class $[\,\overline p\,]$ of
paths $\overline p$ on $\Qbar$ is $[\,\overline
p\,]^{-1}:=[\iota(\,\overline p\,)]$. The covariant functor ${\sf i}:\Psf Q\to
\Psfbar\,Q$ defined by $p\mapsto [p]$ is an embedding which is
\emph{universal}: If $\sf G$ is a groupoid and ${\sf
  v}:\Psf Q\to {\sf G}$ is a covariant functor, then $\sf v$ can be
extended to $\Psf\,\Qbar$ and hence to a unique functorial morphism of groupoids
$\bar{\sf v}:\Psfbar\,Q\to{\sf G}$ such that ${\sf v}=\bar{\sf v}
\circ {\sf i}$.

Using the embedding $\sf i$, we define ${\bf \langle\!\langle} R{\bf \rangle\!\rangle}$ to be the smallest
normal subgroupoid of $\Psfbar\,Q$ containing ${\sf i}(R)$, i.e. for
any $x\in\Psfbar\,Q(v,v'\,)$ with $v,v'\in Q_0$ and $r\in{\bf \langle\!\langle}
R{\bf \rangle\!\rangle}(v)$, the element $x\,r\, x^{-1}$
belongs to ${\bf \langle\!\langle} R{\bf \rangle\!\rangle}(v'\,)$. One can
then construct the quotient groupoid
$\Psfbar\,Q/{\bf\langle\!\langle}R{\bf\rangle\!\rangle}$ in the usual
way, i.e. elements $x,y\in \Psfbar\,Q(v,v'\,)$ are equivalent if and
only if $x^{-1}\,y$ belongs to the subgroup ${\bf \langle\!\langle}
R{\bf \rangle\!\rangle}(v)\subset \Psfbar\,Q(v,v)$. There is
a natural functorial equivalence of factor categories
\beq
\Psfbar\, Q\,\big/\,{\bf\langle\!\langle}R{\bf\rangle\!\rangle} \ \xrightarrow{ \ \approx
  \ } \ \Psf\,\Qbar\, \big/ \, (\, \Rbar\,) \ ,
\label{PQequiv}\eeq
extending the injective map $\Psf Q/(R) \to \Psfbar\,Q/{\bf\langle\!\langle}R{\bf\rangle\!\rangle}$.

For example, the generating set of relations $R^{0,1}$ for the
antifundamental quiver (\ref{Q01quiver}) are given by
(\ref{perels}). By taking inverses in
the path groupoid $\Psfbar\, Q^{0,1}$, one can reverse the steps
in (\ref{R01})--(\ref{perels}) to arrive at the generating set
of relations $\Rbar\,^{0,1}$ given by (\ref{R01})--(\ref{R01conj}). For the general quivers $Q^{k,l}$, the relations
$R^{k,l}$ are obtained by demanding that the arrows in
$Q_1^{k,l}$ around each of the three elementary triangles in
(\ref{Qklbarquiver}) all compose to trivial paths.

\section{Equivariant dimensional reduction and quiver bundles\label{EGTQB}}

\noindent
{\bf $\mbf G$-equivariant gauge fields. \ } In this section we
consider the dimensional reduction of invariant connections on equivariant
vector bundles over the product manifold (\ref{3.1}), and show that it
naturally defines quiver bundles of the quiver $Q$ and its double
$\Qbar$ described in
section~\ref{Doublereps}. The isotopical decomposition (\ref{Ecaldecomp}) associated to a $G$-module
decomposition (\ref{3.2})--(\ref{3.7}) is specifed by $n$ complex
vector bundles $E_1,\dots,E_n$, with trivial $H$-action, over the smooth manifold $M$ of
dimension $d$. We
choose an $H$-invariant hermitian structure on each bundle $E_r\to M$, and let
$A^r$ be a unitary connection on~$E_r$. This data defines a
hermitian vector bundle $\Ecal$ over $\X^{d+h}$ of rank
\beq
N=\sum_{r=1}^n \, N_r\, q_r \ ,
\label{ksumrank}\eeq
where $N_r$ is the rank of $E_r$ and the integers $q_r$ are introduced in (\ref{3.2}). Without loss of generality~\cite{DS} we assume
$c_1(\Ecal)=0$, and that the structure group of $\Ecal$ is $\surm(N)$.

The matrices 
\begin{equation}\label{3.10}
\widetilde I_i:=
\begin{pmatrix}{\bf 1}_{N_1}\otimes I^{q_1}_i &0& \dots &0\\
0&\ddots&&\vdots\\
\vdots&&\ddots&0\\
0&\dots&0&{\bf 1}_{N_n}\otimes I^{q_n}_i
\end{pmatrix}
\end{equation}
for $i=h+1,\dots,\dim G$
are generators of a reducible unitary representation of the
group $H$ on a complex vector space $\widetilde V\cong \C^N$.
Introduce a gauge connection
\begin{equation}\label{3.8}
A:=
\begin{pmatrix}A^1\otimes{\bf 1}_{q_1}&0&\dots &0\\
0&\ddots&&\vdots\\
\vdots&&\ddots&0\\
0&\dots &0&A^n\otimes{\bf 1}_{q_n}
\end{pmatrix}
\end{equation}
on the bundle $\bigoplus_r\, E_r$ over $M$, which by construction satisfies the identity
\begin{equation}\label{3.9}
\big[A\,,\, \widetilde I_i \,\big]=0 \ .
\end{equation}
Its gauge group 
\beq
\Uscr(\widetilde V)=\Big(\, \prod_{r=1}^n\, \urm(N_r) \,\Big) \ \Big/ \ \uo
\label{Ucentral}\eeq
is the centraliser of the image of the homomorphism $H\to\suk$ generated by $\widetilde I_i$. 
Then any $G$-equivariant connection
$\Acal$ on the bundle $\Ecal$ over $\X^{d+h}=M\times G/H$ in
(\ref{Ecaldecomp}) is a natural extension of (\ref{3.8}) given by
\begin{equation}\label{3.11}
\Acal =A+\widetilde I_i\, e^i+X_a\, e^a\ ,
\end{equation}
where $X_a=\varrho(I_a)\in \sukL$ for $a=1,\dots,h$ are
matrices which depend only on the coordinates of
$M$ and locally form the components of a linear map
$\varrho:\mfrak\to\sukL$. The gauge potential (\ref{3.11}) involves the
canonical connection $\widetilde a^0=e^i\, \widetilde I_i$, which will
contribute background $H$-fluxes proportional to $\widetilde I_i$ in the
expressions for the invariant gauge fields below.

For the curvature two-form $\Fcal$, we use the Maurer-Cartan equations (\ref{2.55}) to obtain
\bea
\Fcal &:=& \diff\Acal +\Acal\wedge\Acal \nonumber \\[4pt] &=& F +
\big(\diff X_a+ [A, X_a] \big)\wedge
e^a \nonumber \\ && 
-\, \mbox{$\frac{1}{2}$}\, \big(f^i_{ab}\, \widetilde I_i+f^c_{ab}\, X_c-
[X_a, X_b] \big)\, e^a\wedge e^b +
\big([\widetilde I_i, X_a]-f^b_{ia}\, X_b \big) \, e^i\wedge e^a \ ,
\label{3.12}\eea
where $F=\diff A+A\wedge A$. The last term involving the one-forms
$e^i=e^i_a\, e^a$ in (\ref{3.12})
spoils $G$-invariance and so one should impose the conditions
\begin{equation}\label{3.13}
\big[\, \widetilde I_i\,,\, X_a\big]=f^b_{ia}\, X_b\ ,
\end{equation}
which ensure that the linear map $\varrho:\mfrak\to \sukL$ is
equivariant, i.e. it commutes with the action of~$H$.
For $\widetilde I_i$ given in (\ref{3.10}), the general solution of (\ref{3.13})
has the form
\begin{equation}\label{3.14}
X_a=
\begin{pmatrix}
0&\p_{N_{12}}\otimes I^{q_{12}}_a&\dots&\p_{N_{1n}}\otimes I^{q_{1n}}_a \\
\p_{N_{21}}\otimes I^{q_{21}}_a&0&\ddots&\vdots\\
\vdots&\ddots&\ddots&\p_{N_{n-1\,n}}\otimes I^{q_{n-1\,n}}_a \\
\p_{N_{n1}}\otimes I^{q_{n1}}_a&\dots &\p_{N_{n\,n-1}}\otimes I^{q_{n\,n-1}}_a&0
\end{pmatrix}\ ,
\end{equation}
where $\p_{N_{rs}}$ are any complex $N_r\times N_s$ matrices which
depend only
on the coordinates of $M$ and can be identifed with sections (Higgs
fields) of the bundles $E_r\otimes E_s^\vee$ over $M$.

This solution generically defines a
$\Qbar$-bundle on $M$. The bifundamental scalar fields
$\phi_{N_{rs}}$ in $\Hom(E_s,E_r)$ are situated
on the arrows $\Qbar_1$ of the double of the quiver $Q$. The bundles $E_r$ are sub-bundles of the $H$-equivariant
complex vector bundle $E$ on $M$ with isotopical decomposition
\beq
E=\bigoplus_{r=1}^n\, E_r\otimes V_{q_r} \ ,
\label{Eisodecomp}\eeq
where $V_{q_r}$ is the $H$-equivariant vector bundle $M\times V_{q_r}$; it defines a locally free module over the bundle of path
algebras $\C\, \Qbar$ of the quiver $\Qbar$.
For anti-hermitian matrices $X_a\in\sukL$ we use (\ref{3.7}) to infer
\begin{equation}\label{3.15}
\p_{N_{rs}}^\dagger=\p_{N_{sr}}\ ,
\end{equation}
and the quiver bundle thus induces a hermitian $\Qbar$-bundle of the
double quiver of $Q$.

Thus for the $G$-equivariant gauge field $\Fcal$ we have
\begin{equation}\label{3.17}
\Fcal = F+ DX_a\wedge e^a-\mbox{$\frac{1}{2}$}\, \big(f_{ab}^i\, \widetilde I_i +
f_{ab}^c\, X_c-[X_a, X_b] \big)\, e^a\wedge e^b\ ,
\end{equation}
where
\begin{equation}\label{3.18}
DX_a:=\diff X_a + [A, X_a]
\end{equation}
is the gauge covariant derivative of $X_a$ on $M$.
In a complex basis (\ref{Icomplexbasis}) for $\gfrak^\C$ with redefinitions (\ref{2.34}), one has
\bea
\Fcal &=& F+ DY_{\a}\wedge \Th^{\a}+DY_{\ab}\wedge \Th^{\ab}-
\mbox{$\frac{1}{2}$} \, \big(-\im\,C_{\a\b}^i\, \widetilde I_i + C_{\a\b}^{\g}\, Y_{\g}+
C_{\a\b}^{\bar\g}\, Y_{\bar\g}-[Y_\a, Y_\b]\big)\, \Th^\a\wedge \Th^\b
\nonumber \\ &&
-\, \big(-\im\,C_{\a\bb}^i\, \widetilde I_i + C_{\a\bb}^{\g}\, Y_{\g}+
C_{\a\bb}^{\bar\g}\,Y_{\bar\g}-[Y_\a, Y_{\bb}]\big)\, \Th^\a\wedge
\Th^{\bb} \nonumber \\ &&
-\, \mbox{$\frac{1}{2}$}\,\big(-\im\,C_{\ab\bb}^i\, \widetilde I_i + C_{\ab\bb}^{\g}\,
Y_{\g}+
C_{\ab\bb}^{\bar\g}\, Y_{\bar\g}-[Y_{\ab}, Y_{\bb}]\big)\, \Th^{\ab}\wedge \Th^{\bb}\ ,
\label{3.19}\eea
where
\begin{equation}\label{3.20}
Y_\a :=\mbox{$\frac{1}{2}$}\, (X_{2\a-1}-\im \, X_{2\a})\und Y_{\ab}=-Y_\a^\+\ .
\end{equation}
In the following we will also use a rescaled basis $\widetilde\Th^\a$,
$\widetilde\Th^{\ab}$ (as in (\ref{2.64})--(\ref{Ctildenot0})); in this case one should substitute $Y_\a\to\widetilde Y_\a$,
$Y_{\ab}\to\widetilde Y_{\ab}$ and $C^A_{BC}\to\widetilde C^A_{BC}$ in (\ref{3.19}).

\smallskip

\noindent
{\bf Comparison with coset space dimensional reduction. \ } Suppose
that the structure group of the bundle $\Ecal$ over $\X^{d+h}$
is not the unitary group SU($N$) but an arbitrary compact Lie group $K$ with generators $L_{\hat A}$
and structure constants $l^{\hat A}_{\hat B\hat C}$. In the coset space dimensional
reduction scheme~\cite{5,6,7}, one assumes that the group $H$
appearing in the
coset $G/H$ is embedded in $K$ and may be represented by generators
(\ref{3.10}). The matrices $X_a$ in (\ref{3.11}) must then belong to
the Lie algebra $\mathfrak{k}$ of $K$,
$X_a=X_a^{\hat A}\, L_{\hat A}$, and the equations (\ref{3.13}) yield the
constraint equations
\begin{equation}\label{3.16}
l^{\hat A}_{i\hat B}\, X_a^{\hat B}=f_{ia}^b\, X_b^{\hat A}
\end{equation}
on $X_a$ which ensure that they are intertwining operators connecting the induced representations
of $H$ in $K$ and in $G$~\cite{7}. It is difficult to solve
the equations (\ref{3.16})
in terms of unconstrained scalar fields, except in some special cases including the
case where $G\subseteq K$~\cite{7}. In contrast, the formula
(\ref{3.14}) for the case $K=\suk$ gives an explicit solution
$X_a$ which does not arise in the coset space dimensional
reduction scheme (except for some special cases) and relates it
to a quiver bundle on~$M$. One
can extend all these considerations to arbitrary gauge groups~$K$.

\smallskip

\noindent
{\bf $\mbf{\Qbar\,^{0,1}}$-bundles. \ } For any concrete choice of quiver, gauge
potential (\ref{3.8}), and matrices (\ref{3.10}) and (\ref{3.14}), one can write
the equivariant gauge fields explicitly in terms of $A^r$ and $\p_{N_{rs}}$. For example, by choosing
the matrices (\ref{3.2})--(\ref{3.3}) as generators of the antifundamental $3\times3$
representation $\widehat V^{0,1}$ of SU(3), we obtain from (\ref{3.11}) and (\ref{2.56}) an SU(3)-invariant gauge
connection $\Acal$ as~\cite{21}
\begin{equation}\label{4.16}
\Acal = \begin{pmatrix}A^1\otimes 1 + {\bf 1}_{N_1}\otimes 2b&
\s_2\,\p^{\+}_2\otimes\widetilde\Th^2&-\s_1\,\p_1\otimes\widetilde\Th^{\1}\\[6pt]
-\s_2\,\p_2\otimes\widetilde\Th^{\2}&A^2\otimes 1 + {\bf 1}_{N_2}\otimes (\widehat a_+-b)&
\s_3\,\p^{\+}_3\otimes\widetilde\Th^3 \\[6pt]
\s_1\,\p^{\+}_1\otimes\widetilde\Th^1&-\s_3\,\p_3\otimes\widetilde\Th^{\3}
&A^3\otimes 1 - {\bf 1}_{N_3}\otimes (\widehat a_+ +b)
\end{pmatrix} \ ,
\end{equation}
where $A^1, A^2$ and $A^3$ are $\urmL(N_1)$-, $\urmL(N_2)$- and $\urmL(N_3)$-valued gauge
potentials on hermitian vector bundles $E_1$, $E_2$ and $E_3$ over $M$ with ranks
$N_1$, $N_2$ and $N_3$, respectively, such that
\beq
N_1+N_2+N_3=N={\rm rank}(\Ecal) \ , 
\eeq
while the bundle morphisms
$\phi_1\in \Hom (E_3, E_1)$, $\phi_2\in \Hom (E_1, E_2)$
and $\phi_3\in \Hom (E_2, E_3)$ are
bifundamental scalar fields on $M$. 
The bundles $E_r$ are sub-bundles of the quiver bundle
\begin{equation}\label{4.17}
E^{0,1}=E_1\otimes\C \ \oplus \ E_2\otimes\C \ \oplus \ E_3\otimes\C
\end{equation}
over $M$ for the quiver $\Qbar\,^{0,1}$
in (\ref{Qbar01quiver}), where the factors $\C$ denote trivial
$H$-equivariant line bundles over $M$ arising from the decomposition
of the representation $\widehat V^{0,1}\cong\C^3$ into irreducible representations
of $H=\uo\times \uo$. In (\ref{Qbar01quiver}) the bundle $E_1$ is assigned to the vertex $(-1,-1)_1$, $E_2$ to $(1,-1)_1$, and $E_3$ to $(0,2)_0$.

For the curvature $\Fcal =\diff\Acal+\Acal\wedge\Acal=(\Fcal^{rs})$
of the invariant connection (\ref{4.16}) we obtain
\begin{eqnarray}
\Fcal^{11}&=&F^{1}-\s_1^2\,\big({\bf 1}_{N_1}-\p_1\, \p_1^\+ \big)\, 
\widetilde\Th^1\wedge\widetilde\Th^{\1}+\s_2^2\, \big({\bf
  1}_{N_1}-\p_2^\+\, \p_2\big) \,
\widetilde\Th^2\wedge\widetilde\Th^{\2}\ ,\label{4.18}\\[4pt]
\Fcal^{22}&=&F^{2}-\s_2^2\,\big({\bf 1}_{N_2}-\p_2\, \p_2^\+ \big)\,
\widetilde\Th^2\wedge\widetilde\Th^{\2}+\s_3^2\,\big({\bf
  1}_{N_2}-\p_3^\+\, \p_3 \big) \,
\widetilde\Th^3\wedge\widetilde\Th^{\3}\ ,\label{4.19}\\[4pt]
\Fcal^{33}&=&F^{3}- \s_3^2\, \big({\bf
  1}_{N_3}-\p_3 \, \p_3^\+\big) \,
\widetilde\Th^3\wedge\widetilde\Th^{\3}+\s_1^2\, \big({\bf 1}_{N_3}-\p_1^\+\, \p_1
\big)\, 
\widetilde\Th^1\wedge\widetilde\Th^{\1} \ ,\label{4.20}\\[4pt]
\Fcal^{13}&=& -\s_1\, \big(\diff\phi_{1}+A^{1}\, \phi_{1}-
\phi_{1}\, A^{3} \big)\wedge\widetilde\Th^{\1} - \s_2\,
\s_3\, \big(\p_1-\p_2^\+\, \p^\+_3\big)\, 
\widetilde\Th^2\wedge\widetilde\Th^{3}\ ,\label{4.21}\\[4pt]
\Fcal^{21}&=& -\s_2\,\big(\diff\phi_2 + A^{2}\, \phi_{2}-
\phi_{2}\, A^{1} \big)\wedge\widetilde\Th^{\2}-\s_1\,
\s_3\, \big(\p_2-\p_3^\+\, \p_1^\+\big)\, 
\widetilde\Th^3\wedge\widetilde\Th^{1}\ ,\label{4.22}\\[4pt]
\Fcal^{32}&=& -\s_3\,\big(\diff\phi_3 + A^{3}\, \phi_{3}-
\phi_{3}\, A^{2}\big)\wedge\widetilde\Th^{\3}-\s_1\,
\s_2\, \big(\p_3-\p_1^\+\, \p_2^\+\big)\, 
\widetilde\Th^1\wedge\widetilde\Th^{2}\ ,\label{4.23}
\end{eqnarray}
plus their hermitian conjugates $\Fcal^{sr}=-(\Fcal^{rs})^{\+}$ for $r\ne s$.
In (\ref{4.18})--(\ref{4.23}) the superscripts $r,s$ label $N_r\times N_s$
blocks in $\Fcal$, and we have suppressed tensor products in order to simplify
notation. Here $F^r=\diff A^r + A^r\wedge A^r$
is the curvature of the connection $A^r$ on the complex vector bundle
$E_r\to M$.

\smallskip

\noindent
{\bf $\mbf{\Qbar\,^{1,0}}$-bundles. \ } Let us now describe how the analogous constructions for the fundamental quiver (\ref{Q10quiver}) are related to those of the $\Qbar\,^{0,1}$-bundles above. The three-dimensional representations $\widehat V^{1,0}$ and $\widehat V^{0,1}$ of $\sut$ are related as $\widehat V^{1,0}=-\big(\widehat V^{0,1}\big)^\top$. The constructions of section~\ref{F3Kahler} are all based on the representation $\widehat V^{0,1}$ and the quiver $Q^{0,1}$. To describe the quiver $Q^{1,0}$, one reverses the orientation of $\F_3$; this amounts to interchanging coordinates on $\F_3$ in all formulas of section~\ref{F3Kahler} with their complex conjugates, the forms $\Th^\a$ with $\Th^\ab$, and the matrices $I_\a^-$ with $I_\ab^+$ in (\ref{2.58}) (but keeping the same signs in (\ref{2.31}) and (\ref{2.23})). Then via the replacement $\widehat\Acal\to -\widehat\Acal\,^\top$ one gets the analogue of the flat connection (\ref{2.56}) for the irreducible representation $\widehat V^{1,0}$ and the quiver $Q^{1,0}$.

In this way the $\sut$-invariant gauge connection (\ref{4.16}) is modified to
\beq
\Acal'=\begin{pmatrix}A^1\otimes 1 - {\bf 1}_{N_1}\otimes 2b&
\s_2\,\psi_2\otimes\widetilde\Th^{\2}&-\s_1\,\psi^{\+}_1\otimes\widetilde\Th^1\\[6pt]
-\s_2\,\psi^{\+}_2\otimes\widetilde\Th^2&A^2\otimes 1 - {\bf 1}_{N_2}\otimes (\widehat a_+-b)&
\s_3\,\psi_3\otimes\widetilde\Th^{\3} \\[6pt]
\s_1\,\psi_1\otimes\widetilde\Th^{\1}&-\s_3\,\psi^{\+}_3\otimes\widetilde\Th^3&
A^3\otimes 1 + {\bf 1}_{N_3}\otimes (\widehat a_+ +b)
\end{pmatrix} \ ,
\label{Acalprime}\eeq
where $\psi_1\in\Hom(E_1,E_3)$, $\psi_2\in\Hom(E_2,E_1)$ and $\psi_3\in\Hom(E_3,E_2)$. The vector bundle $E_1$ is now associated with the vertex $(1,1)_1$ of the quiver (\ref{Q10quiver}), $E_2$ with $(-1,1)_1$, and $E_3$ with $(0,-2)_0$. The corresponding curvature two-form $\Fcal'=\diff\Acal'+\Acal'\wedge \Acal'$ is given by
\begin{eqnarray}
\Fcal'\,^{11}&=&F^{1}+\s_1^2\,\big({\bf 1}_{N_1}-\psi_1^\+\, \psi_1 \big)\, 
\widetilde\Th^1\wedge\widetilde\Th^{\1}-\s_2^2\, \big({\bf
  1}_{N_1}-\psi_2\, \psi_2^\+\big) \,
\widetilde\Th^2\wedge\widetilde\Th^{\2}\ ,\label{4.18p}\\[4pt]
\Fcal'\,^{22}&=&F^{2}+\s_2^2\,\big({\bf 1}_{N_2}-\psi_2^\+\, \psi_2 \big)\,
\widetilde\Th^2\wedge\widetilde\Th^{\2}-\s_3^2\,\big({\bf
  1}_{N_2}-\psi_3\, \psi_3^\+ \big) \,
\widetilde\Th^3\wedge\widetilde\Th^{\3}\ ,\label{4.19p}\\[4pt]
\Fcal'\,^{33}&=&F^{3}+\s_3^2\, \big({\bf
  1}_{N_3}-\psi_3^\+ \, \psi_3\big) \,
\widetilde\Th^3\wedge\widetilde\Th^{\3}-\s_1^2\, \big({\bf 1}_{N_3}-\psi_1\, \psi_1^\+
\big)\, 
\widetilde\Th^1\wedge\widetilde\Th^{\1} \ ,\label{4.20p}\\[4pt]
\Fcal'\,^{31}&=& \s_1\, \big(\diff\psi_{1}+A^{3}\, \psi_{1}-
\psi_{1}\, A^{1} \big)\wedge\widetilde\Th^{\1}+\s_2\,
\s_3\, \big(\psi_1-\psi_3^\+\, \psi^\+_2\big)\, 
\widetilde\Th^2\wedge\widetilde\Th^{3}\ ,\label{4.21p}\\[4pt]
\Fcal'\,^{12}&=& \s_2\,\big(\diff\psi_2 + A^{1}\, \psi_{2}-
\psi_{2}\, A^{2} \big)\wedge\widetilde\Th^{\2}+\s_1\,
\s_3\, \big(\psi_2-\psi_1^\+\, \psi_3^\+\big)\, 
\widetilde\Th^3\wedge\widetilde\Th^{1}\ ,\label{4.22p}\\[4pt]
\Fcal'\,^{23}&=& \s_3\,\big(\diff\psi_3 + A^{2}\, \psi_{3}-
\psi_{3}\, A^{3}\big)\wedge\widetilde\Th^{\3}+ \s_1\,
\s_2\, \big(\psi_3-\psi_2^\+\, \psi_1^\+\big)\, 
\widetilde\Th^1\wedge\widetilde\Th^{2}\ . \label{4.23p}
\end{eqnarray}

\smallskip

\noindent
{\bf $\mbf{\Qbar\,^{k,l}}$-bundles. \ } The previous examples can be
generalised to the quivers $Q^{k,l}$ associated to higher irreducible
representations $\widehat V^{k,l}$ of $\sut$ given by
(\ref{I78Bied})--(\ref{lambdaklnm}). Let $E_{(q,m)_n}\to M$ for
$(q,m)_n\in W^{k,l}$ be hermitian vector bundles of rank $N_{(q,m)_n}$
with $\sum_{(q,m)_n\in W^{k,l}}\, N_{(q,m)_n}= N$. Let $A^{(q,m)_n}$
be a unitary connection on $E_{(q,m)_n}$, and choose bifundamental scalar
fields 
$$
\phi^{1 \ (\pm)}_{(q,m)_n} \ \in \ \Hom\big(E_{(q-1,m-3)_{n\pm1}} \,,\,
E_{(q,m)_n} \big) \ , \qquad \phi^2_{(q,m)_n} \ \in \
\Hom\big(E_{(q+2,m)_{n}} \,,\,
E_{(q,m)_n}\big)
$$ 
\beq
\mbox{and} \qquad \phi^{3 \ (\pm)}_{(q,m)_n} \ \in \
\Hom\big(E_{(q-1,m+3)_{n\pm1}} \,,\,
E_{(q,m)_n}\big) \ .
\label{bifundkl}\eeq
Let $\widehat\Pi_{(q,m)_n}$ be the
  hermitian projection of the $H$-restriction of $\widehat V^{k,l}$ onto the
  one-dimensional representation of $H=\uo\times\uo$ with weight vector
  $(q,m)_n\in W^{k,l}$, and let $\Pi_{(q,m)_n}=\phi\big(1_{({q,m})_n}\big)$ be
    the hermitian projection onto the sub-bundle $E_{(q,m)_n}$ of the quiver
    bundle
\beq
E^{k,l}=\bigoplus_{(q,m)_n\in W^{k,l}}\, E_{(q,m)_n}\otimes\C
\label{Eklquiverbun}\eeq
over $M$ for the quiver $\Qbar\,^{k,l}$.

Then an $\sut$-equivariant gauge connection $\Acal$ on the
corresponding bundle (\ref{Ecaldecomp}) over $\X^{d+6}$ is given by
\bea
\Acal &=& \sum_{(q,m)_n\in W^{k,l}}\, \bigg[\, A^{(q,m)_n}\otimes
\widehat\Pi_{(q,m)_n}+ \Pi_{(q,m)_n}\otimes \Big(2q\,b-\mbox{$\frac12$}\,
(q-m)\,\big(\widehat a_++b\big)\Big)\,\widehat\Pi_{(q,m)_n} \nonumber \\ && +\,
\sum_\pm\, \s_1\,\Big(\phi^{1 \ (\pm)}_{(q,m)_n}{}^\dag \otimes
\widehat\Pi_{(q-1,m-3)_{n\pm1}} \, I_1^-\, \widehat\Pi_{(q,m)_n}\,
\widetilde\Theta^1 \nonumber \\ && \qquad\qquad\qquad -\,\phi^{1 \ (\pm)}_{(q,m)_n} \otimes
\widehat\Pi_{(q,m)_n} \, I_\1^+ \, \widehat\Pi_{(q-1,m-3)_{n\mp1}}\,
\widetilde\Theta^\1\Big) \nonumber\\ && +\, \s_2 \,
\Big(\phi^{2}_{(q,m)_n}{}^\dag \otimes
\widehat\Pi_{(q+2,m)_{n}} \, I_2^-\, \widehat\Pi_{(q,m)_n}\,
\widetilde\Theta^2 - \phi^{2}_{(q,m)_n} \otimes
\widehat\Pi_{(q,m)_n} \, I_\2^+ \, \widehat\Pi_{(q+2,m)_{n}}\,
\widetilde\Theta^\2\Big) \nonumber\\ && +\,
\sum_\pm\, \s_3\, \Big( \phi^{3 \ (\pm)}_{(q,m)_n}{}^\dag \otimes
\widehat\Pi_{(q-1,m+3)_{n\pm1}} \, I_3^-\, \widehat\Pi_{(q,m)_n}\,
\widetilde\Theta^3 \nonumber \\ && \qquad\qquad\qquad -\,\phi^{3 \ (\pm)}_{(q,m)_n}\otimes
\widehat\Pi_{(q,m)_n} \, I_\3^+\, \widehat\Pi_{(q-1,m+3)_{n\mp1}}\,
\widetilde\Theta^\3\Big) \, \bigg] \ .
\label{AcalQkl}\eea
The diagonal curvature matrix elements of (\ref{3.19}) at each vertex
$(q,m)_n\in W^{k,l}$ of the weight diagram for $\widehat V^{k,l}$ are
given by
\bea
\Fcal^{(q,m)_n\, (q,m)_n} &=& F^{(q,m)_n} \label{Fcalkldiag} \\ && +\,
\frac{\s_1^2}{24}\,
\widetilde\Th^1\wedge \widetilde\Th^\1 \, \sum_\pm\, \bigg[\, \mbox{$\frac{n\pm
    q+1\pm1}{n+1\pm1}$}\, \lambda_{k,l}^\mp(n\pm 1,m+3)^2\,\Big({\bf
  1}_{N_{(q,m)_n}}- \phi^{1 \ (\pm)}_{(q,m)_n}\, \phi^{1 \
  (\pm)}_{(q,m)_n}{}^\dag\Big) \nonumber\\ && \qquad \qquad\qquad -\, \mbox{$\frac{n\mp
    q\mp1}{n+1}$}\, \lambda_{k,l}^\mp(n,m)^2\, \Big({\bf
  1}_{N_{(q,m)_n}}- \phi^{1 \ (\pm)}_{(q+1,m+3)_{n\mp1}}{}^\dag\, \phi^{1 \
  (\pm)}_{(q+1,m+3)_{n\mp1}} \Big)\, \bigg] \nonumber \\ && +\,
\frac{\s_2^2}{48}\, \widetilde\Th^2\wedge
\widetilde\Th^\2 \, \bigg[ (n-q)\, (n+q+2)\, \Big({\bf
  1}_{N_{(q,m)_n}}- \phi^{2}_{(q,m)_n}\, \phi^{2}_{(q,m)_n}{}^\dag \Big) \nonumber\\ && \qquad \qquad \qquad \qquad -\, (n+q)\, (n-q+2)\, \Big({\bf
  1}_{N_{(q,m)_n}}- \phi^{2}_{(q-2,m)_n}{}^\dag \,
\phi^{2}_{(q-2,m)_n}\Big) \bigg] \nonumber \\ && +\, \frac{\s_3^2}{24}\,
\widetilde\Th^3\wedge \widetilde\Th^\3 \, \sum_\pm\, \bigg[\, \mbox{$\frac{n\mp
    q\pm1}{n+1}$}\, \lambda_{k,l}^\pm(n,m)^2\,\Big({\bf
  1}_{N_{(q,m)_n}}- \phi^{3 \ (\pm)}_{(q,m)_n}\, \phi^{3 \
  (\pm)}_{(q,m)_n}{}^\dag \Big) \nonumber\\ && \qquad -\, \mbox{$\frac{n\pm
    q+1\pm1}{n+1\pm1}$}\, \lambda_{k,l}^\pm(n\mp1,m-3)^2\, \Big({\bf
  1}_{N_{(q,m)_n}}- \phi^{3 \ (\pm)}_{(q+1,m-3)_{n\mp1}}{}^\dag \, \phi^{3 \
  (\pm)}_{(q+1,m-3)_{n\mp1}} \Big)\, \bigg] \nonumber
\eea
where $F^{(q,m)_n}= \diff A^{(q,m)_n}+ A^{(q,m)_n}\wedge
A^{(q,m)_n}$. The remaining non-vanishing off-diagonal matrix elements of the
curvature two-form $\Fcal$ are given by
\bea
\Fcal^{(q,m)_n\, (q-1,m-3)_{n\pm1}} &=&
\mbox{$\frac{\lambda_{k,l}^\pm(n\pm1,m-3)}{\sqrt{48}}$}\,
  \bigg\{\s_1\,\sqrt{\mbox{$\frac{2(n\pm q+1\pm1)}{n+1\pm1}$}} ~ D \phi^{1 \
    (\pm)}_{(q,m)_n} \wedge\widetilde\Th^\1 \label{Fcalkloffdiag1}
  \\ && +\, \s_2\, \s_3\, \widetilde\Th^2\wedge \widetilde\Th^3 \,\bigg[\, 
  \sqrt{\mbox{$\frac{((n+1\pm1)^2-q^2)\, (n\pm
        q+1\pm1)}{24(n+1\pm1)}$}}\nonumber \\ && \qquad \qquad \qquad \qquad \qquad \times \, \Big( \phi^{1 \
    (\pm)}_{(q,m)_n}- \phi^{3 \
  (\mp)}_{(q+1,m-3)_{n\pm 1}}{}^\dag \, \phi^{2}_{(q-1,m-3)_{n\pm1}}{}^\dag \Big) \nonumber\\ && - \,
\sqrt{\mbox{$\frac{(n+q) \, (n-q+2)\, (n\pm
        q+1\pm 1)}{24(n+1\pm1)}$}}\, \Big( \phi^{1 \
    (\pm)}_{(q,m)_n}- \phi^{2}_{(q-2,m)_{n}}{}^\dag\, \phi^{3 \
  (\mp)}_{(q-1,m-3)_{n\pm 1}}{}^\dag
\Big) \, \bigg]\bigg\}  \ , \nonumber \\[4pt]
\Fcal^{(q,m)_n\, (q+2,m)_{n}} &=& \s_2\, \sqrt{\mbox{$\frac{(n+q)\,
      (n+q+2)}{48}$}} ~ D\phi^{2}_{(q,m)_n} \wedge\widetilde\Th^\2
\nonumber\\ && +\, \mbox{$\frac{\s_1\, \s_3}{24}$}\,
\widetilde\Th^3\wedge\widetilde\Th^1 \, \sum_\pm\, \bigg[\,
\mbox{$\frac{n\mp q+1\mp1}{n+1}$}\, \lambda_{k,l}^\pm(n,m)^2
\label{Fcalkloffdiag2} \\ && \qquad \qquad \qquad \qquad \qquad \times
\, \Big(\phi^{2}_{(q,m)_n}- \phi^{1 \
  (\pm)}_{(q+1,m+3)_{n\mp 1}}{}^\dag \,\phi^{3 \
  (\mp)}_{(q+2,m)_{n}}{}^\dag  \Big) \nonumber \\ && -\,
\mbox{$\frac{n\pm q+1\pm1}{n+1\pm1}$} \,
\lambda^\mp_{k,l}(n\pm1,m-3)^2 \, \Big( \phi^{2}_{(q,m)_n}-  \phi^{3 \
  (\pm)}_{(q+1,m-3)_{n\mp 1}}{}^\dag \, \phi^{1 \
  (\mp)}_{(q+2,m)_{n}}{}^\dag \Big) \, \bigg]\ , \nonumber \\[4pt]
\Fcal^{(q,m)_n\, (q-1,m+3)_{n\pm1}} &=&
\mbox{$\frac{\lambda_{k,l}^\pm(n,m)}{\sqrt{48}}$}\,
  \bigg\{\s_3\,\sqrt{\mbox{$\frac{2(n\mp q+1\pm1)}{n+1}$}} ~ D\phi^{3 \
    (\pm)}_{(q,m)_n} \wedge\widetilde\Th^\3 \label{Fcalkloffdiag3}
  \\ && + \, \s_1\, \s_2\, \widetilde\Th^1\wedge \widetilde\Th^2 \,\bigg[\, 
  \sqrt{\mbox{$\frac{(n+q) \, (n-q+2)\, (n\mp
        q+1\mp 1)}{24(n+1)}$}}\nonumber \\ && \qquad \qquad \qquad \qquad \times \, \Big( \phi^{3 \
    (\pm)}_{(q,m)_n}- \phi^{2}_{(q-2,m)_{n}}{}^\dag \, \phi^{1 \
  (\mp)}_{(q-1,m+3)_{n\pm 1}}{}^\dag\Big) \nonumber\\ && - \,
\sqrt{\mbox{$\frac{((n+1\pm1)^2-q^2)\, (n\mp
        q+1\mp 1)}{24(n+1)}$}}\, \Big( \phi^{3 \
    (\pm)}_{(q,m)_n}- \phi^{1 \
  (\mp)}_{(q+1,m+3)_{n\pm 1}}{}^\dag \, \phi^{2}_{(q-1,m+3)_{n\pm1}}{}^\dag
\Big) \, \bigg]\bigg\}  \ , \nonumber
\eea
and
\bea
\Fcal^{(q+1,m-3)_{n\mp1}\, (q+1,m+3)_{n\pm1}} &=& \mbox{$\frac{\s_1\, \s_3}{24}\, \frac{n\mp q+1\mp1}{\sqrt{(n+1)\, (n+1\mp1)}}$}\, \lambda_{k,l}^\pm(n,m)\, \lambda_{k,l}^\pm(n\mp1,m-3)\, \widetilde\Th^1\wedge \widetilde\Th^\3 \nonumber \\ && \times\, \Big(\phi^{3 \ (\pm)}_{(q+1,m-3)_{n\mp1}} \, \phi^{1 \
  (\mp)}_{(q+1,m+3)_{n\pm1}}{}^\dag - \phi^{1 \
  (\mp)}_{(q+2,m)_{n}}{}^\dag \, \phi^{3 \
  (\pm)}_{(q+2,m)_{n}} \Big) \ , \nonumber \\ && \label{Fcalkloffdiag13} \\[4pt]
\Fcal^{(q-2,m)_{n\mp1}\, (q+1,m+3)_{n}} &=& \mbox{$\frac{\s_1\,\s_2}{24}\, \sqrt{\frac{((n+1)^2-q^2)(n\mp q+1)}{2(n+1\mp1)}}$}\, \lambda_{k,l}^\pm(n\mp1,m)\, \widetilde\Th^1\wedge\widetilde\Th^\2 \label{Fcalkloffdiag12} \\ &&\times\, \Big(\phi^{2}_{(q-2,m)_{n\mp1}} \, \phi^{1 \
  (\mp)}_{(q+1,m+3)_{n}}{}^\dag - \phi^{1 \
  (\mp)}_{(q-1,m+3)_{n}}{}^\dag \, \phi^{2}_{(q-1,m+3)_{n}} \Big) \ , \nonumber \\[4pt]
\Fcal^{(q+2,m)_{n}\, (q-1,m+3)_{n\pm1}} &=& \mbox{$\frac{\s_2\, \s_3}{24}\, \sqrt{\frac{(n-q)\, (n+q+2)\, (n\mp q+1\pm1)}{2(n+1)}}$}\, \lambda_{k,l}^\pm(n,m)\, \widetilde\Th^2\wedge \widetilde\Th^\3 \label{Fcalkloffdiag23} \\ && \times\, \Big(\phi^{3 \
  (\pm)}_{(q+2,m)_{n}} \, \phi^{2}_{(q-1,m+3)_{n\pm1}}{}^\dag -
\phi^{2}_{(q,m)_{n}}{}^\dag \, \phi^{3 \
  (\pm)}_{(q,m)_{n}} \Big) \ , \nonumber
\eea
plus their hermitian conjugates $\Fcal^{(q',m'\, )_{n'}\,
  (q,m)_n}=-\big( \Fcal^{(q,m)_n\, (q',m'\, )_{n'}}\big)^\dag$ for $(q',m'\,)_{n'} \neq (q,m)_n$. Here
\bea
D\phi^{1 \
    (\pm)}_{(q,m)_n} &=& \diff \phi^{1 \
    (\pm)}_{(q,m)_n}+ A^{(q,m)_n} \,\phi^{1 \
    (\pm)}_{(q,m)_n}- \phi^{1 \
    (\pm)}_{(q,m)_n}\, A^{(q-1,m-3)_{n\pm1}} \ , \label{Qklbifund1}\\[4pt]
D \phi^{2}_{(q,m)_n} &=& \diff\phi^{2}_{(q,m)_n}+A^{(q,m)_n} \, \phi^{2}_{(q,m)_n}-\phi^{2}_{(q,m)_n}\, A^{(q+2,m)_n} \ , \label{Qklbifund2} \\[4pt]
D \phi^{3 \
    (\pm)}_{(q,m)_n} &=& \diff \phi^{3 \
    (\pm)}_{(q,m)_n}+ A^{(q,m)_n} \,\phi^{3 \
    (\pm)}_{(q,m)_n}- \phi^{3 \
    (\pm)}_{(q,m)_n}\, A^{(q-1,m+3)_{n\pm1}}
\label{Qklbifund3}\eea
are bifundamental covariant derivatives of the Higgs fields on $M$.

\section{Quiver gauge theory\label{QGT}}

\noindent
{\bf Generalized instanton equations. \ } In this section we construct
natural gauge theories on the quiver bundles of section~\ref{EGTQB}. Fix a riemannian structure on the manifold $M$, and let $\Sigma$ be a differential
form of degree $d+h-4$ on the manifold $\X^{d+h}=M\times G/H$.
When the coset space $G/H$ is a nearly K\"ahler six-manifold, a natural choice for such a form on $\X^{d+6}$ is given by
\begin{equation}\label{3.21}
\Sigma=\mbox{$\frac{1}{3}$}\,\vk\,{\rm vol}_d\wedge\ome \ ,
\end{equation}
where $\ome$ is the fundamental two-form on $G/H$, $\vk\in\R$ is a constant and
vol$_d$ is the volume form on~$M$.

Let $\Ecal$ be a complex vector bundle over $\X^{d+h}$ endowed with a connection
$\Acal$. The $\Sigma$-anti-self-dual Yang-Mills equations are
defined as the first order equations~\cite{8} 
\begin{equation}\label{3.22}
\star\Fcal =-\Sigma\wedge\Fcal
\end{equation}
on the connection $\Acal$ with curvature $\Fcal =\diff\Acal +
\Acal\wedge\Acal$. Here $\star$ is the Hodge duality operator on~$\X^{d+h}$.

Taking the exterior derivative of (\ref{3.22}) and using the Bianchi identity, we obtain
\begin{equation}\label{3.23}
\diff \star \Fcal+\Acal\wedge\star\Fcal -\star\Fcal\wedge\Acal+\star\Hcal\wedge\Fcal =0\ ,
\end{equation}
where the three-form $\Hcal$ is defined by the formula
\begin{equation}\label{3.24}
\star\,\Hcal:=\diff\Sigma\ .
\end{equation}
The second order equations (\ref{3.23}) differ from the standard Yang-Mills equations
by the last term involving a three-form $\Hcal$ which can be identified with a totally
antisymmetric torsion on $\X^{d+h}$. This torsion term naturally appears in
string theory compactifications with fluxes where it is identified
with a supergravity three-form field~\cite{9,10}.\footnote{For recent discussions of heterotic string theory
with torsion see e.g.~\cite{11,12,13} and references therein.} When $\Fcal$ solves the first order equations (\ref{3.22}), the torsion term
in the Yang-Mills equations (\ref{3.23}) can vanish in certain instances. For example, on
nearly K\"ahler six-manifolds $\X^{6}$ the three-form $\diff\Sigma$ from
(\ref{3.21}) (with $d=0$) is a sum of $(3,0)$-
and $(0,3)$-forms, and $\Fcal$ is a $(1,1)$-form, hence their exterior product vanishes.

The equations (\ref{3.22}) are BPS-type instanton equations in $D>4$
dimensions. For various classes of {\it closed} forms $\Sigma$ (the
integrable case) they
were introduced and studied in~\cite{14,15,16,17,18}. Many of these equations,
such as the hermitian Yang-Mills equations~\cite{15}, naturally appear in
superstring theory
as the conditions for survival of at least one unbroken supersymmetry
in the low-energy effective field theory in $D \le 4$ dimensions. Some solutions of
the first order gauge field equations (\ref{3.22}) were described e.g. in~\cite{19,20,HILP,HP,BILL}.

\smallskip

\noindent
{\bf Reduction of the Yang-Mills action functional. \ } If $M$ is a
closed manifold, the Yang-Mills equations with torsion (\ref{3.23})
are variational equations for the action~\cite{HP,BILL}
\beq
S=-\frac{1}{4}\, \int_{\X^{d+h}}\, \tr \big(\Fcal\wedge\star\Fcal +
\Fcal \wedge\Fcal\wedge\Sigma\big) \ ,
\label{YMactionSigma}\eeq
where $\tr$ denotes the trace in the fundamental representation of the
gauge group.
Then $S=0$ whenever the gauge field $\Fcal$
satisfies the $\Sigma$-anti-self-duality equation
(\ref{3.22}). Integrating the second term in (\ref{YMactionSigma}) by parts
using (\ref{3.24}), we can write the action
in the form
\bea
S &=& -\frac{1}{4}\, \int_{\X^{d+h}}\, \tr \Big(\Fcal\wedge\star\Fcal +
\big(\Acal\wedge\diff\Acal
+\mbox{$\frac{2}{3}$}\, \Acal\wedge\Acal\wedge\Acal
\big)\wedge\star\Hcal \Big) \nonumber\\ && -\, \frac{1}{4}\, \int_{\X^{d+h}}\, \diff\Big(\tr\big(\Acal\wedge\diff\Acal
+\mbox{$\frac{2}{3}$}\, \Acal\wedge\Acal\wedge\Acal
\big) \wedge\Sigma\Big) \ ,
\label{3.25}\eea
and the second integral in (\ref{3.25}) vanishes by our assumption
that $\partial M=\emptyset$.

We will explicitly work out the dimensional reduction of
(\ref{YMactionSigma})
over $\X^{d+6}=M\times G/H$ to a quiver gauge
theory action on a $\Qbar$-bundle over $M$, in the case where $G/H$ is a
six-dimensional nearly K\"ahler coset space with $\Sigma$ given in (\ref{3.21}). The $G$-equivariant
gauge field $\Fcal$ from (\ref{3.17}) has components
\begin{equation}\label{3.27}
\Fcal_{\m\n}=F_{\m\n}\ , \qquad\Fcal_{\m a}=D_{\m}X_a\und
\Fcal_{ab}=-f^i_{ab}\, \widetilde I_i - f^c_{ab}\, X_c + [X_a,X_b] \ ,
\end{equation}
where the indices $\m ,\n ,\ldots=1,\dots,d$ label components along the
manifold $M$ and $a,b,\ldots=1,\dots,6$ label components along the
internal coset space $G/H$. Substituting (\ref{3.27}) into the 
term involving the Yang-Mills form in (\ref{3.25}), and using the structure constant
identities (\ref{2.50}) together with the equivariance conditions
(\ref{3.13}), we obtain
\bea
S_{\rm YM} &:=& -\frac{1}{4}\, \int_{\X^{d+6}}\, \tr
\big(\Fcal\wedge\star\Fcal \big) \nonumber \\[4pt] &=& -\frac{1}{8}\,
\int_{\X^{d+6}}\,
{\rm vol}_{d+6}\ \tr\Big( \Fcal_{\m\n}\, \Fcal^{\m\n}+2\Fcal_{\m a}\,
  \Fcal^{\m a}
+ \Fcal_{ab}\, \Fcal^{ab}\Big) \label{3.28} \\[4pt]
&=& -\frac{1}{8}\,{\rm Vol}(G/H)\, \int_{M}\, {\rm vol}_{d}\
\tr\Big(F_{\m\n}\, F^{\m\n}+
2D_{\m} X_a \,D^{\m} X_a + f_{iab}\, f_{jab}\, \widetilde I_i\, \widetilde I_j
\nonumber \\ && \qquad \qquad \qquad \qquad \qquad \qquad -\, \mbox{$\frac{1}{3}$}\,
X_a\, X_a-2f_{abc}\,X_a\, [X_b,X_c]+[X_b,X_c]\, [X_b,X_c] \Big) \ , \nonumber
\eea
where ${\rm vol}_{d+6}={\rm vol}_d\wedge e^{123456}$ and ${\rm
  Vol}(G/H)=\int_{G/H}\, e^{123456}$.
To reduce the term involving the Chern-Simons three-form in
(\ref{3.25}), we use (\ref{3.11}) to get
\begin{equation}\label{3.29}
\Acal_\m = A_\m\ \und \Acal_a= e^i_a \, \widetilde I_i + X_a \ ,
\end{equation}
and from the relation $\star\diff\omega=\frac12\, f_{abc}\, e^{abc}$ we obtain
\bea
S_{\Sigma{\rm-CS}} &:=& -\frac{1}{4}\,\int_{\X^{d+6}}\, \tr\big(\Acal\wedge\diff\Acal
+\mbox{$\frac{2}{3}$}\, \Acal\wedge\Acal\wedge\Acal \big)\wedge \star\Hcal \nonumber\\[4pt]
&=& -\frac{1}{24}\,\vk\,{\rm Vol}(G/H)\, \int_{M}\, {\rm vol}_{d}\ f_{abc}\,\tr
\Big(\Acal_{a}\, \Fcal_{bc}-\mbox{$\frac{1}{3}$} \,\Acal_{a}\,
[\Acal_{b},\Acal_{c}]\Big) \label{3.30} \\[4pt]
&=& -\frac{1}{24}\,\vk\,{\rm Vol}(G/H)\, \int_{M}\, {\rm vol}_{d}\
\tr \Big(X_a\, X_{a}-2f_{abc}\, X_{a}\, [X_{b},X_{c}] - f_{iab}\, f_{jab}\, \widetilde I_i\, \widetilde I_j \Big) \ . \nonumber
\eea

Thus the reduction of the total action $S=S_{\rm YM}
+S_{\Sigma{\rm-CS}}$ is given by
\bea
S&=& -\frac{1}{8}\,{\rm Vol}(G/H)\,\int_{M}\, {\rm vol}_{d}\
\tr\Big( F_{\m\n}\, F^{\m\n}+2D_{\m} X_a \, D^{\m} X_a -\mbox{$\frac13$}\, (\vk-3)\, 
f_{iab}\, f_{jab}\, \widetilde I_i\, \widetilde I_j \label{3.31}\\ && \qquad
\qquad \qquad +\, \mbox{$\frac{1}{3}$}\, (\vk -1)\, X_a\, X_a -
\mbox{$\frac{2}{3}$}\, (\vk +3)\, f_{abc}\, X_a\, [X_b,X_c]+
[X_b,X_c]\, [X_b,X_c]\Big) \ .\nonumber
\eea
Note that the value of $\vk\in\R$ is not arbitrary but defined by the
torsion three-form components $\vk\, f_{abc}$
entering into the structure equations on $G/H$. For the nearly
K\"ahler case, from (\ref{2.54}) one
has $\vk =-1$. This explains the choice of multiplier $\frac{1}{3}$
in (\ref{3.21}). Below we also consider more general situations with
$\vk\neq -1$, corresponding to non-canonical metric connections on the
cotangent bundle of $G/H$ with holonomy group larger than $H$. In particular, the choice $\vk=3$
will be singled out by BPS conditions.

\smallskip

\noindent
{\bf Quiver matrix models. \ }
Dimensional reduction of the action (\ref{3.31}) further to a point
truncates the quiver gauge theory to a quiver matrix model in $d=0$
dimensions with action $S=\frac18\, {\rm Vol}(G/H)\,\widehat\Scal_\vk$, which is
naturally associated with a homogeneous vector
bundle $\widehat \Vcal=G\times_H \widehat V$ over $G/H$ corresponding to a $G$-module
decomposition (\ref{3.2})--(\ref{3.7}). 
Upon substituting the equivariant decomposition (\ref{3.14}) for
$X_a$, the action of the matrix model can be regarded as a Higgs potential
\beq
\widehat\Scal_\vk(\phi)=-\tr\Big([X_a,X_b]^2 -
\mbox{$\frac{2}{3}$}\, (\vk +3)\, f_{abc}\, X_a\, [X_b,X_c]
+\mbox{$\frac{1}{3}$}\, (\vk -1)\, X_a^2 - \mbox{$\frac13$}\, (\vk-3)\, 
f_{iab}\, f_{jab}\, \widetilde I_i\, \widetilde I_j \Big)
\label{Wphi}\eeq
in the original quiver gauge theory for the collection of
bifundamental scalar fields
$\phi=\big(\phi_{N_{rs}}\in\Hom(E_s,E_r)\big)$. The critical points of
this potential are determined by solutions of the matrix equations
\beq
\mbox{$\frac16$}\, (\vk-1)\, X_a-\mbox{$\frac12$}\, (\vk+3)\,
f_{abc}\,[X_b,X_c]- \big[X_b\,,\,[X_b,X_a] \big]=0 \ .
\label{Higgscritpts}\eeq

The first term of (\ref{Wphi}) is the standard Yang-Mills matrix model
action from reduction of Yang-Mills gauge theory in flat space. The
Chern-Simons and mass deformations owe to the curvature of
the original homogeneous manifold $G/H$. The last term is a constant
$H$-flux which is due to the topologically non-trivial gauge fields on
$G/H$; its appearence ensures e.g. that there is non-trivial dynamical mass generation from the vacuum state
of the matrix model~\cite{DS}. To compute it explicitly, we normalise the traces
over the carrier spaces $V_{q_r}$ using the quadratic Dynkin indices
$\chi_r$ of the representations so that
\beq
\tr^{~}_{V_{q_r}}\big(I_i^{q_r}\, I_j^{q_r}\big) = -\chi_r\, \delta_{ij} \
,
\label{Dynkinnorm}\eeq
for $r=1,\dots,n$ and $i,j=7,\dots,\dim G$. Then using (\ref{2.50}) the constant
$H$-flux term in the action (\ref{Wphi}) evaluates to
\beq
\mbox{$\frac13$}\, (\vk-3)\, f_{iab}\, f_{jab}\,\tr^{~}_{\widehat V}\big( \widetilde I_i\, \widetilde I_j \big) = - \mbox{$\frac13$}\, (\vk-3)\, 
f_{iab}\, f_{iab}\, \sum_{r=1}^n\, N_r\, \chi_r = - \mbox{$\frac23$}\, (\vk-3)\, \sum_{r=1}^n\,
N_r\, \chi_r \ .
\label{Hfluxcomp}\eeq

For the special value $\vk=3$ (so that $\Sigma=\omega$ in
(\ref{3.21}) and the constant term (\ref{Hfluxcomp}) vanishes), the action (\ref{Wphi}) is non-negative and can be
written in the complex parametrization (\ref{3.20}), with structure
constants (\ref{Cnot0})--(\ref{2.62}), as a sum of
squares
\bea
\widehat\Scal_{\vk=3}(\phi) &=& \tr\Big| \ve_{\bar\delta\ab\bb}\, \big(
-\im\,C_{\a\b}^i\, \widetilde I_i + C_{\a\b}^{\g}\, Y_{\g}+
C_{\a\b}^{\bar\g}\, Y_{\bar\g}-[Y_\a, Y_\b] \big) \Big|^2
\nonumber\\ && +\, \tr\Big|\delta^{\alpha\bar\alpha}\, \big(
-\im\,C_{\a\ab}^i\, \widetilde I_i + C_{\a\ab}^{\b}\, Y_{\b}+
C_{\a\ab}^{\bar\b}\,Y_{\bar\b}-[Y_\a, Y_{\ab}] \big)\Big|^2 \ ,
\label{Sk3}\eea
where we use the matrix notation $|Y|^2:= \frac12\, \big(Y^\dag\, Y+Y\,Y^\dag\big)$.
The vacuum solutions of the quiver matrix model in this case are thus
determined by the equations
\bea
\phi(\widehat r_{\bar\delta}) \ := \ \ve_{\bar\delta\ab\bb}\, \big( [Y_\a, Y_\b]- C_{\a\b}^{\g}\, Y_{\g}-
C_{\a\b}^{\bar\g}\, Y_{\bar\g}
+\im\, C_{\a\b}^i\, \widetilde I_i\big) &=& 0  \ ,
\label{F02d0} \\[4pt] \mu_{\widehat V} \ := \ 
\sum_{\a=1}^3\, \big( [Y_\a, Y_{\ab}]- C_{\a\ab}^{\b}\, Y_{\b}-
C_{\a\ab}^{\bar\b}\,Y_{\bar\b} +\im\, 
C_{\a\ab}^i\, \widetilde I_i\big) &=& 0  \ .
\label{F11d0}\eea
This feature follows
from the more general fact that the $\Sigma$-anti-self-duality equations
(\ref{3.22}) on a nearly K\"ahler six-manifold with $\Sigma =\omega$ are equivalent to the hermitian Yang-Mills
equations~\cite{15,8,HILP}
\begin{equation}\label{4.6}
\Fcal^{0,2}=0 \und \ome\lrc\Fcal =0 \ ,
\end{equation}
whose solutions saturate the absolute minimum value $S=0$ of the action
functional (\ref{YMactionSigma}) in
this case. In the present instance, the curvature
two-form $\Fcal$ is given by (\ref{3.19}) with $F=DY_{\a}=0$.

The system of equations (\ref{F02d0}) expresses the generating set of
relations $\Rbar=\{\widehat r_{\bar\delta}\}$ (plus their $*$-conjugates
$\phi(\widehat r_{\bar\delta}^{\,*}):= \phi(\widehat r_{\bar\delta})^\dag$) in the $\Qbar$-module $(\widehat V,\phi)$ which is
equivalent to pseudo-holomorphicity of the homogeneous vector bundle $\widehat \Vcal\to G/H$, while
(\ref{F11d0}) is the moment map equation for modules in
$\Rep\big(\,\Qbar\,,\,\widehat V\, \big)$ over the deformed preprojective algebra
(\ref{calPlambda}) with deformation vector $\lambda=(\lambda_1,\dots,\lambda_n)$ fixed by the background fluxes
$I^{q_r}_i$ and the structure constants $C^i_{\a\ab}$. The presence of constant terms in these expressions only reflects the non-holonomic nature of the frame $\{\Th^\a,\Th^\ab\}$, i.e. they vanish in a holonomic frame. Moreover, the constant term in (\ref{F02d0}) vanishes for the quasi-K\"ahler case and the constant term in (\ref{F11d0}) vanishes for the nearly K\"ahler case.

\smallskip

\noindent
{\bf Gauge theory on $\mbf{\Qbar\,^{0,1}}$-bundles. \ }
For an $\sut$-invariant connection $\Acal$ on a hermitian quiver
bundle (\ref{4.17}) over the flag manifold $\F_3$, we substitute into
(\ref{3.27})--(\ref{3.31}) the equivariant decompositions (\ref{4.16})
and (\ref{4.18})--(\ref{4.23}) in the nearly K\"ahler limit
$\s_1=\s_2=\s_3=:\s$, with the complex parametrization (\ref{3.20}) and structure
constants (\ref{2.65}). After a bit of calculation, we arrive at the
quiver gauge theory action
\bea
S^{0,1} &=& -\frac12\,{\rm Vol}(\F_3)\, \int_M\, {\rm vol}_d\, \Big( \,
\frac14\, \sum_{s=1}^3\, \tr_{N_s}\big( (F^s)^\dag_{\mu\nu}\,
(F^s)^{\mu\nu}+ \s^2\,
(D_\mu\phi_s)\, (D^\mu\phi_s)^\dag \nonumber\\ && \qquad \qquad \qquad \qquad\qquad +\, \s^2\,
(D_\mu\phi_{s+1})^\dag\,(D^\mu\phi_{s+1}) \big) +
\widehat\Scal^{0,1}_{\s,\vk}(\phi_1,\phi_2,\phi_3) \Big) \ ,
\label{S01action}\eea
where we identify $\phi_s:= \phi_{s\, {\rm mod}\,3}\in
\Hom(E_{s+2},E_{s})$, and
\beq
D\phi_s=\diff\phi_s+A^{s}\, \phi_s-\phi_s\, A^{s+2}
\label{Dphi01}\eeq
are bifundamental covariant derivatives of the Higgs fields. The Higgs
potential is given by
\bea
\widehat\Scal^{0,1}_{\s,\vk}(\phi_1,\phi_2,\phi_3) &=& - \s^4\, \sum_{s=1}^3\,
\tr_{N_s}\Big( \mbox{$\frac{1}4$}\, \big({\bf
  1}_{N_s} -\phi_s\, \phi^\dag_s\, \big)^2 + \mbox{$\frac{1}4$}\, \big({\bf
  1}_{N_s} -\phi_{s+1}^\dag\, \phi_{s+1}\big)^2-4\vk\, \phi_{s+1}^\dag\,
\phi_{s+1} \nonumber \\ && \qquad \qquad +\,
\mbox{$\frac{1}4$}\, \big| \phi_{s+1}-\phi_{s+2}^\dag\,
\phi_{s}^\dag\, \big|^2  +16\vk\,
\big(\phi_{s}\, \phi_{s+2}\, \phi_{s+1} +
\phi_{s+1}^\dag\, \phi_{s+2}^\dag\, \phi_{s}^\dag\, \big) \Big) \nonumber \\ &&
+\, \mbox{$\frac16$}\,\vk\,\s^4\, (N_1+N_2+N_3) \ .
\label{Higgspot01}\eea

In the reduction to a quiver matrix model in $d=0$ dimensions at the
special torsion value $\vk=3$, the action (\ref{Higgspot01}) reads
\beq
\widehat\Scal^{0,1}_{\s,\vk=3}(\phi_1,\phi_2,\phi_3) = \s^4\, \sum_{s=1}^3\,
\tr_{N_s}\Big( 4\big| \phi_s\,
\phi^\dag_s-\phi_{s+1}^\dag\, \phi_{s+1}\big|^2
+\big|\phi_{s+1}-\phi_{s+2}^\dag\, \phi_{s}^\dag\, \big|^2 \Big) \ .
\label{matrix01}\eeq
The corresponding vacuum equations are
\beq
\phi(r_s) := \phi_{s}-\phi_{s+1}^\dag\, \phi_{s+2}^\dag =0 \und \mu_s:= \phi_s\,
\phi_s^\dag-\phi^\dag_{s+1}\, \phi_{s+1} =0
\label{vaceqs01}\eeq
for $s=1,2,3$, which coincide respectively with the relations
(\ref{R01}) (plus their $*$-conjugates (\ref{R01conj})) and the moment
map equations (\ref{muV011})--(\ref{muV013}). The quiver modules in
this case are thus identified with representations of the (undeformed)
preprojective algebra $\Pcal_0^{0,1}$. As we discuss in
section~\ref{vortex}, this is in marked contrast to the quiver
representations associated to the K\"ahler geometry of $\F_3$ that
generically correspond to deformations determined by the background
(monopole) fluxes, which arise from contraction of $\omega$ and $\diff\widehat a_+$. In the nearly K\"ahler case this contraction vanishes because both monopole fields on $\F_3$ satisfy the hermitian Yang-Mills equations.

\smallskip

\noindent
{\bf Gauge theory on $\mbf{\Qbar\,^{k,l}}$-bundles. \ } For any given pair of non-negative integers $(k,l)$, the $Q^{k,l}$ quiver gauge theory can be described using the formulas (\ref{AcalQkl})--(\ref{Qklbifund3}). The general equations are rather lengthy and complicated, and not very informative. We will therefore satisfy ourselves by briefly commenting on the structure of the vacuum equations for the corresponding quiver matrix model, with $\vk=3$ and $\s_1=\s_2=\s_3=:\s$, which follow from (\ref{4.6}), (\ref{Fcalkloffdiag1})--(\ref{Fcalkloffdiag3}), and (\ref{Fcalkldiag}) with constant Higgs fields and vanishing gauge potentials. One then finds a representation of the set of relations $\Rbar\,^{k,l}$ given by
\bea
\phi\big(r_{(q,m)_n}^{1\, (\pm)}\big)&=& \sqrt{\big((n+1\pm1)^2-q^2\big)\, (n\pm
        q+1\pm1)}\, \Big( \phi^{1 \
    (\pm)}_{(q,m)_n}- \phi^{3 \
  (\mp)}_{(q+1,m-3)_{n\pm 1}}{}^\dag\, \phi^{2}_{(q-1,m-3)_{n\pm1}}{}^\dag \Big) \nonumber\\ && - \,
\sqrt{(n+q) \, (n-q+2)\, (n\pm
        q+1\pm 1)}\, \Big( \phi^{1 \
    (\pm)}_{(q,m)_n}- \phi^{2}_{(q-2,m)_{n}}{}^\dag\, \phi^{3 \
  (\mp)}_{(q-1,m-3)_{n\pm 1}}{}^\dag
\Big) \ , \nonumber\\ && \label{r1qmn} \\[4pt]
\phi\big(r_{(q,m)_n}^{2}\big)&=& \sum_\pm\, \bigg[\, \mbox{$\frac{n\mp
    q+1\mp1}{n+1}$}\, \lambda_{k,l}^\pm(n,m)^2 \,
\Big(\phi^{2}_{(q,m)_n}-\phi^{1 \ (\pm)}_{(q+1,m+3)_{n\mp 1}}{}^\dag \, \phi^{3 \
  (\mp)}_{(q+2,m)_{n}}{}^\dag \Big) \nonumber \\ && -\,
\mbox{$\frac{n\pm q+1\pm1}{n+1\pm1}$} \,
\lambda^\mp_{k,l}(n\pm1,m-3)^2 \, \Big( \phi^{2}_{(q,m)_n}- \phi^{3 \
  (\pm)}_{(q+1,m-3)_{n\mp 1}}{}^\dag \, \phi^{1 \
  (\mp)}_{(q+2,m)_{n}}{}^\dag \Big) \, \bigg] \ , \label{r2qmn}\\[4pt]
\phi\big(r_{(q,m)_n}^{3\, (\pm)}\big)&=& \sqrt{(n+q) \, (n-q+2)\, (n\mp
        q+1\mp 1)} \, \Big( \phi^{3 \
    (\pm)}_{(q,m)_n}- \phi^{2}_{(q-2,m)_{n}}{}^\dag \, \phi^{1 \
  (\mp)}_{(q-1,m+3)_{n\pm 1}}{}^\dag \Big) \nonumber\\ && - \,
\sqrt{\big((n+1\pm1)^2-q^2\big)\, (n\mp
        q+1\mp 1)}\, \Big( \phi^{3 \
    (\pm)}_{(q,m)_n}- \phi^{1 \
  (\mp)}_{(q+1,m+3)_{n\pm 1}}{}^\dag\, \phi^{2}_{(q-1,m+3)_{n\pm1}}{}^\dag
\Big) \ , \nonumber\\ &&
\label{r3qmn}\eea
and the moment map
\bea
\mu_{(q,m)_n}&=& \sum_\pm\, \bigg[\, \mbox{$\frac{n\pm
    q+1\pm1}{n+1\pm1}$}\, \lambda_{k,l}^\mp(n\pm 1,m+3)^2\,\phi^{1 \
  (\pm)}_{(q,m)_n} \, \phi^{1 \
  (\pm)}_{(q,m)_n}{}^\dag \label{muqmn} \\ && -\, \mbox{$\frac{n\mp
    q\mp1}{n+1}$}\, \lambda_{k,l}^\mp(n,m)^2\, \phi^{1 \
  (\pm)}_{(q+1,m+3)_{n\mp1}}{}^\dag \, \phi^{1 \
  (\pm)}_{(q+1,m+3)_{n\mp1}} +\mbox{$\frac{n\mp
    q\pm1}{n+1}$}\, \lambda_{k,l}^\pm(n,m)^2\,\phi^{3 \
  (\pm)}_{(q,m)_n} \, \phi^{3 \
  (\pm)}_{(q,m)_n}{}^\dag \nonumber\\ && -\, \mbox{$\frac{n\pm
    q+1\pm1}{n+1\pm1}$}\, \lambda_{k,l}^\pm(n\mp1,m-3)^2\, \phi^{3 \
  (\pm)}_{(q+1,m-3)_{n\mp1}}{}^\dag \, \phi^{3 \
  (\pm)}_{(q+1,m-3)_{n\mp1}} \, \bigg] \nonumber\\ && +\,
(n-q)\, (n+q+2)\, \phi^{2}_{(q,m)_n} \, \phi^{2}_{(q,m)_n}{}^\dag -
(n+q)\, (n-q+2)\, \phi^{2}_{(q-2,m)_n}{}^\dag \,
\phi^{2}_{(q-2,m)_n} \nonumber
\eea
at each vertex $(q,m)_n\in W^{k,l}$ of the weight diagram for the $\sut$-module $\widehat V^{k,l}$. Note that the remaining field strength components (\ref{Fcalkloffdiag13})--(\ref{Fcalkloffdiag23}) play no role in the vacuum sector of the quiver gauge theory.

\section{Double quiver vortex equations\label{vortex}}

\noindent
{\bf Spin(7)-instanton equations. \ } In this section we consider the $\Sigma$-anti-self-dual Yang-Mills
equations (\ref{3.22}) on eight-dimensional manifolds and their reduction to
quiver vortex equations in two dimensions. Consider an oriented riemannian manifold
$(\X^8, g)$ of dimension eight endowed with a four-form $\Sigma$
which defines an almost Spin(7)-structure\footnote{One can omit the word
`almost' if $\Sigma$ is closed~\cite{17,18}.} on $\X^8$.
A Spin(7)-instanton is defined to be a connection $\Acal$ on a complex vector
bundle $\Ecal$ over $\X^8$ whose curvature $\Fcal$ satisfies the
$\Sigma$-anti-self-dual gauge field equations (\ref{3.22}). For more details
see e.g.~\cite{8, 17, 18}. As previously we assume that $\Ecal$ has
degree zero, i.e. $c_1(\Ecal)=\frac{\im}{2\pi}\,\tr(\Fcal)=0$.

Suppose that an (almost) Spin(7)-manifold $\X^8$ allows an almost complex
structure $\widehat\J$ and an SU(4)-structure, i.e. $c_1(\X^8)=0$. Then
there exists a non-degenerate $(4,0)$-form $\Ow$ and a $(1,1)$-form $\ot$ on $\X^8$
such that
\begin{equation}\label{4.1}
\Ow=\Th^1\wedge\Th^2\wedge\Th^3\wedge\Th^4\und
\ot=\sfrac{\im}{2}\,\big(\Th^1\wedge\Th^{\1}+\Th^2\wedge\Th^{\2}+\Th^3\wedge\Th^{\3}
+\Th^4\wedge\Th^{\4}\,\big)\ .
\end{equation}
The one-forms $\Th^{\hat\a}$ are defined by
\begin{equation}\label{4.2}
\widehat\J\,\Th^{\hat\a}=\im\,\Th^{\hat\a} \quad \for {\hat\a}=1,2,3,4\ ,
\end{equation}
i.e. they are $(1,0)$-forms with respect to $\widehat\J$.\footnote{For
an integrable almost complex structure $\widehat\J$ this defines a Calabi-Yau four-fold.}

Given any SU(4)-structure $\big(\,\ot \,,\, \Ow\, \big)$ on an eight-manifold $\X^8$,
there is a compatible Spin(7)-structure determined by
\begin{equation}\label{4.3}
\Sigma = \sfrac{1}{2}\,\ot\wedge\ot - \mbox{Re}\,\Ow\ .
\end{equation}
When the Spin(7)-structure is defined via the four-form (\ref{4.3}), the
Spin(7)-instanton equations (\ref{3.22}) can be reduced via the
inclusion ${\rm SU}(4)\subset{\rm Spin}(7)$ to the equations
\begin{equation}\label{4.4}
\ot\lrc\Fcal =0\und
\Fcal^{0,2}_+=0\ .
\end{equation}
Here we have used the fact that the complex conjugate of $\Ow$ induces an anti-linear
involution $\star_{\Ow}: \bigwedge^{0,2}T^*\X^8 \to\bigwedge^{0,2}T^*\X^8$ on $\X^8$, so that one can
introduce the corresponding self-dual part
\beq
\Fcal^{0,2}_+:=\sfrac{1}{2}\,\big(\Fcal^{0,2}+\star^{}_{\Ow}\,\Fcal^{0,2}
\big)
\label{F+02}\eeq
of $\Fcal^{0,2}$ in the $+1$ eigenspace of $\star_{\Ow}$~\cite{18}.
In the basis of $(1,0)$-forms $\Th^{\hat\a}$ and $(0,1)$-forms
$\Th^{\bar{\hat\a}}:=\overline{\Th^{\hat\a}}$, the equations
(\ref{4.4}) read
\begin{equation}\label{4.5}
\de^{\hat\a\bar{\hat\a}}\, \Fcal_{\hat\a\bar{\hat\a}}=0\und
\Fcal_{\bar{\hat\a}\bar{\hat\b}}=-\sfrac{1}{2}\,
\ve_{\bar{\hat\a}\bar{\hat\b}\bar{\hat\g}\bar{\hat\r}} \,
\Fcal_{{\hat\g}{\hat\r}}\ .
\end{equation}

The Spin(7)-instanton equations (\ref{4.4}) are weaker than the
hermitian Yang-Mills equations
\begin{equation}\label{4.6a}
\ot\lrc\Fcal =0 \und \Fcal^{0,2}=0 \ .
\end{equation}
Any solution of the hermitian Yang-Mills equations is automatically
a solution of the Spin(7)-instanton equations, but not conversely. Recall that~\cite{4} any connection $\Acal$ on the bundle $\Ecal$ which
satisfies (\ref{4.6a}) defines a pseudo-holomorphic structure
$\overline\pa_{\Acal}$ on $\Ecal$. In the case of an integrable almost complex structure
$\widehat\J$, such hermitian Yang-Mills connections $\Acal$ define (semi)stable
holomorphic vector bundles $\Ecal\to \X^8$~\cite{15}.

\smallskip

\noindent
{\bf SU(4)-structures on $G/H\times\R^2$. \ } Consider the manifold $G/H\times\R^2$,
where the reductive six-dimensional coset $G/H$ is endowed with an SU(3)-structure
$(\ome , \Om )$, an almost hermitian metric $g$, and a compatible never integrable almost
complex structure $\J$. The ensuing considerations hold not only on
the plane $M=\R^2$ but also on the cylinder $\R\times S^1$ and on the
torus $T^2=S^1\times S^1$. On the eight-manifold
\begin{equation}\label{4.7}
\X^8=G/H \times\R^2
\end{equation}
we introduce an almost complex structure $\widehat\J =(\J, \mathfrak{j})$,
where $\mathfrak{j}$ is the canonical (integrable) almost complex
structure on the plane $\R^2$ with coordinates $x^7, x^8$, i.e.
$\mathfrak{j}$ is defined so that
\begin{equation}\label{4.8}
\Th^4:=\diff z^4 =\diff x^7 + \im\,\diff x^8
\end{equation}
is a $(1,0)$-form on $\R^2\cong\C$.

On (\ref{4.7}) we introduce non-degenerate forms
\bea\label{4.9}
\Ow \= \Om\wedge\Th^4 &=& \Th^1\wedge\Th^2 \wedge\Th^3 \wedge\Th^4 \ ,
\\[4pt] \label{4.10}
\ot \= \ome +\sfrac{\im}{2}\,\Th^4\wedge\Th^{\4} &=&
\sfrac{\im}{2}\,\big(\Th^1\wedge\Th^{\1} + \Th^2 \wedge\Th^{\2} +
\Th^3\wedge\Th^{\3} + \Th^4\wedge\Th^{\4}\,\big) \ ,
\eea
and the metric
\begin{equation}\label{4.11}
\widehat g=g+\Th^4\otimes \Th^{\4}=\Th^1\otimes \Th^{\1}+\Th^2\otimes \Th^{\2}+\Th^3\otimes \Th^{\3}+
\Th^4\otimes \Th^{\4}\ .
\end{equation}
This makes $G/H\times\R^2$ into an eight-dimensional riemannian manifold with an
SU(4)-structure. Hence on the manifold (\ref{4.7}) one can introduce both
the Spin(7)-instanton equations (\ref{4.4}) as well as the hermitian Yang-Mills
equations (\ref{4.6a}).

\smallskip

\noindent
{\bf Quiver vortex equations on $\R^2$. \ } Let $\Ecal\to G/H \times\R^2$ be a
$G$-equivariant complex vector bundle of rank $N$ and degree zero,
and let $\Acal$ be an $\surmL(N)$-valued
$G$-equivariant connection on $\Ecal$ with curvature $\Fcal$, described
by (\ref{3.11}) and (\ref{3.17}). After substitution of the components from
(\ref{3.19}) into the Spin(7)-instanton equations (\ref{4.5}),
we obtain non-abelian vortex-type equations
\bea\label{4.12}
F_{z\zb}&=&\Big(-\im\,\widetilde C_{\a\ab}^i\, \widetilde I_i +
\widetilde C_{\a\ab}^{\g}\, \widetilde Y_{\g}+
\widetilde C_{\a\ab}^{\bar\g}\, \widetilde Y_{\bar\g}-\big[\,\widetilde
Y_\a \,,\,
\widetilde Y_{\ab}\big]\Big)\, \de^{\a\ab}\ , \\[4pt]
\label{4.13}
\pa_{\zb}\widetilde Y_{\ab}+\big[A_{\zb} \,,\, \widetilde Y_{\ab}\big] &=&
-\mbox{$\frac{1}{2}$}\, \ve_{\ab\bb\bar\g}\,\Big(-\im\,\widetilde
C_{\b\g}^i\, \widetilde I_i +
\widetilde C_{\b\g}^{\de}\, \widetilde Y_{\de}+
\widetilde C_{\b\g}^{\bar\de}\, \widetilde Y_{\bar\de}-\big[\,
\widetilde Y_\b \,,\,\widetilde Y_{\g}\big]\Big)\ ,
\eea
where we use a rescaled basis $\widetilde\Th^\a$ and the structure constants
(\ref{Ctildenot0}). Here $z:=z^4=x^7+\im\,x^8$. Note that the
right-hand side of (\ref{4.12}) is just the moment map of
(\ref{F11d0}) after rescaling, while the right-hand side of
(\ref{4.13}) coincides with the relations in (\ref{F02d0}).

In the example $G/H=\F_3=\sut/\uo\times \uo$, we substitute the explicit values
(\ref{2.65}) of
$\widetilde C^A_{BC}$ into (\ref{4.12})--(\ref{4.13}) to get
\bea\label{4.14}
F_{z\zb}&=& \mbox{$\frac{\im}{4\sqrt{3}}$} \,\big(\s_1^2+\s_2^2- 2\s_3^2 \big)\,\widetilde I_7
-\mbox{$\frac{\im}{4}$}\, \big(\s_1^2-\s_2^2\big)\,\widetilde I_8 -
\de^{\a\ab} \,
\big[\, \widetilde Y_\a \,,\,\widetilde Y_{\ab}\big]\ , \\[4pt]
\label{4.15}
\pa_{\zb}\widetilde Y_{\ab}+\big[A_{\zb}\,,\, \widetilde Y_{\ab}\big] &=&
-\mbox{$\frac{1}{2}$}\,\ve_{\ab\bb\bar\g}\,
\Big(\mbox{$\frac{2}{\sqrt{3}}$} \,\ve_{\b\g\de}\,
\mbox{$\frac{\s_\b \, \s_\g}{\s_{\de}}$} \,\widetilde Y_{\bar\de}+\big[\,
\widetilde Y_\b \,,\,\widetilde Y_{\g}\big] \Big) \ .
\eea
In the nearly K\"ahler limit $\s_1=\s_2=\s_3=:\s$, these equations
reduce to
\bea\label{4.14a}
F_{z\zb} &=& - \de^{\a\ab}\, [Y_\a, Y_{\ab}]\ , \\[4pt] \label{4.15a}
\pa_{\zb}Y_{\ab}+[A_{\zb} , Y_{\ab}] &=&
-\mbox{$\frac{1}{2}$}
\,\ve_{\ab\bb\bar\g}\,\big(\, \mbox{$\frac{2\s}{\sqrt{3}}$} \,\ve_{\b\g\de}\,
Y_{\bar\de}+[Y_\b, Y_{\g}] \big)\ .
\eea
It is instructive to compare these equations with those which arise 
when using a K\"ahler structure on $\F_3$; in that case one should
instead substitute the structure
constants (\ref{2.61a})--(\ref{2.61b}) into (\ref{4.12})--(\ref{4.13}) and use unscaled Higgs fields $Y_{\a}$. Then for the vortex equations
in the K\"ahler case we obtain
\bea\label{4.14b}
F_{z\zb} &=& \mbox{$\frac{\im\, \sqrt{3}}{4}$}\,\widetilde I_7-
\de^{\a\ab}\, [Y_\a, Y_{\ab}]\ , \\[4pt] \label{4.15b}
\pa_{\zb}Y_{\ab}+[A_{\zb}, Y_{\ab}] &=& 0\ , \\[4pt] \label{4.15c}
\mbox{$\frac{1}{\sqrt{6}}$} \,Y_{3}+[Y_1, Y_2] &=& 0 \ .
\eea
Here we have taken into account that for the K\"ahler case there is no
SU(4)-structure on the manifold $\F_3\times\R^2$, and therefore one can
impose only the hermitian Yang-Mills equations (\ref{4.6a}) on the gauge fields.
This results in separate holomorphicity equations (\ref{4.15b}) and
holomorphic relations (\ref{4.15c}). For any concrete choice of quiver, gauge
potential (\ref{3.8}), and matrices (\ref{3.10}) and (\ref{3.14}), one can write
all of these equations explicitly in terms of $A^r$ and $\p_{N_{rs}}$. 

\smallskip

\noindent
{\bf $\mbf{\Qbar\,^{0,1}}$-vortex equations. \ } For the special case
(\ref{4.16})--(\ref{4.23}), the non-abelian coupled vortex equations (\ref{4.12})--(\ref{4.13}) are
\begin{eqnarray}
F^{1}_{z\bar z}&=&\big(\s_1^2-\s_2^2\big)\, {\bf 1}_{N_1}-\s_1^2\, \p_1\, \p_1^\+
+\s_2^2\, \p_2^\+\, \p_2\ ,\label{4.24}\\[4pt]
F^{2}_{z\bar z}&=&\big(\s_2^2-\s_3^2\big)\, {\bf 1}_{N_2} -\s_2^2\, \p_2\, \p_2^\+
+\s^2_3\, \p_3^\+\, \p_3\ ,\label{4.25}\\[4pt]
F^{3}_{z\bar z}&=&\big(\s_3^2-\s_1^2\big)\, {\bf 1}_{N_3} -\s_3^2\, \p_3\, \p_3^\+
+\s_1^2\, \p_1^\+\, \p_1\ ,\label{4.26}\\[4pt]
\pa_{\bar z}\phi_{1}+A^{1}_{\bar z}\,\phi_{1}-\phi_{1}\,A^{3}_{\bar z}&=&
\mbox{$\frac{\s_2\,\s_3}{\s_1}$} \,\big(\p_1-\p_2^\+\,\p_3^\+\, \big)\ ,\label{4.27}\\[4pt]
\pa_{\bar z}\phi_{2}+A^{2}_{\bar z}\,\phi_{2}-\phi_{2}\,A^{1}_{\bar z}&=&
\mbox{$\frac{\s_1\,\s_3}{\s_2}$} \,\big(\p_2-\p_3^\+\,\p_1^\+\, \big)\ ,\label{4.28}\\[4pt]
\pa_{\bar z}\phi_{3}+A^{3}_{\bar z}\,\phi_{3}-\phi_{3}\,A^{2}_{\bar z}&=&
\mbox{$\frac{\s_1\,\s_2}{\s_3}$} \,\big(\p_3-\p_1^\+\,\p_2^\+\, \big)\ .\label{4.29}
\end{eqnarray}
Let us denote by $\overline\pa_A$ the Dolbeault operator of the vector
bundle $E^{0,1}$ over $M=\R^2$, acting on $\p_r$ in
(\ref{4.27})--(\ref{4.29}). From these equations we see that the Higgs fields
$\p_r$, defining homomorphisms between the bundles $E_r\to M$ for $r=1,2,3$, are {\it not}
holomorphic. In fact, $\overline\pa_A\p_s$ is proportional to the quiver relation
$\phi(r_s)$ defined in (\ref{vaceqs01}), i.e. one has
\begin{equation}\label{4.31}
\overline\pa_A\p_s = \phi(r_s)\, \kappa_s \ ,
\end{equation}
with $\kappa_1=\s_1^{-1}\, \s_2\, \s_3\, \diff\zb$, $\kappa_2=\s_1\,
\s_2^{-1}\, \s_3\, \diff\zb$,
and $\kappa_3=\s_1\, \s_2\, \s_3^{-1}\, \diff\zb$. This reflects the fact that the almost
complex structure $\widehat\J$ on $\X^8= \F_3\times \R^2$ is not
integrable, the bundle $\Ecal\to \X^8$ is not holomorphic, and therefore
$\Fcal^{0,2}$ does not vanish. Of course, for the SU(3)-equivariant
gauge connection
$\Acal$ on $\Ecal$ one can impose the hermitian Yang-Mills equations (\ref{4.6a})
which are stronger than the Spin(7)-instanton equations.
Then one obtains the quiver vortex equations with relations
\begin{equation}\label{4.32}
\overline\pa_A\p_s = 0\und \phi(r_s) =0
\end{equation}
plus the equations (\ref{4.24})--(\ref{4.26}). However, there exist solutions to
(\ref{4.24})--(\ref{4.29}) which do not reduce to solutions of
(\ref{4.24})--(\ref{4.26}) and (\ref{4.32})~\cite{HP}.

In the nearly K\"ahler case we have
$\s_1=\s_2=\s_3=:\s$ and the equations (\ref{4.24})--(\ref{4.29}) become
\begin{eqnarray}
F^{1}_{z\bar z}&=&-\s^2\,\big(\p_1\,\p_1^\+ -\p_2^\+\,\p_2\big)\ ,\label{4.24a}\\[4pt]
F^{2}_{z\bar z}&=&-\s^2\,\big(\p_2\,\p_2^\+-\p_3^\+\,\p_3\big)\ ,\label{4.25a}\\[4pt]
F^{3}_{z\bar z}&=&-\s^2\,\big(\p_3\,\p_3^\+-\p_1^\+\, \p_1\big)\ ,\label{4.26a}\\[4pt]
\pa_{\bar z}\phi_{1}+A^{1}_{\bar z}\,\phi_{1}-\phi_{1}\,A^{3}_{\bar z}&=&
\s\,\big(\p_1-\p_2^\+\,\p_3^\+\, \big)\ ,\label{4.27a}\\[4pt]
\pa_{\bar z}\phi_{2}+A^{2}_{\bar z}\,\phi_{2}-\phi_{2}\,A^{1}_{\bar z}&=&
\s\,\big(\p_2-\p_3^\+\,\p_1^\+\, \big)\ ,\label{4.28a}\\[4pt]
\pa_{\bar z}\phi_{3}+A^{3}_{\bar z}\,\phi_{3}-\phi_{3}\,A^{2}_{\bar z}&=&
\s\,\big(\p_3-\p_1^\+\,\p_2^\+\, \big)\ .\label{4.29a}
\end{eqnarray}
For comparison, we note that in the K\"ahler case the vortex equations
determined by the structure constants (\ref{2.61a}) and the hermitian
Yang-Mills equations are given by
\begin{eqnarray}
F^{1}_{z\bar z}&=&-\p_1\, \p_1^\+
+\p_2^\+\, \p_2\ ,\label{4.24ak}\\[4pt]
F^{2}_{z\bar z}&=&-\mbox{$\frac18$}\, {\bf 1}_{N_2} - \p_2\, \p_2^\+
+\p_3^\+\, \p_3\ ,\label{4.25k}\\[4pt]
F^{3}_{z\bar z}&=&\mbox{$\frac18$}\, {\bf 1}_{N_3} -\p_3\, \p_3^\+
+\p_1^\+\, \p_1\ ,\label{4.26k}\\[4pt]
\pa_{\bar z}\phi_{1}+A^{1}_{\bar z}\,\phi_{1}-\phi_{1}\,A^{3}_{\bar z}&=&
0 \ ,\label{4.27k}\\[4pt]
\pa_{\bar z}\phi_{2}+A^{2}_{\bar z}\,\phi_{2}-\phi_{2}\,A^{1}_{\bar z}&=&
0 \ ,\label{4.28k}\\[4pt]
\pa_{\bar z}\phi_{3}+A^{3}_{\bar z}\,\phi_{3}-\phi_{3}\,A^{2}_{\bar z}&=&
0 \ , \label{4.29k}\\[4pt]
\p_3-\sqrt6\, \p_2\,\p_1 &=&0 \ . \label{4.29kk}
\end{eqnarray}
This system, called a holomorphic triangle in~\cite{LPS3}, is the basic building block for all quiver vortices associated to the K\"ahler geometry of the flag variety $\F_3$.

\smallskip

\noindent
{\bf $\mbf{\Qbar\,^{k,l}}$-vortex equations. \ } In the nearly K\"ahler limit of a general $\Qbar\,^{k,l}$-bundle over $M=\R^2$, the vortex equations are obtained by combining the moment map (\ref{muqmn}) with the natural moment map on the invariant symplectic space of unitary connections on the vector bundle $E_{(q,m)_n}\to M$, and also the anti-holomorphic components of the bifundamental covariant derivatives (\ref{Qklbifund1})--(\ref{Qklbifund3}) with the relations (\ref{r1qmn})--(\ref{r3qmn}). Using the Spin(7)-instanton equations (\ref{4.5}) and the field strength components (\ref{Fcalkldiag})--(\ref{Fcalkloffdiag3}), one thus finds
\bea
F_{z\zb}^{(q,m)_n}&=& \mbox{$\frac{\s^2}{24}$} \, \mu_{(q,m)_n} \ , \label{Fzqmn} \\[4pt]
D_{\zb}\phi^{1 \
    (\pm)}_{(q,m)_n} &=& \mbox{$\frac{\s}{\lambda_{k,l}^\pm(n\pm1,m-3)}$}\, \sqrt{\mbox{$\frac{24(n+1\pm1)}{n\pm q+1\pm1}$}}\, \phi\big(r_{(q,m)_n}^{1\, (\pm)}\big) \ , \label{Dz1qmn} \\[4pt]
D_{\zb} \phi^{2}_{(q,m)_n} &=& \mbox{$\frac{\s}{\sqrt{12(n+q)\, (n+q+2)}}$}\, \phi\big(r^{2}_{(q,m)_n}\big) \ , \label{Dz2qmn} \\[4pt]
D_{\zb} \phi^{3 \
    (\pm)}_{(q,m)_n} &=& \mbox{$\frac{\s}{\lambda_{k,l}^\pm(n,m)}$}\, \sqrt{\mbox{$\frac{24(n+1)}{n\mp q+1\pm1}$}}\, \phi\big(r^{3 \
    (\pm)}_{(q,m)_n} \big)
\label{Dz3qmn}\eea
at each vertex $(q,m)_n\in W^{k,l}$. As in
(\ref{r1qmn})--(\ref{r3qmn}), the right-hand sides of the equations
(\ref{Dz1qmn})--(\ref{Dz3qmn}) all vanish in the case that the
hermitian Yang-Mills equations (\ref{4.6a}) are satisfied. Then the
double quiver relations $\Rbar\,^{k,l}$ are automatically fulfilled,
and all Higgs fields are holomorphic. These equations can be compared
with the non-abelian coupled quiver vortex equations corresponding to
the K\"ahler geometry of $\F_3$~\cite[eqs.~(4.19)--(4.23)]{LPS3}.

\section*{Acknowledments}

We would like to thank Olaf Lechtenfeld for fruitful discussions. A.D.P. also
wishes to thank the Institute for Theoretical Physics of Hannover University,
where this work was completed, for hospitality. The work of A.D.P. was partially
supported by the Deutsche Forschungsgemeinschaft (grant 436 RUS 113/995) and the
Russian Foundation for Basic Research (grant 09-02-91347). The work of R.J.S.
was supported in part by grant ST/G000514/1 ``String Theory Scotland'' from the UK Science and Technology
Facilities Council.

\bigskip
%\newpage

\end{document}